\documentclass[a4paper,11pt]{article}
\pdfoutput=1 % if your are submitting a pdflatex (i.e. if you have
\usepackage{jheppub} % for details on the use of the package, please
\usepackage[T1]{fontenc} % if needed

\usepackage[utf8]{inputenc}
\usepackage{amssymb}
\usepackage{amsmath}
\usepackage{amsfonts}
\usepackage{graphicx}
\usepackage{color}
\usepackage{xspace}
\usepackage{comment}
\usepackage{hyperref}
\usepackage[normalem]{ulem}
\usepackage[section]{placeins}
\usepackage{afterpage}
\usepackage{slashed}
\usepackage{enumerate}
\usepackage{enumitem}
\usepackage{caption}
\usepackage{subfigure}
\usepackage{float}
\usepackage{ulem}
\usepackage{verbatim}
\usepackage{multirow,rotating}
\usepackage[dvipsnames]{xcolor}
\usepackage{amsfonts}
\usepackage{amssymb}
\usepackage{amsmath}

\newcommand{\be}{\begin{equation}}
\newcommand{\ee}{\end{equation}}
\newcommand{\bea}{\begin{eqnarray}}
\newcommand{\eea}{\end{eqnarray}}

\newcommand{\met}{\slashed{E}_{\rm T}}

\newcommand{\iab}{\rm ab^{-1}}
\newcommand{\ifb}{\rm fb^{-1}}
\newcommand{\mltp}{{\mkern-2mu\times\mkern-2mu}}

\title{\boldmath Sensitivities to New Resonance Couplings to $W$-Bosons at the LHC}

\author[a,1]{Ying-nan Mao\note{Corresponding author.}}
\author[a]{, Kechen Wang}
\author[a,2]{and Yiheng Xiong\note{Corresponding author.}}

\affiliation[a]{Department of Physics, School of Physics and Mechanics, Wuhan University of Technology, 430070 Wuhan, Hubei, China}

\emailAdd{ynmao@whut.edu.cn}
\emailAdd{kechen.wang@whut.edu.cn}
\emailAdd{yiheng@whut.edu.cn}

\abstract{
We propose a search strategy at the HL-LHC for a new neutral particle \( X \) that couples to \( W \)-bosons, using the process \( p p \to W^{\pm} X (\to W^{+} W^{-}) \) with a tri-\( W \)-boson final state. Focusing on events with two same-sign leptonic \( W \)-boson decays into muons and a hadronically decaying \( W \)-boson, our method leverages the enhanced signal-to-background discrimination achieved through a machine-learning-based multivariate analysis. Using the heavy photophobic axion-like particle (ALP) as a benchmark, we evaluate the discovery  sensitivities on both production cross section times branching ratio \( \sigma(p p \to W^{\pm} X) \times \text{Br}(X \to W^{+} W^{-}) \) and the coupling \( g_{aWW} \) for the particle mass over a wide range of 170–3000 GeV at the HL-LHC with center-of-mass energy \( \sqrt{s} = 14 \, \text{TeV} \) and integrated luminosity \( \mathcal{L} = 3 \, \text{ab}^{-1} \). Our results show significant improvements in discovery sensitivity, particularly for masses above 300 GeV, compared to existing limits derived from CMS analyses of
Standard Model (SM) tri-\( W \)-boson production at \( \sqrt{s} = 13 \, \text{TeV} \). This study demonstrates the potential of advanced selection techniques in probing the coupling of new particles to \( W \)-bosons and highlights the HL-LHC's capability to explore the physics beyond the SM.
}

\begin{document}
\maketitle
%\生成目录
\flushbottom

%%%%%%%%%%%%%%%%%%%%%%%%%%%%%%%%%%%%%%%%%%%%%%%%%%%%%%%%%%%%%%%%%%%%%
%%%%%%%%%%%%%%%%%%%%%%%%%%%%%%%%%%%%%%%%%%%%%%%%%%%%%%%%%%%%%%%%%%%%%
\section{Introduction}
\label{sec:Intro}

The search for new particles beyond the Standard Model (SM) is a cornerstone of high-energy physics, driven by unresolved questions such as the nature of dark matter, the origin of the neutrino mass, the baryogenesis, and the dark energy. Among the various extensions to the SM, new particles coupling to \( W \)-bosons represent a particularly intriguing avenue for exploration.
In physics beyond the SM (BSM), there exist various types of neutral particles \( X \) which couple to \( W \)-bosons. Examples include:
(i) extended gauge models with additional gauge bosons (\( Z' \))~\cite{Langacker:2008yv}, which couples to \( W^+W^- \) via mixing with SM gauge bosons $\gamma$ or $Z$;
(ii) extended scalar sectors like the Two-Higgs-Doublet Model (2HDM)~\cite{Branco:2011iw} and supersymmetric models~\cite{Djouadi:2005gj}, where the new CP-even scalar couple to \( W^+W^- \) via the mixing with the SM Higgs boson, and it may also be related to some recent hints at LHC \cite{Bhattacharya:2023lmu};
(iii) pseudo-scalars, such as axions~\cite{Peccei:1977hh,Peccei:1977ur,Weinberg:1977ma, Wilczek:1977pj,Kim:2008hd} or axion-like particles (ALPs) \( a \), which couple to \( W^+W^- \) via a dimension-five operator \( a W^+_{\mu\nu}\widetilde{W}^{-,\mu\nu} \).
Besides, new neutral particles coupling to \( W \)-bosons also emerge in theoretical frameworks such as composite Higgs models \cite{Panico:2015jxa}, extra-dimensional scenarios \cite{Csaki:2000zn}, and dark matter models \cite{Kahlhoefer:2017dnp}.

Since the new neutral particle \( X \) couples to di-\( W \)-bosons, it can be produced and decay via the \( X \)-\( W \)-\( W \) vertex at the Large Hadron Collider (LHC). In this study, we consider \( X \) produced in association with a \( W \)-boson through the \( s \)-channel exchange of a \( W^{(*)} \) boson, followed by its decay into a \( W^+W^- \) pair, resulting in a tri-\( W \)-boson final state. The mass of \( X \) is assumed to be greater than 170 GeV, ensuring its di-\( W \)-boson decay products are both on-shell. We choose the heavy photophobic ALP \cite{Craig:2018kne} as a benchmark model in this work.

Axion models were originally proposed to address the strong CP problem in QCD~\cite{Dine:2000cj, Kim:2008hd}. In these models, the axion arises as a pseudo-Nambu-Goldstone boson from a spontaneously broken U(1) symmetry at a high energy scale, a mechanism known as the Peccei-Quinn mechanism~\cite{Peccei:1977hh,Peccei:1977ur}.
In such original models~\cite{Peccei:1977hh,Peccei:1977ur,Weinberg:1977ma, Wilczek:1977pj, Kim:1979if, Shifman:1979if, Zhitnitsky:1980tq, Dine:1981rt, Hook:2019qoh, Quevillon:2019zrd, DiLuzio:2020wdo}, couplings of the axion is strictly related to its mass $m_a$.
For the extended concept of ALPs, the mass and couplings of ALPs are treated as independent parameters. This flexibility significantly broadens the ALP parameter space, making them promising candidates for astrophysical and collider-based searches~\cite{Galanti:2022ijh,Choi:2020rgn,Qiu:2024muo}. At colliders, ALPs are typically studied through their interactions with SM particles~\cite{Bauer:2017ris, Dolan:2017osp, Bauer:2018uxu, Zhang:2021sio, dEnterria:2021ljz, Agrawal:2021dbo, Tian:2022rsi, Ghebretinsaea:2022djg, Antel:2023hkf, Biswas:2023ksj, Lu:2024fxs}.
Among these interactions, previous experiments have primarily focused on the ALP's coupling to diphoton, \( g_{a\gamma\gamma} \).
The constraints on \( g_{a\gamma\gamma} \) are stringent across most of the ALP mass range, with limits coming from astrophysical phenomena for \( m_a \lesssim \mathcal{O}(\text{MeV}) \)~\cite{ParticleDataGroup:2024cfk,ALPlimits}, low-energy \( e^-e^+ \) collisions for \( 0.1 \, \text{GeV} \lesssim m_a \lesssim 10 \, \text{GeV} \)~\cite{Belle-II:2020jti,BESIII:2022rzz,Jiang:2023lnw}, Pb-Pb collisions at the LHC for \( 10 \, \text{GeV} \lesssim m_a \lesssim 100 \, \text{GeV} \)~\cite{CMS:2018erd,ATLAS:2020hii}, and $p$-$p$ collisions at the LHC for \( 100 \, \text{GeV} \lesssim m_a \lesssim 2 \, \text{TeV} \)~\cite{dEnterria:2021ljz}. Consequently, the ALP-photon coupling is expected to be small, motivating the study of photophobic ALPs~\cite{Craig:2018kne}.
Photophobic ALPs are characterized by suppressed couplings to diphoton, with their primary interactions occurring with other SM electroweak bosons.

Previous studies~\cite{Craig:2018kne,Bonilla:2022pxu,Aiko:2024xiv} on heavy photophobic ALPs at the LHC have derived limits based on tri-$W$-boson final state by reinterpreting experimental analyses at center-of-mass energy \( \sqrt{s} = 8 \) or 13 TeV, with limited luminosities and mass ranges. Ref.~\cite{Craig:2018kne} derives
constraints on ALP masses in the range of $40-500$ GeV by reinterpreting triboson searches in the SM framework at \( \sqrt{s} = 8 \, \text{TeV} \), where the ATLAS analyses of \( pp \to W^\pm(\to\mu^\pm \nu) \, W^\pm(\to\mu^\pm \nu) \, W^\mp(\to jj) \) with \( \mathcal{L} =\) 20.3 fb\(^{-1}\)~\cite{ATLAS:2016jeu} were utilized as a ALP-mediated signal process \( p p \to  W^\pm \, a (\to W^\pm W^\mp ) \).
Ref.~\cite{Bonilla:2022pxu} reinterprets the CMS analyses of the SM process \(pp \to jj\, W^\pm(\to \ell^\pm \nu)\, W^\pm(\to \ell^\pm \nu)\) at \(\sqrt{s} = 13\) TeV~\cite{CMS:2020gfh} as a non-resonant ALP-mediated vector boson scattering process, with the ALP serving as an off-shell mediator, and the results reveal discovery sensitivities for a photophobic ALP with a mass below 100 GeV at the High-Luminosity LHC (HL-LHC) with \( \sqrt{s} = 14 \) TeV and \( \mathcal{L} = 3 \) ab\(^{-1}\).
Ref.~\cite{Aiko:2024xiv} reinterpreted the CMS analyses for SM process \( pp \to W^\pm W^\pm W^\mp \) at \( \sqrt{s} = 13 \) TeV with \( \mathcal{L} =\) 35.9 fb\(^{-1}\)~\cite{CMS:2019mpq} as a ALP-mediated signal process \( pp \to W^\pm\, a(\to W^\pm W^\mp) \), setting limits on ALP mass up to 1 TeV.

More recently, some of our authors conducted detailed analyses of the \( \gamma Z \) decay mode for heavy photophobic ALPs at the HL-LHC with \( \sqrt{s} = 14 \) TeV~\cite{Ding:2024djo}. In that study, ALPs are produced with two jets through \( s \)-channel vector boson exchange and vector boson fusion, resulting in the process \( pp \to jj\, a (\to \gamma\, Z (\to \ell^+ \ell^-)) \), where \( \ell = e, \mu \). Machine-learning-based multivariate analyses were used to optimize background rejection. Discovery sensitivities for the ALP coupling to di-\( W \)-bosons, \( g_{aWW} \), were evaluated over masses from 100 to 4000 GeV with \( \mathcal{L} = 3 \) ab\(^{-1} \) and 140 fb\(^{-1} \). Other studies on heavy photophobic ALPs at the LHC are also reviewed in Ref.~\cite{Ding:2024djo}.
Besides, heavy photophobic ALPs have also been studied through a global fit analysis of electroweak precision observables~\cite{Aiko:2023trb}.

We note that since the referred experimental analyses were designed to target SM production processes, rather than the  heavy photophobic ALPs, discrepancies in signal kinematics and background characteristics frequently result in conservative or potentially unreliable discovery sensitivities for previous reinterpreting studies. In this study, we accomplish detailed analyses and derive discovery sensitivities at the HL-LHC for a new heavy neutral particle \( X \) that couples to \( W \)-bosons, using the process \( p p \to W^{\pm} X (\to W^{+} W^{-}) \) with a tri-\( W \)-boson final state.
Compared with previous reinterpreting studies based on tri-$W$-boson final state, key advancements include:
(i) performing simulations for signal process of \( p p \to W^{\pm} X (\to W^{+} W^{-}) \) and related background processes to fully capture signal and background kinematics at the detector-level;
(ii) using machine learning-based multivariate analysis (MVA) for optimal signal-to-background discrimination;
(iii) presenting results over a wide mass range $170-3000$ GeV at \( \sqrt{s} = 14 \, \text{TeV} \) with \( \mathcal{L} = 3 \, \text{ab}^{-1} \);
(iv) providing the model-independent discovery sensitivities for the production cross section times branching ratio, \( \sigma(p p \to W^{\pm} X) \times \text{Br}(X \to W^{+} W^{-}) \), which can be applied to general models involving a new heavy neutral particle \( X \) that couples to \( W \)-bosons.

This paper is organized as follows. In Sec.~\ref{sec:SigProd}, we introduce the signal process under study. Sec.~\ref{sec:SM_BKG} outlines the main SM backgrounds relevant to our signal. In Sec.~\ref{sec:Data_Sim}  and Sec.~\ref{sec:Sear_Stra}, we describe the simulation setup and search strategy, respectively. Our results are presented in Sec.~\ref{sec:Results}, and we conclude in Sec.~\ref{sec:Conc}. Additional details supporting the main text are provided in the appendices.

\section{The Signal Production}
\label{sec:SigProd}

\begin{figure}[h]
\centering
\includegraphics[width=10cm, height=6cm]{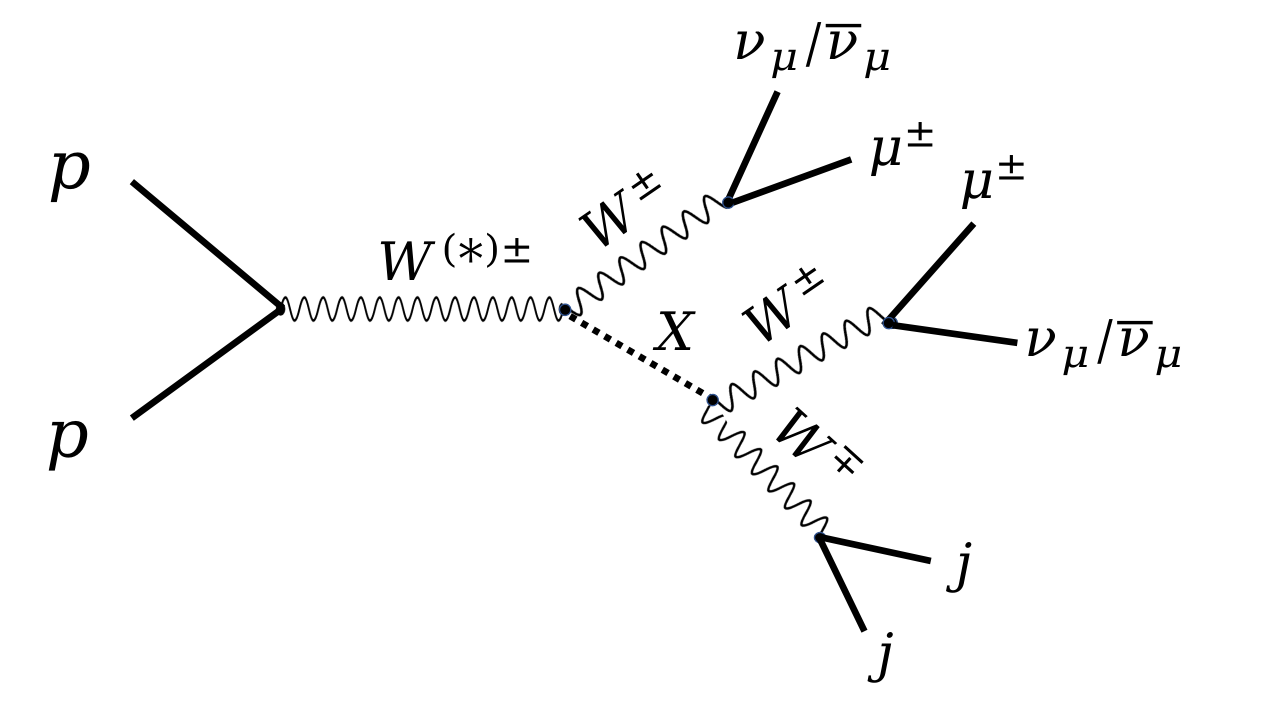}
\caption{
Production and decay of new particle $X$ coupling to di-$W$ boson at $pp$ colliders, leading to the tri-$W$ final state.
The same-charged $W^{\pm}$ bosons are considered to decay into $\mu^{\pm}$ and $\nu_{\mu}$ (or $\overline{\nu}_{\mu}$), while the remaining $W^\mp$ boson decays into di-jet.
}
\label{fig:Feyn_Diag}
\end{figure}

As shown in Fig.~\ref{fig:Feyn_Diag},
we consider the production of $X$ in association with a $W$-boson at the HL-LHC with center-of-mass energy $\sqrt{s}$ = 14 TeV.
$X$ then decays to di-$W$ boson.
The corresponding signal process is
$p p \to {W^{\pm}} \, X (\to W^{+} W^{-})$,
leading to the tri-$W$-boson final state.
To eliminate the background effectively, two same charged $W$-bosons are required to decay leptonically into muons, leading to the existence of same-sign muons in the final state.
The remanent $W$-boson decays hadronically to di-jet which has larger branching ratio and can increase the signal production cross section.

To accomplish a concrete study, in this work, we take the ALP theory model as an example. The neutral particle $X$ is assumed to be the heavy photophobic ALP which couples to electroweak gauge bosons only. The effective linear ALP Lagrangian is~\cite{Brivio:2017ije}
\begin{equation}
\begin{aligned}
\mathcal{L}_{\rm eff}
\supset \,\frac{1}{2}\,(\partial_{\mu}a)\,(\partial^{\mu}a) - \,\frac{1}{2}\,m^2_{a}\,a^2 - \, c_{\tilde{W}}\,\frac{a}{f_{a}}\,W^b_{\mu\nu}\,\widetilde{W}^{b,\mu\nu}\, - \, c_{\tilde{B}}\,\frac{a}{f_{a}}\,B_{\mu\nu}\,\widetilde{B}^{\mu\nu}\, ,
\label{eqn:Eff_Lag}
\end{aligned}
\end{equation}
where $W^b_{\mu\nu}$ and $B_{\mu\nu}$ represent the field strength tensors of the $\textrm{SU}(2)_L$ and $\textrm{U}(1)_Y$ gauge groups, and $\widetilde{O}^{\mu\nu} \equiv \frac{1}{2}\epsilon^{\mu\nu\alpha\beta} O_{\alpha\beta}$ $(O = W,B)$ is the dual field strength tensor.
$m_a$ and $f_a$ denote the mass of ALP and its decay constant~\footnote{In ALP models, $f_a$ always affects on physical observables together with the coefficients $c_i$ [such as the coefficients $c_{\tilde{W},\tilde{B}}$ in Eq.~(\ref{eqn:Eff_Lag})], thus people can only set constraints on the combinations $c_i/f_a$. As a new physics scale, $f_a$ is expected to be much higher than the electro-weak scale, thus we expect $f_a\gtrsim\mathcal{O}(\textrm{TeV})$.}, respectively, and they are assumed to be independent parameters in this work.

After electroweak symmetry breaking, the interactions between ALP and gauge bosons are generally expressed as
\begin{equation}
	\begin{aligned}
		{\mathcal L}_{\rm eff}
		\supset &\,-\frac{1}{4}\,g_{a\gamma\gamma}\,a\,F_{\mu\nu}\,\widetilde{F}^{\mu\nu}
		- \frac{1}{2}\,g_{a{\gamma}Z}\,a\,Z_{\mu\nu}\,\widetilde{F}^{\mu\nu} \\
		&- \frac{1}{4}\,g_{aZZ}\,a\,Z_{\mu\nu}\,\widetilde{Z}^{\mu\nu}
		- \frac{1}{2}\,g_{_{aWW}}\,a\,W^+_{\mu\nu}\,\widetilde{W}^{-,\mu\nu},
		\label{eqn:el_Lag}
	\end{aligned}
\end{equation}
where $F_{\mu\nu}$, $Z_{\mu\nu}$ and $W^+_{\mu\nu}$ are the field strength tensors of the electromagnetic field , $Z$- and $W$-fields. In addition, the coupling constants are expressed as
\begin{equation}
	\begin{aligned}
		 g_{a\gamma\gamma} &= \frac{4}{f_a}\,(s^2_{\theta}\,c_{\tilde{W}}\,+c^2_{\theta}\,c_{\tilde{B}}),\\
		 g_{aZZ} &= \frac{4}{f_a}\,(c^2_{\theta}\,c_{\tilde{W}}\,+s^2_{\theta}\,c_{\tilde{B}}),\\
		 g_{a{\gamma}Z} &= \frac{2\,s_{2\theta}}{f_a}\,(c_{\tilde{W}} - c_{\tilde{B}}),\\
		 g_{_{aWW}} &= \frac{4}{f_a}\,c_{\tilde{W}},
		\label{eqn:ayy_couling}
	\end{aligned}
\end{equation}
where $c_{\tilde{W}}$ and $c_{\tilde{B}}$ are the same constants in Eq.~(\ref{eqn:Eff_Lag}), and $c_{\theta} = \cos{\theta}_W$, $s_{\theta} = \sin{\theta}_W$, $s_{2\theta} = \sin2{\theta}_W$ are all trigonometric functions of
the Weinberg angle ${\theta}_W$.
In this work, focusing on the photophobic ALP scenario, we assume the coupling between ALP and di-photon $g_{a\gamma\gamma} = \frac{4}{f_a}\,(s^2_{\theta}\,c_{\tilde{W}}\,+c^2_{\theta}\,c_{\tilde{B}}) = 0$.
As a consequence, $s^2_{\theta}\,c_{\tilde{W}}\,+c^2_{\theta}\,c_{\tilde{B}} = 0$, making a constant proportional relationship between $c_{\tilde{W}}$ and $c_{\tilde{B}}$. Thus, $g_{a{\gamma}Z}$ and $g_{aZZ}$ can be expressed as functions of $g_{_{aWW}}$:
\begin{equation}
g_{a\gamma Z} = t_\theta\, g_{aWW},\quad\quad\textrm{and}\quad\quad g_{aZZ} = (1 - t_\theta^2)\, g_{aWW},
\end{equation}
where $t_\theta = \tan \theta_W$.

\section{The SM Background}
\label{sec:SM_BKG}

Since the signal final state contains two same-sign muons plus di-jet and moderate missing energy, relevant SM processes which can mimic the signal are listed as follows,
\begin{enumerate}[label*=(\roman*)]
\item one-boson production with two jets: $W^\pm jj$ and $Z jj$;
\item two-boson production with two jets: $W^\pm W^\pm jj$, $W^+ W^-jj$, $W^\pm Zjj$ and $ZZjj$;
%\item Two-boson production with only one jet, $W^\pm Z j$.
\item three-boson production: $W^\pm W^\pm W^\mp$;
\item
top quark pair production:
$t\bar{t}$.
\end{enumerate}

After $W$-boson decays, the processes $W^\pm W^\pm jj$ and $W^\pm W^\pm W^\mp$ can result in the same final state as the signal and serve as irreducible backgrounds, despite their relatively low production cross sections. The process $W^{\pm}jj$, which has a significantly higher production cross section, also contributes to the background due to the potential misidentification of muons. Additionally, charge mis-measurement of final state muons can lead to backgrounds from processes like $Zjj$, $W^+ W^- jj$, and $t \bar{t}$. Furthermore, the $W^{\pm} Z jj$ process becomes relevant when one of the muons in the final state remains undetected, and the $ZZjj$ process can contribute both through missed muons and mis-measured muon charges.

\section{Event Simulation}
\label{sec:Data_Sim}

We firstly use MadGraph5\_aMC$@$NLO program with version 2.6.7~\cite{Alwall:2014hca} to simulate the proton-proton collision events at the parton level where the \textsc{nnpdf23} parton distribution function (PDF)~\cite{Ball:2012cx} of the proton is utilized.
At the parton level, to produce data as close to the experimental results as possible,
following relatively loose thresholds are applied for both the signal and background processes:
(a) the minimal transverse momentum of jets, photons and charged leptons is set to 0.5 GeV, i.e. $p_{T}(j/\gamma/l^{\pm})>$ 0.5 GeV;
(b) the maximal pseudorapidity of jets is set to 10, i.e. $|\eta(j)| < 10$, and it is set to 5 for the photons and charged leptons, i.e. $|\eta(\gamma/l^{\pm})| < 5$;
(c) the solid angular difference $\Delta R=\sqrt{{\Delta\eta}^2+{\Delta\phi}^2}$ between objects is set to 0.1\,.
These thresholds criteria are set in the ``run\_card.dat'' file of the MadGraph program.

All signal and background events are then passed through \textsc{pythia} program with version 8.2~\cite{Sjostrand:2014zea} for parton showering, hadronization and decay of unstable particles.
Besides, we use the Delphes program with version 3.4.2~\cite{deFavereau:2013fsa} to perform the detector simulation where the ATLAS detector configuration is adopted
and the minimal transverse momenta of leptons and photons accepted by the detector is set to be 2 GeV.
Due to limited computing resources, when simulating background processes of $W^\pm jj$, $Z jj$ and $W^\pm W^\mp jj$, decay mode of one gauge boson is fixed to decay to the muon, i.e. $W^\pm (\to \mu^\pm \nu_\mu)\, jj$, $Z(\to \mu^+\mu^-)\, jj$ and $W^\pm(\to \mu^\pm \nu_\mu)\, W^\mp jj$, respectively,
so that these processes can still have enough number of events after all selection criteria.

\begin{figure}[h]
\centering
\includegraphics[width=12cm, height=8cm]{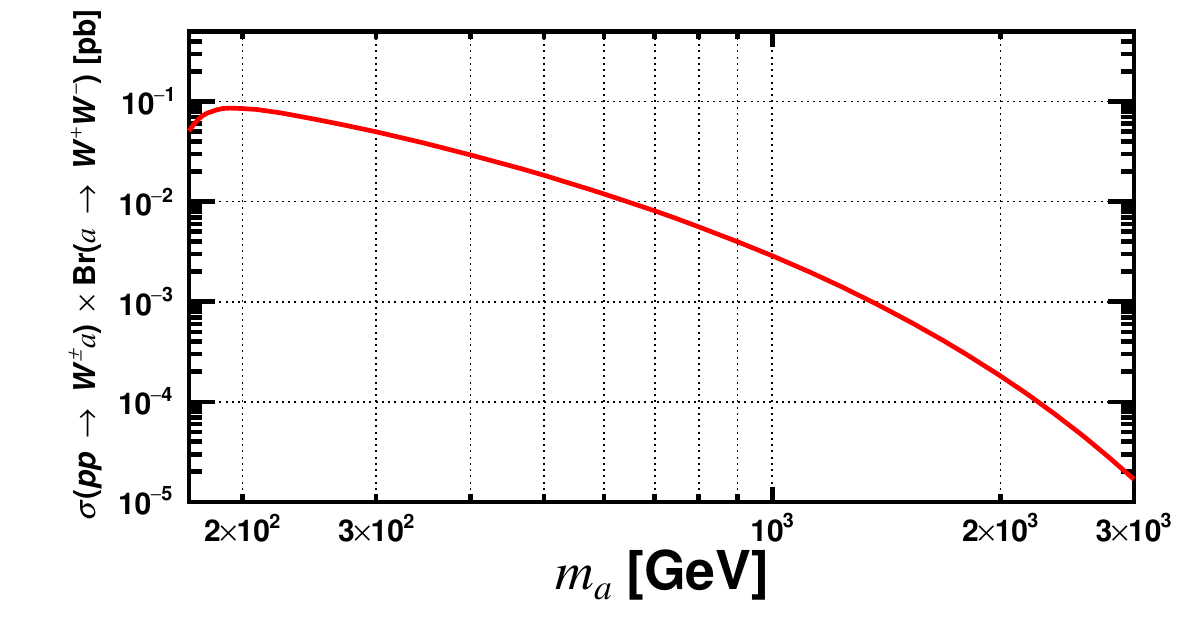}
\caption{
Signal production cross section $\sigma(p \, p \,\rightarrow {W^{\pm}} \, a)$
multiplied by the branching ratio
${\rm Br}(a \to W^+W^-)$
as a function of ALP mass $m_{_{a}}$ from 170 GeV to 3000 GeV at the HL-LHC with $\sqrt{s} = 14$ TeV, assuming the coupling $g_{_{aWW}} =$ 1 TeV$^{-1}$.
}
\label{fig:Cross_vs_ma}
\end{figure}

For the signal, we apply
the ALP model file with the linear Lagrangian~\cite{Brivio:2017ije} in the Universal FeynRules Output (UFO) format~\cite{Degrande:2011ua} into the MadGraph5 program
and generate the $p \, p \,\rightarrow {W^{\pm}} \, a(\rightarrow W^+ W^-)$ events.
A scan of the ALP mass, $m_{a}$, is done in the following fashion: individual mass 170 GeV, 25 GeV increments in the mass range 200-300 GeV, 100 GeV increments in the mass range 300-600 GeV, individual mass 750GeV, 800GeV, 900 GeV,  and 400 GeV increments in the mass range 1000-3000 GeV. For each ALP mass, the coupling
$g_{_{aWW}} $  is set to
1 TeV$^{-1}$.
For each iteration of ALP mass $m_a$, at least $10^6$ events have been generated to ensure that statistical uncertainties are minimized as much as possible within the bounds of computational resources.

To ensure consistency in our analysis, we use the production cross sections calculated by MadGraph5 program to estimate the event yields for both signal and background processes. Figure~\ref{fig:Cross_vs_ma} illustrates the signal production cross section \(\sigma(p \, p \,\rightarrow W^{\pm} \, a)\), multiplied by the branching ratio \({\rm Br}(a \to W^+W^-)\), as a function of the ALP mass \(m_a\) in the range of 170 GeV to 3000 GeV at the HL-LHC with \(\sqrt{s} = 14~\mathrm{TeV}\), where the coupling \(g_{aWW}\) is fixed to \(1~\mathrm{TeV}^{-1}\).
As \(m_a\) increases beyond 160 GeV, the phase space for the decay mode \(a \to W^+W^-\) expands rapidly, leading to a sharp rise in the branching ratio, which reaches approximately 50\% at \( \sim 190~\mathrm{GeV}\)~\cite{Ding:2024djo}. Beyond this point, the branching ratio increases more gradually, with only minor changes at higher masses.
At the same time, production cross section for process \(p \, p \,\rightarrow W^{\pm} \, a\) decreases steadily with increasing \(m_a\). Consequently, the production cross section times branching ratio peaks when \(m_a \sim 190~\mathrm{GeV}\), just above the \(WW\) decay threshold of 160 GeV.

\section{Search Strategy}
\label{sec:Sear_Stra}

Following preselection criteria have been employed before a multivariate analysis is done, where final state muons and jets are ordered based on their transverse momenta and labeled as \(O_i\) (\(i = 1, 2, \ldots\)), with \(O = \mu, \, j\), respectively.
\begin{enumerate}[label*=(\arabic*)]
\item
Events are required to have exactly two muons,
i.e. $N (\mu) =$ 2.
\item The minimal $p_{_{\rm T}}$ of the two muons is 10 GeV, i.e. $p_{_{\rm T}} (\mu) > 10$ GeV.
\item The two muons have the same charge, i.e. $\mu^\pm \mu^\pm$.

\item The minimal $p_{_{\rm T}}$ of all jets is 30 GeV,
i.e. $p_{_{\rm T}}(j) > 30$ GeV.

\item
Event are required to have at least two jets, i.e. $N(j) \geq$ 2, and the number of $b-$tagged jets are required to be zero, i.e. $N(j_b) = 0$.

 \label{last-item}

\end{enumerate}

\begin{table}[h]
\centering
\begin{tabular}{ccccccc}
\hline
\hline
& initial  & (1)-(2) &  (3) & (4)-(5) \\ 	
\hline
$W^\pm a(\rightarrow W^\pm W^\mp)$ & $5.51 \mltp 10^{4}$ & $1.27 \mltp 10^{3}$ & $4.26 \mltp 10^{2}$ & $2.54 \mltp 10^{2}$  \\
\hline
$W^{\pm}(\rightarrow\mu^{\pm}\nu_{\mu}) jj$ & $3.62 \mltp 10^{11}$ & $7.24 \mltp 10^{6}$ & $2.69 \mltp 10^{6}$ & $4.89 \mltp 10^{5}$ \\
$Z(\rightarrow\mu^{\pm}\mu^{\mp})jj$ & $3.31 \mltp 10^{10}$ & $1.13 \mltp 10^{10}$ & $3.33 \mltp 10^{5}$ & $6.67 \mltp 10^{4}$  \\
$W^\pm W^\pm jj$ & $1.10 \mltp 10^{6}$ & $5.22 \mltp 10^{3}$ & $5.19 \mltp 10^{3}$ & $3.88 \mltp 10^{3}$  \\
$W^\pm(\rightarrow\mu^{\pm}\nu_{\mu}) W^\mp jj$ & $5.05 \mltp 10^{8}$ & $2.43 \mltp 10^{7}$ & $8.81 \mltp 10^{3}$ & $4.53 \mltp 10^{3}$  \\
$W^\pm Zjj$ & $1.03 \mltp 10^{9}$ & $1.42 \mltp 10^{7}$ & $6.17 \mltp 10^{5}$ & $1.99 \mltp 10^{5}$ \\
$ZZjj$ & $3.10 \mltp 10^{8}$ & $8.42 \mltp 10^{6}$ & $4.32 \mltp 10^{4}$ & $1.55 \mltp 10^{4}$ \\
$W^\pm W^\pm W^\mp$ & $3.79 \mltp 10^{5}$ & $5.92 \mltp 10^{3}$ & $1.73 \mltp 10^{3}$ & $8.55 \mltp 10^{2}$ \\
$t\bar{t}$ & $1.79 \mltp 10^{9}$ & $1.22 \mltp 10^{7}$ & $8.71 \mltp 10^{4}$ & $2.91 \mltp 10^{4}$ \\
\hline
\hline
\end{tabular}
\caption{
Expected number of events for the signal with benchmark mass $m_a$ = 500 GeV and coupling $g_{_{aWW}} =$ 1 TeV$^{-1}$ and background processes after applying preselection criteria (1)-(5) sequentially at the HL-LHC with $\sqrt{s} =$ 14 TeV and $\mathcal{L} =$ 3 $\iab$.
}
\label{tab:Crsc}
\end{table}

Table~\ref{tab:Crsc} presents the expected number of events, $N_{\rm exp}$, for the signal, with a benchmark mass of \(m_a = 500~\mathrm{GeV}\) and coupling \(g_{aWW} = 1~\mathrm{TeV}^{-1}\), as well as for the background processes.
Here, $N_{\rm exp}$ is calculated
by the formula,
\begin{equation}
N_{\rm exp} = \sigma_{\rm pro} \times \mathcal{L} \times \epsilon_{\rm pre}
\label{eqn:expected_eventsnumber},
\end{equation}
where $\sigma_{\rm pro}$ is the production cross section of the signal or background process;
$\mathcal{L}$ is the integrated luminosity;
and $\epsilon_{\rm pre}$ is the preselection efficiency which is evaluated based on our analyses.
The numbers are shown after applying the preselection criteria (1)-(5) sequentially at the HL-LHC, with \(\sqrt{s} = 14~\mathrm{TeV}\) and \(\mathcal{L} = 3~\mathrm{ab}^{-1}\).
As shown in Table~\ref{tab:Crsc}, after applying all preselection criteria, the number of background events for most processes is reduced by at least \(10^4\)-fold. Nevertheless, the total background rate remains much larger than the expected number of signal events.

After applying the preselection criteria, the Toolkit for Multivariate Analysis (TMVA) package~\cite{TMVA:2007ngy} is employed to carry out a multivariate analysis (MVA), which enhances the ability to distinguish between signal and background events. The events that pass the preselection steps are then subjected to analysis using the Boosted Decision Trees (BDT) algorithm within the TMVA framework. In this process, the discrimination between signal and background is achieved by utilizing the following set of kinematic observables.

\begin{enumerate}[label=(\alph*)]

\item
Missing transverse energy and its azimuth angle: $\met$, $\phi ({\met})$.

\item
The $x$, $y$ and $z$-component of momentum ($p_x$, $p_y$, $p_z$) and energy ($E$) of the first two leading jets and muons:
$p_x (j_1)$, $p_y (j_1)$, $p_z (j_1)$, $E (j_1)$; $p_x (j_2)$, $p_y (j_2)$, $p_z (j_2)$, $E (j_2)$;
$p_x (\mu_1)$, $p_y (\mu_1)$, $p_z (\mu_1)$, $E (\mu_1)$; $p_x (\mu_2)$, $p_y (\mu_2)$, $p_z (\mu_2)$, $E (\mu_2)$.

\item
The number of charged tracks($N_{\rm track}$) and
the ratio ($R_{\rm E}$) of the hadronic versus electromagnetic energy deposited in the calorimeter cells
for the first two leading jets: $N_{\rm track}(j_1)$, $N_{\rm track}(j_2)$; $R_{\rm E}(j_1)$, $R_{\rm E}(j_2)$.
$R_{\rm E}$ is typically greater than one for a jet.

\item
The solid angular separation $\Delta R$
between $j_1$ and $j_2$, the invariant mass ($m$) of system of ($j_1+j_2$), and the number ($N$) of jets:
$\Delta R(j_1,\, j_2)$, $m(j_{1}+j_{2})$, $N(j)$.

\item
Observables related to the isolation quality of muons.
(i) The summed $p_{\rm T}$ of other objects excluding the muon in a $R = 0.4$ cone around the muon:
$p^{\rm iso}_{\rm T}(\mu_1)$, $p^{\rm iso}_{\rm T}(\mu_2)$, and the bigger of the two $p^{\rm iso}_{\rm T,\,{\rm max}}(\mu)$.
(ii) The ratio of the transverse energy in a $3 \times 3$ grid surrounding the muon to the $p_{_{\rm T}}$ of the muon (the ``etrat'' in Ref.~\cite{lhcoFormat}), which is a percentage between 0 and 0.99:
$R_{\rm grid}(\mu_1)$, $R_{\rm grid}(\mu_2)$.
(iii) We make all possible combination between $\mu_1 / \mu_2$ and every jet, compare the $\Delta R$ and find the minimal value:
$\Delta R_{\rm min}(\mu,\, j)$.
For well-isolated muons, $p^{\rm iso}_{\rm T}$, $R_{\rm grid}$ should be small, while $\Delta R_{\rm min}(\mu,\, j)$  is large.

\item
To reconstruct the hadronically decaying $W$-boson ($jj_{_W}$),
we make all di-jet combinations and select the pair with the invariant mass closest to 80 GeV. The corresponding two jets are labeled as $j_{_{W1}}$ and $j_{_{W2}}$ sorted by their $p_{_{\rm T}}$.
We input the following observables of these two jets: $p_x (j_{_{W1}})$, $p_y(j_{_{W1}})$, $p_z(j_{_{W1}})$, $E(j_{_{W1}})$;
$p_x (j_{_{W2}})$, $p_y(j_{_{W2}})$, $p_z(j_{_{W2}})$, $E(j_{_{W2}})$;
$N_{\rm track}(j_{_{W1}})$, $N_{\rm track}(j_{_{W2}})$; $R_{\rm E}(j_{_{W1}})$, $R_{\rm E}(j_{_{W2}})$;
$\Delta R(j_{_{W1}},\, j_{_{W2}})$, $m(j_{_{W1}}+j_{_{W2}})$.

\item
The transverse mass $m_{\rm T}$~\cite{Han:2005mu} of the system, that include one of muons and the missing transverse energy: $m_{\rm T}(\mu_{1}+\met)$, $m_{\rm T}(\mu_{2}+\met)$.
Here, the transverse mass $m_{\rm T} \equiv \sqrt{ (E_{\rm T}^{\rm vis.}+\met)^2 - ( \vec{p}_{\rm T}^{\rm \,\,vis.} + \vec{\slashed{p}}_{\rm T} )^2 }$\,, where $E_{\rm T}^{\rm vis.}$ ($\vec{p}_{\rm T}^{\rm \,\,vis.}$) is the transverse energy (momentum) of the visible object or system, while $\met$ ($\vec{\slashed{p}}_{\rm T}$) is the missing transverse energy (momentum). $E_{\rm T}^{\rm vis.} = \sqrt{ (\vec{p}_{\rm T}^{\rm \,\,vis.})^2 + (m^{\rm vis.})^2 }$\,, where $m^{\rm vis.}$ is the invariant mass of the visible object or system. $\met = |\vec{\slashed{p}}_{\rm T}|$, assuming the invariant mass of the invisible object is zero.

\item
We calculate and compare $\Delta R$ values of two combinations between the $jj_{_W}$ and $\mu_1 / \mu_2$. The muon with smaller (bigger) $\Delta R$ value is labeled as $\mu_a$ ($\mu_b$).
We input the minimal value $\Delta R(j_{_{W1}}+j_{_{W2}},\, \mu_a)$ and maximal value $\Delta R(j_{_{W1}}+j_{_{W2}},\, \mu_b)$.

\item
The observables related to the reconstruction of the ALP mass: $m(j_{_{W1}}+j_{_{W2}}+\mu_a)$, $m(j_{_{W1}}+j_{_{W2}}+\mu_b)$; $m_{\rm T}(j_{_{W1}}+j_{_{W2}}+\mu_a+\met)$, $m_{\rm T}(j_{_{W1}}+j_{_{W2}}+\mu_b+\met)$.

\item
The ovservables related to the reconstruction of the off-shell $W^{(*)\pm}$ boson:
$m_{\rm T}(j_{_{W1}}+j_{_{W2}}+\mu_a+\mu_b+\met)$.

\end{enumerate}

Details of the observables $N_{\rm track}(j_{i})$, $R_E(j_{i})$, $R_{\rm grid}(\mu_i)$ and $p^{\rm iso}_{\rm T}(\mu)$ can be found in~\cite{lhcoFormat}.
The observables \(N_{\text{track}}(j_{i})\), \(R_E(j_{i})\), \(R_{\text{grid}}(\mu_i)\), and \(p^{\text{iso}}_{\text{T}}(\mu)\) are described in detail.
The hadronic \(W\)-boson (\(jj_{_W}\)) is reconstructed using the method described above.
Appendix~\ref{app:observables} presents distributions of kinematical observables after applying preselection criteria for signal and background processes at the HL-LHC with \(\sqrt{s} = 14~\textrm{TeV}\).
%kinematical observables for \(j_{_{W1}}\) and \(j_{_{W2}}\), after preselection for signal and background processes, assuming \(m_{a} = 400~\textrm{GeV}\) in Fig.~\ref{fig:jjW};
%observables related to the ALP mass and off-shell \(W^{(*)\pm}\) boson reconstruction, after preselection for signal and background processes, assuming \(m_{a} = 400~\textrm{GeV}\) in Fig.~\ref{fig:ALPandWstar};
%observables for ALP mass and off-shell \(W^{(*)\pm}\) boson reconstruction, after preselection for the signal only, for various \(m_{a}\) values in Fig.~\ref{fig:ALPandWstarMasses}.
%
%The distributions of \( m( j_{_{W1}} + j_{_{W2}} + \mu_a) \), \( m(j_{_{W1}} + j_{_{W2}} + \mu_b) \), \( m_{\text{T}}(j_{_{W1}} + j_{_{W2}} + \mu_a + \met ) \), and \( m_{\text{T}}(j_{_{W1}} + j_{_{W2}} + \mu_b + \met) \), as shown in Fig.~\ref{fig:ALPandWstarMasses}.
Among them,
for \( m_a \) values below 500 GeV, the endpoint positions of the distributions for both \( m(j_{_{W1}} + j_{_{W2}} + \mu_a) \) and \( m_{\rm T}(j_{_{W1}} + j_{_{W2}} + \mu_a + \met) \) align closely with \( m_a \);
when \( m_a \) exceeds 500 GeV, the endpoint positions of the \( m_{\rm T}(j_{_{W1}} + j_{_{W2}} + \mu_a + \met) \) distribution are still related to \( m_a \), but this phenomenon becomes less obvious.
%Among them, when \( m_a \) is less than 500 GeV, positions of the end edge of distributions of \( m( j_{_{W1}} + j_{_{W2}} + \mu_a) \) and \( m_{\text{T}}(j_{_{W1}} + j_{_{W2}} + \mu_a + \met ) \) are generally around \( m_a \), while when \( m_a \) is above 500 GeV, positions of the end edge of distributions of \( m_{\text{T}}(j_{_{W1}} + j_{_{W2}} + \mu_a + \met ) \) are more related to \( m_a \).
Appendix~\ref{app:MVAdetails} provides detailed information about the MVA analyses performed using the TMVA package~\cite{TMVA:2007ngy}.

\begin{figure}[h]
\centering
\includegraphics[width=7.3cm, height=4.5cm]{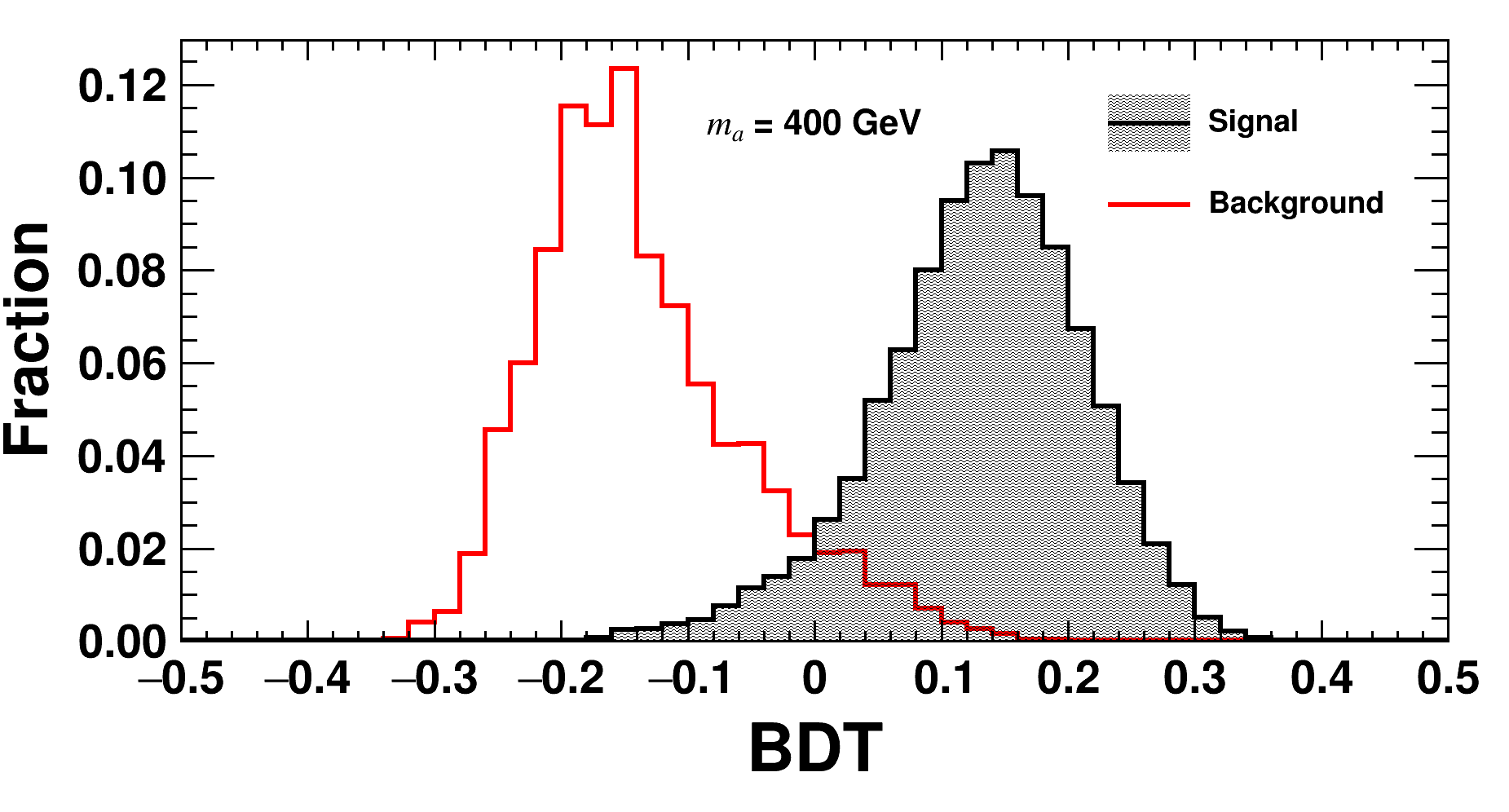}\,\,\,\,\,
\includegraphics[width=7.3cm, height=4.5cm]{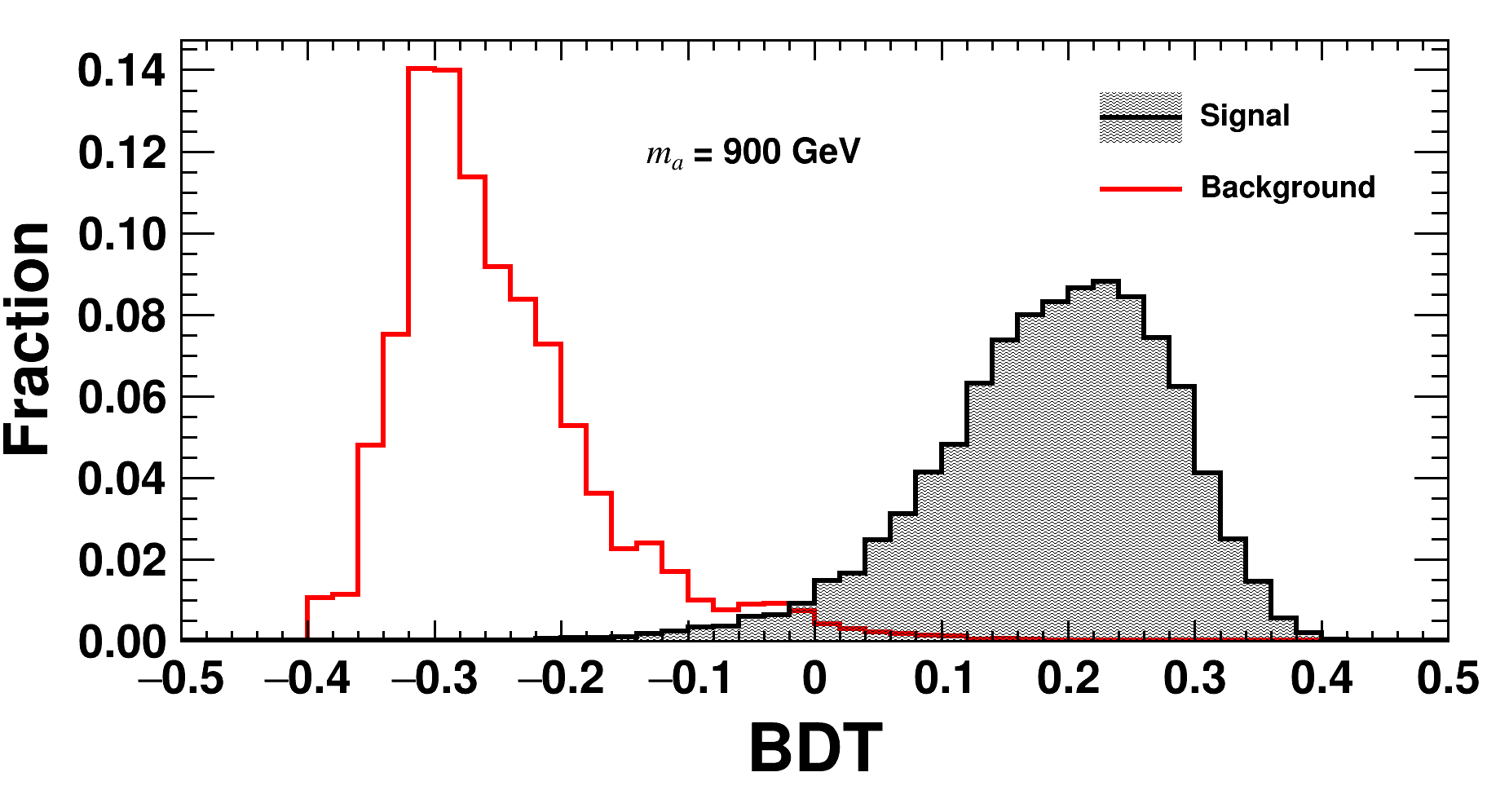}
\caption{
BDT response distributions for total SM background and the signal when ALP mass $m_a$ = 400 GeV (left) and 900 GeV (right) at the HL-LHC with $\sqrt{s} =$ 14 TeV.
}
\label{fig:BDT_400_900}
\end{figure}

Figure~\ref{fig:BDT_400_900} presents the BDT response distributions for the total background and the signal at the HL-LHC, with benchmark masses of \(m_a = 400~\text{GeV}\) and \(m_a = 900~\text{GeV}\). The left and right plots show similar trends in the distributions for both mass points. The clear separation between the signal and background distributions indicates that the BDT criteria are effective for background rejection at the HL-LHC. Additionally, the separation is more pronounced for \(m_a = 900~\text{GeV}\) compared to \(m_a = 400~\text{GeV}\), suggesting improved discriminating power at higher masses. Appendix~\ref{app:BDT} presents distributions of BDT responses after applying preselection criteria for the signal and background processes at the HL-LHC with $\sqrt{s} =$ 14 TeV, assuming various $m_a$ cases.

After the preselection, the BDT cut is optimized according to the signal statistical significance calculated by Eq.~(\ref{eqn:Stat_Sgf}) for each mass case.
\begin{equation}
\sigma_{\rm stat} = \sqrt{2 \left[\left(N_s+N_b\right) {\rm ln}\left(1+\frac{N_s}{N_b}\right) - N_s \right] },
\label{eqn:Stat_Sgf}
\end{equation}
where $N_s$ and $N_b$ are the expected numbers of events for signal and total background after applying both the preselection and BDT criteria.
Appendix~\ref{app:Sel_Eff} shows selection efficiencies of preselection and BDT criteria for signal and background processes at the HL-LHC with $\sqrt{s} =$ 14 TeV assuming different ALP masses, where ``$-$" means the number of events can be reduced to be negligible with $\mathcal{L} =$ 3 ab$^{-1}$.

\section{Results}
\label{sec:Results}

\begin{figure}[h]
\centering
\includegraphics[width=12cm, height=8cm]{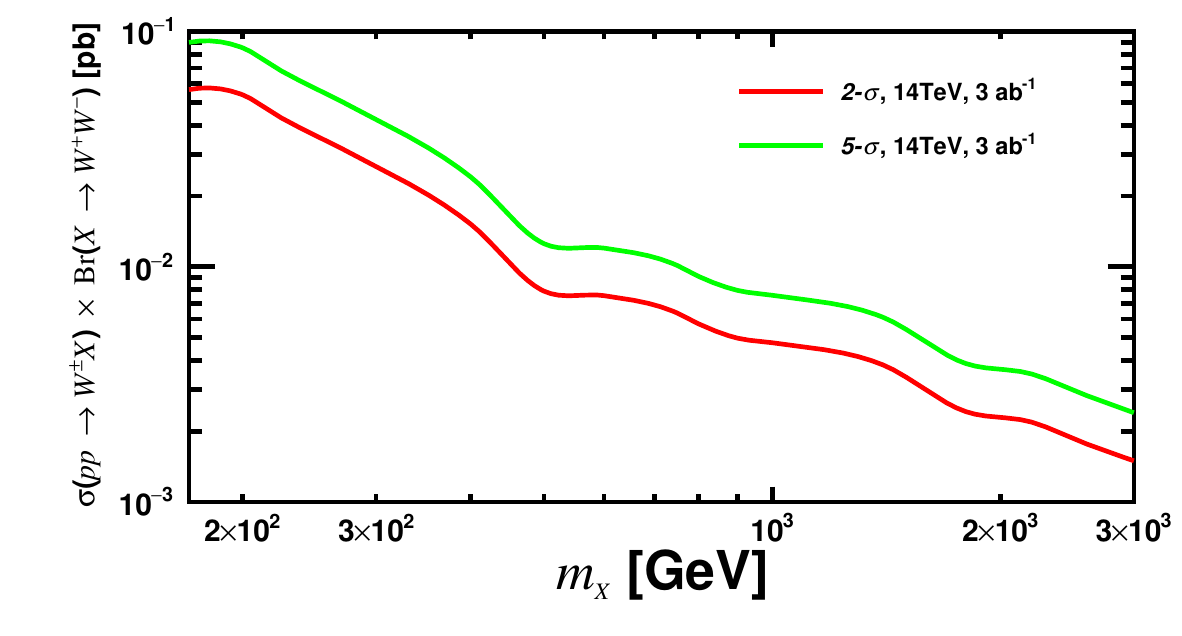}
\caption{
Discovery sensitivities on the production cross section $\sigma(p p \to W^{\pm} \, X)$ times branching ratio Br$(X \to W^{+} W^{-})$
in the mass range of 170 - 3000 GeV at the HL-LHC with $\sqrt{s} =$ 14 TeV and $\mathcal{L} =$ 3 $\iab$.
Red and green curves correspond to 2-$\sigma$ and 5-$\sigma$ significances, respectively.
}
\label{fig:sensSigmaBr}
\end{figure}

Using our search strategy, in Fig.~\ref{fig:sensSigmaBr}, we present the discovery sensitivities for the production cross section \( \sigma(p p \to W^{\pm} X) \) multiplied by the branching ratio \( \text{Br}(X \to W^{+} W^{-}) \) as a function of \( m_X \) in the mass range of 170–3000 GeV at the HL-LHC, with \( \sqrt{s} = 14~\text{TeV} \) and an integrated luminosity of \( \mathcal{L} = 3~\text{ab}^{-1} \). The red and green curves correspond to the 2-\(\sigma\) and 5-\(\sigma\) significances, respectively.
As shown in the BDT distributions in Appendix~\ref{app:BDT}, the separation between the signal and background improves significantly for larger masses, leading to more effective background rejection. Consequently, the sensitivities for heavier masses do not decrease very rapidly, demonstrating strong discovery potential across a wide mass range.

\begin{figure}[h]
\centering
\includegraphics[width=12cm, height=8cm]{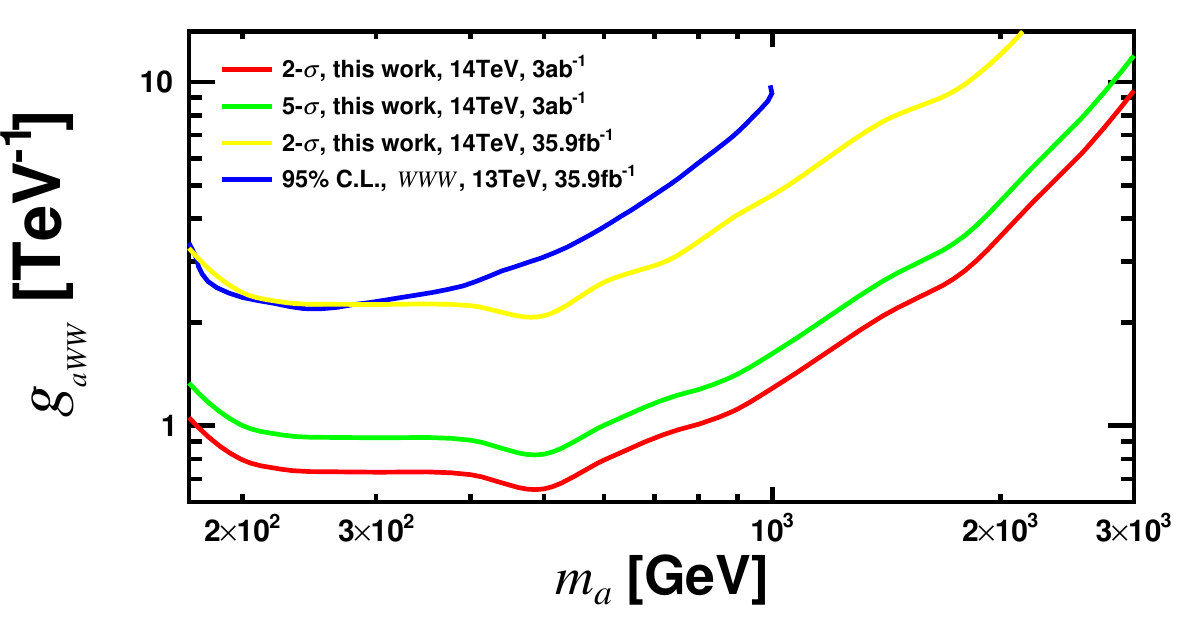}
\caption{
Discovery sensitivities with 2-$\sigma$ and 5-$\sigma$ significances on the coupling $g_{_{aWW}}$
in the mass range of 170 - 3000 GeV at the HL-LHC with $\sqrt{s} =$ 14 TeV and $\mathcal{L} =$ 3 $\iab$ and 35.9 fb$^{-1}$.
The 95\% C.L. limit (blue curve)~\cite{Aiko:2024xiv} derived from reinterpreting CMS analyses of SM production $ pp \to W^\pm W^\pm W^\mp$ at $\sqrt{s} =$ 13 TeV and $\mathcal{L} =$ 35.9 fb$^{-1}$ is displayed for comparison.
}
\label{fig:sensALP}
\end{figure}

In Fig.~\ref{fig:sensALP}, for the concrete case of heavy photophobic ALPs, we show the discovery sensitivities on the coupling \( g_{aWW} \) with 2-\(\sigma\) and 5-\(\sigma\) significances as a function of the ALP mass \( m_a \) in the range of 170–3000 GeV at the HL-LHC, with \( \sqrt{s} = 14~\text{TeV} \) and integrated luminosities of \(3~\text{ab}^{-1} \) and 35.9 fb\(^{-1}\). For comparison, we also display the 95\% C.L. limit (blue curve) from Ref.~\cite{Aiko:2024xiv} which is derived from reinterpreting the CMS analyses of the SM production \( pp \to W^\pm W^\pm W^\mp \)  at \( \sqrt{s} = 13~\text{TeV} \) with \( \mathcal{L} = 35.9~\text{fb}^{-1} \).
Our results demonstrate that at the HL-LHC with \( \mathcal{L} = 3~\text{ab}^{-1} \), the 2-\(\sigma\) sensitivity for \( g_{aWW} \) decreases from \( 1.06~\text{TeV}^{-1} \) to \( 0.75~\text{TeV}^{-1} \) as \( m_a \) increases from 170 GeV to 225 GeV, remains nearly stable for \( m_a \) in the range of 225–600 GeV, and then rises sharply to \( 9.42~\text{TeV}^{-1} \) at \( m_a = 3000~\text{GeV} \).
This pattern reflects the varying sensitivity of the search strategy across different mass ranges.

One observes that with the same luminosity of 35.9 $\ifb$, our strategy achieves better sensitivities for \( m_a \) above 300 GeV compared to the referenced approach~\cite{Aiko:2024xiv}. This improvement is primarily due to differences in the optimization of the analysis strategies. The referenced studies focus on the SM process \( pp \to W^\pm W^\pm W^\mp \), while our strategy is specifically tailored to the tri-\( W \) final state produced by a heavy resonance. For larger masses, the kinematics of these two processes differ significantly, allowing our approach to reject background more effectively, as demonstrated by the BDT distributions in Appendix~\ref{app:BDT}.

\section{Conclusion}
\label{sec:Conc}

In this study, we proposed a search strategy at the HL-LHC for a new neutral particle \( X \) that couples to a pair of \( W \)-bosons. The particle \( X \) is produced with a \( W \)-boson and decays into two \( W \)-bosons, resulting in the process \( p p \to W^{\pm} X (\to W^{+} W^{-}) \) and a tri-\( W \)-boson final state. To suppress background, we focus on events where two same-charge \( W \)-bosons decay leptonically into muons, producing same-sign muons in the final state. The third \( W \)-boson decays hadronically into jets, leveraging its higher branching ratio to enhance the signal. As a case study, we use the heavy photophobic ALP as an example.

Signal and background events are simulated at the detector level. The signal production cross section times branching ratio, \( \sigma(p \, p \rightarrow W^{\pm} \, a) \times \text{Br}(a \to W^+W^-)\), is evaluated as a function of the ALP mass \(m_a\) at the HL-LHC with \(\sqrt{s} = 14~\text{TeV}\). Events are selected with exactly two same-sign di-muons and at least two non-\(b\)-tagged jets. A machine-learning-based MVA is applied to enhance signal-background discrimination. Distributions of key input variables and the corresponding BDT responses are presented, and selection efficiencies for both preselection and BDT criteria are provided for various ALP masses.

We present the model-independent discovery sensitivities for the production cross section times branching ratio \( \sigma(p p \to W^{\pm} X) \times \text{Br}(X \to W^{+} W^{-}) \) as a function of the particle mass \( m_X \) in the range of 170–3000 GeV at the HL-LHC with \( \sqrt{s} = 14~\text{TeV} \) and \( \mathcal{L} = 3~\text{ab}^{-1} \), which can be applied to general models involving a new neutral particle \( X \) that couples to \( W \)-bosons. The sensitivities for the coupling \( g_{aWW} \) of heavy photophobic ALPs are also shown for 2-\(\sigma\) and 5-\(\sigma\) significances in the same mass range under luminosities of \( 3~\text{ab}^{-1} \) and 35.9 \(\text{fb}^{-1}\). At \( \mathcal{L} = 3~\text{ab}^{-1} \), the 2-\(\sigma\) sensitivity for \( g_{aWW} \) decreases from \( 1.06~\text{TeV}^{-1} \) to \( 0.75~\text{TeV}^{-1} \) as \( m_a \) increases from 170 GeV to 225 GeV, remains nearly constant for \( m_a \) between 225 and 600 GeV, and then sharply rises to \( 9.42~\text{TeV}^{-1} \) at \( m_a = 3000~\text{GeV} \). Compared to previous limits derived from reinterpreting CMS analyses of SM production \( pp \to W^\pm W^\pm W^\mp \) at \( \sqrt{s} = 13~\text{TeV} \), our strategy provides improved sensitivity for \( m_a \) above 300 GeV.

\appendix

\section{Distributions of Representative Observables}
\label{app:observables}

\begin{figure}[H]
\centering
\subfigure{
\includegraphics[width=7.3cm, height=4.5cm]{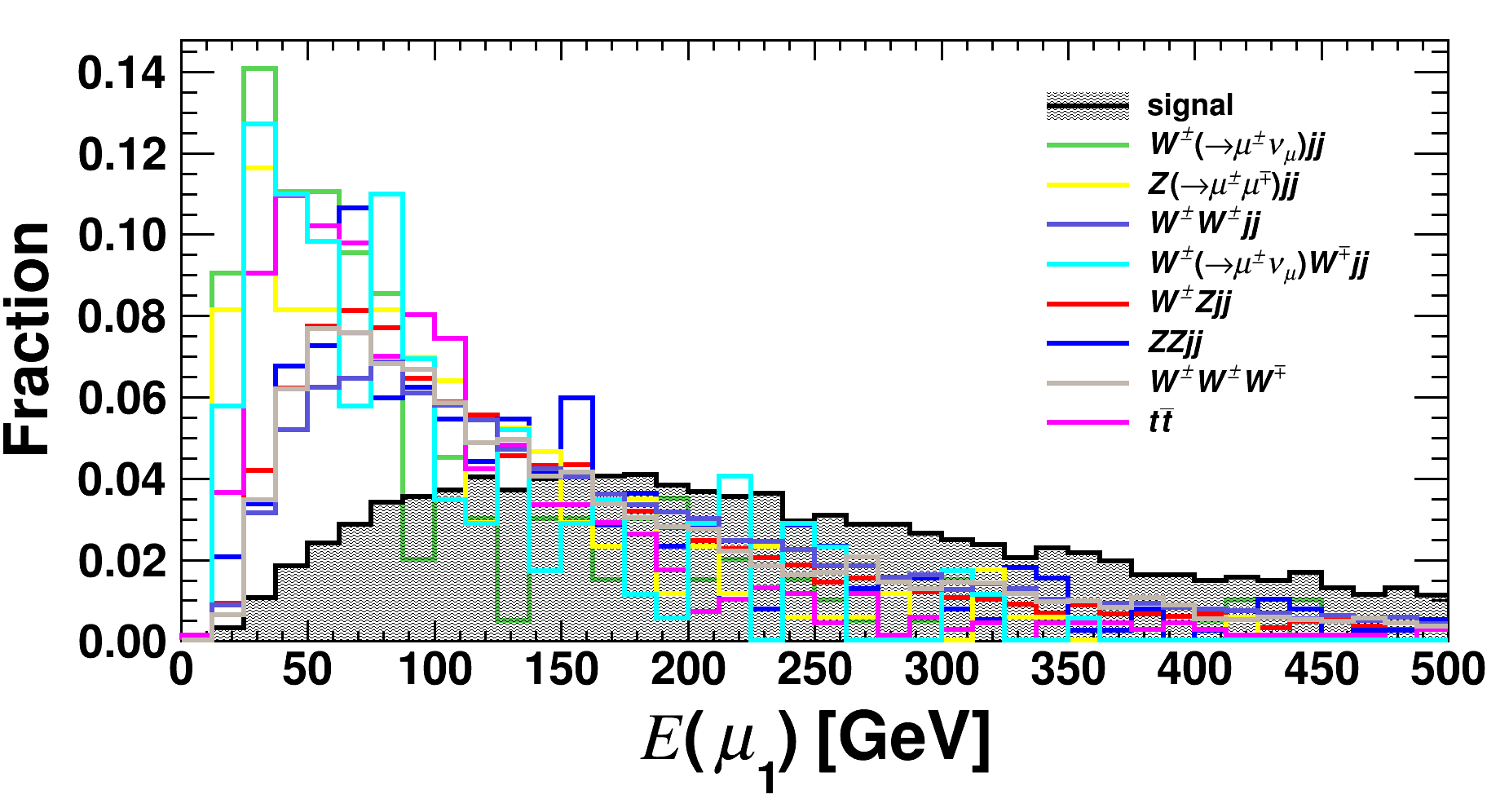}\,\,\,\,\,
\includegraphics[width=7.3cm, height=4.5cm]{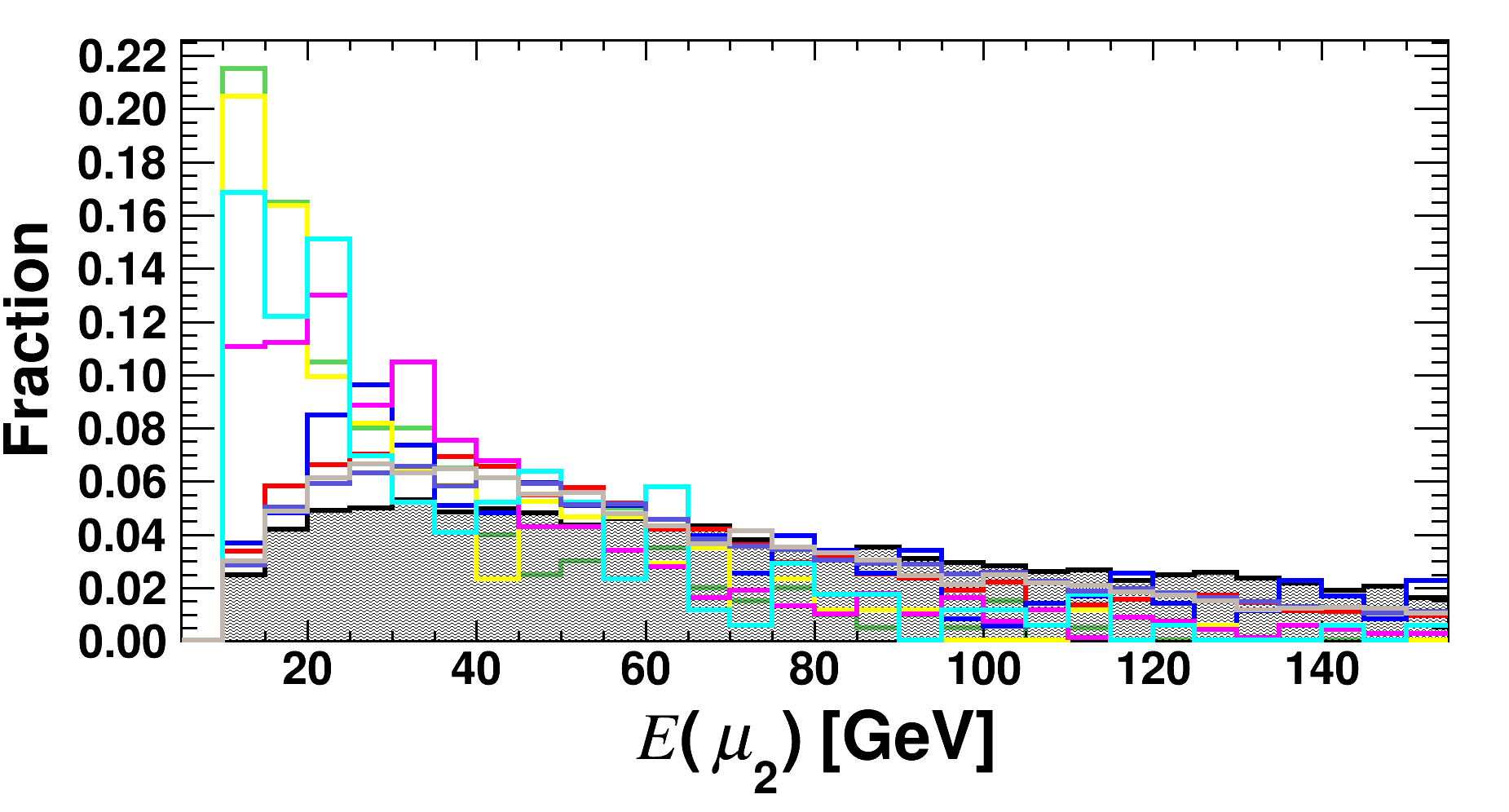}
}
\end{figure}
%\addtocounter{figure}{-1}
%
\vspace{-1.0cm}
\begin{figure}[H]
\centering
\subfigure{
\includegraphics[width=7.3cm, height=4.5cm]{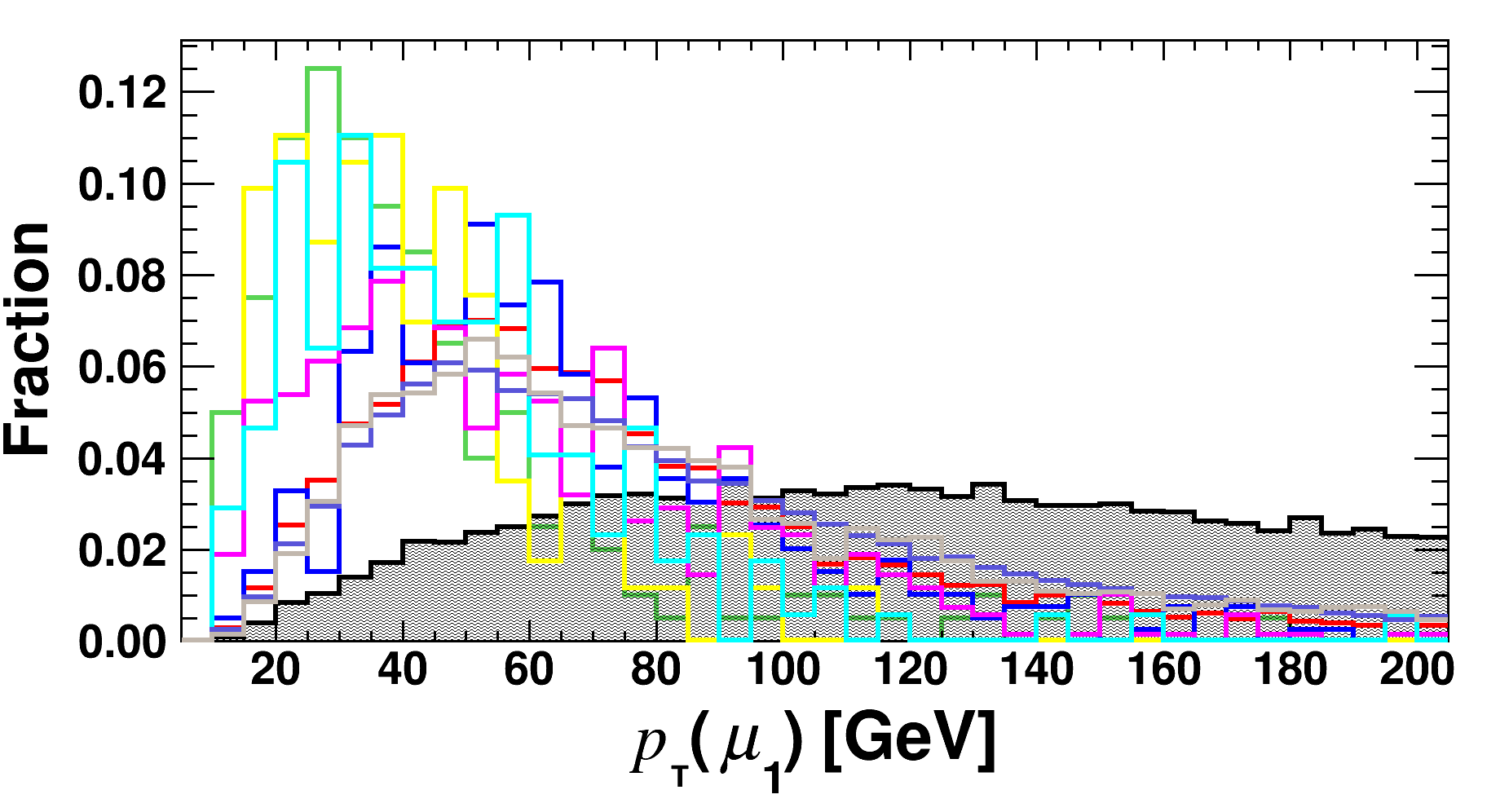}\,\,\,\,\,
\includegraphics[width=7.3cm, height=4.5cm]{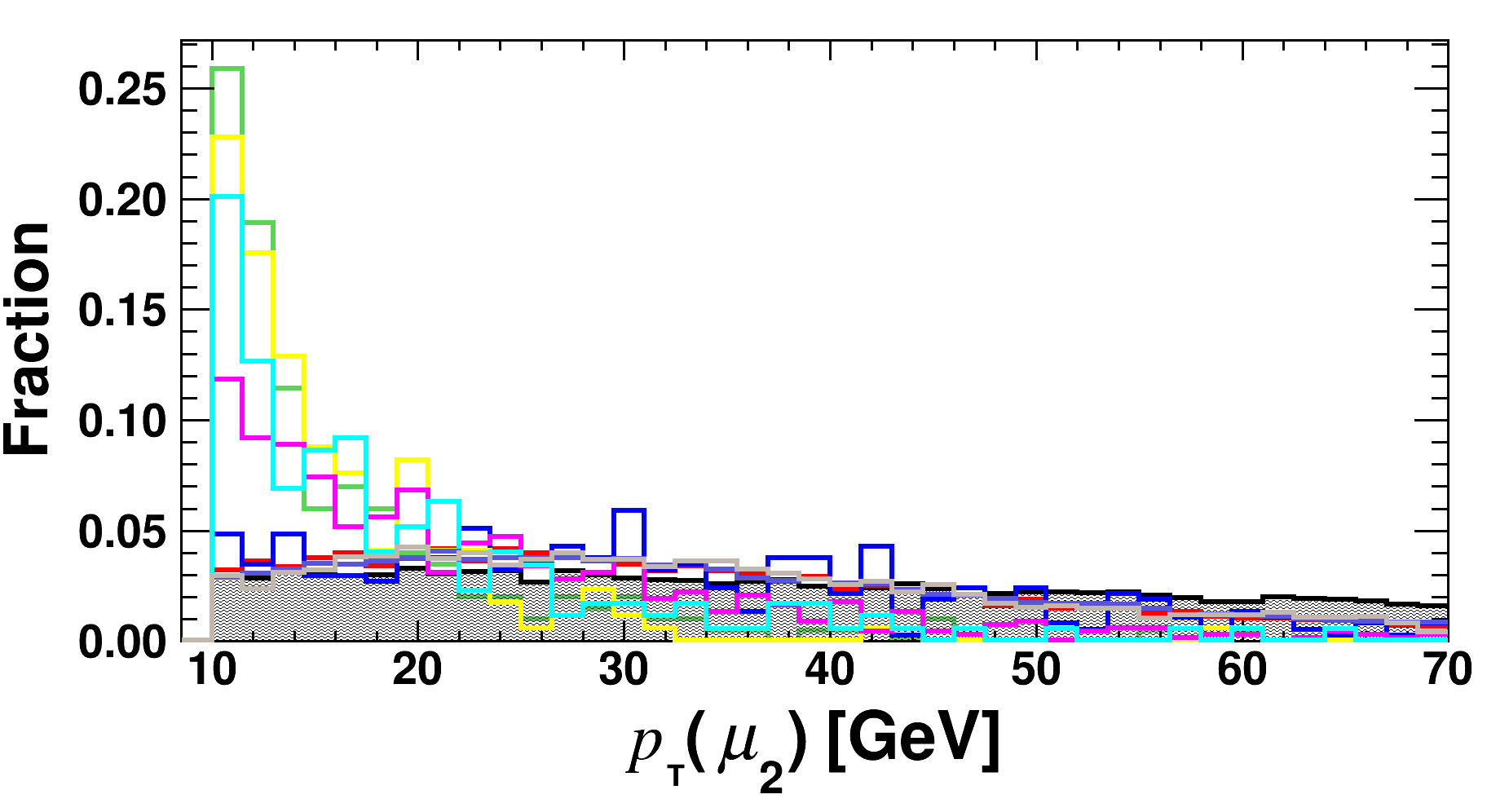}
}
\end{figure}
%\addtocounter{figure}{-1}
%
\vspace{-1.0cm}
\begin{figure}[H]
\centering
\subfigure{
\includegraphics[width=7.3cm, height=4.5cm]{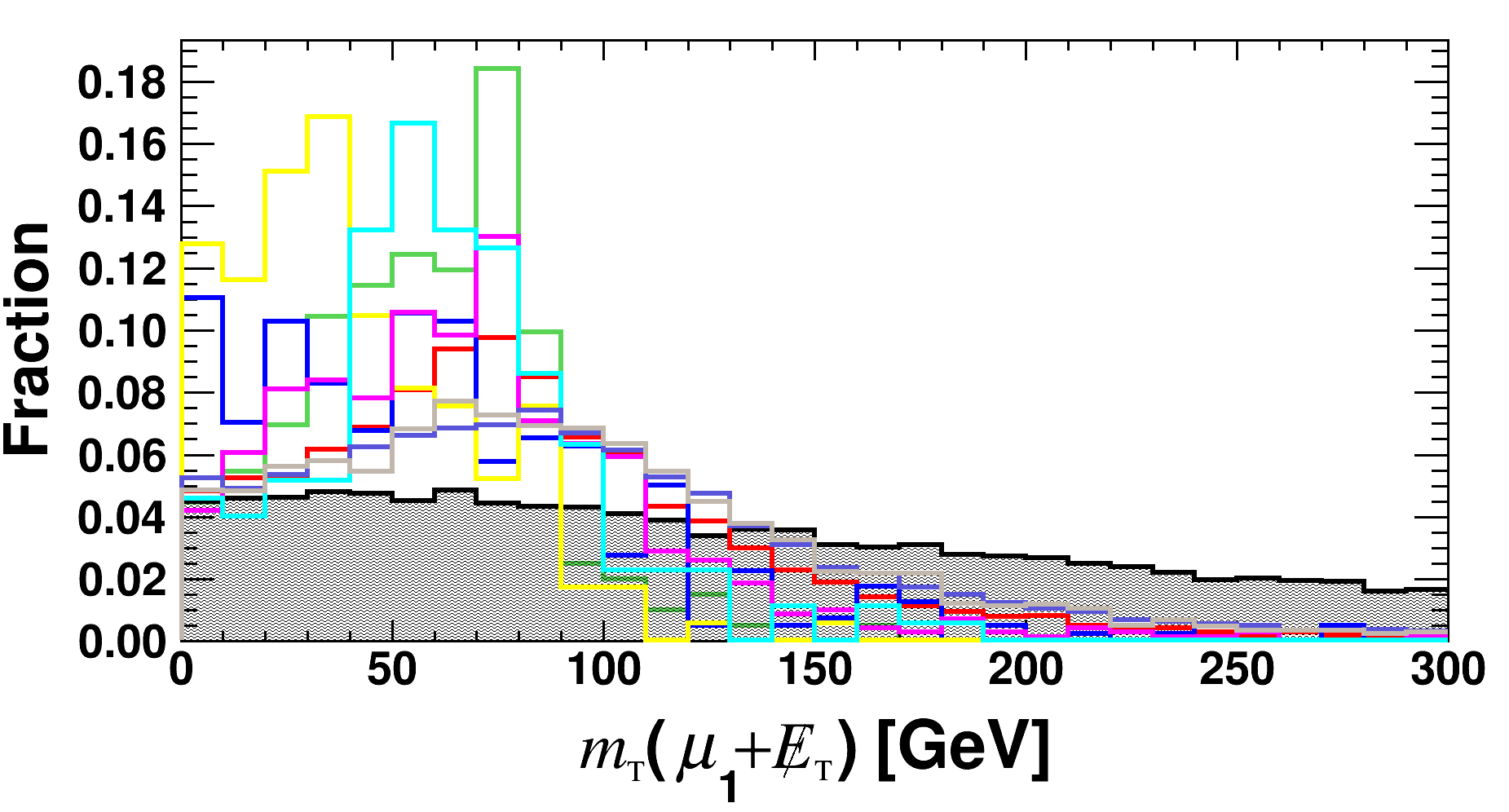}\,\,\,\,\,
\includegraphics[width=7.3cm, height=4.5cm]{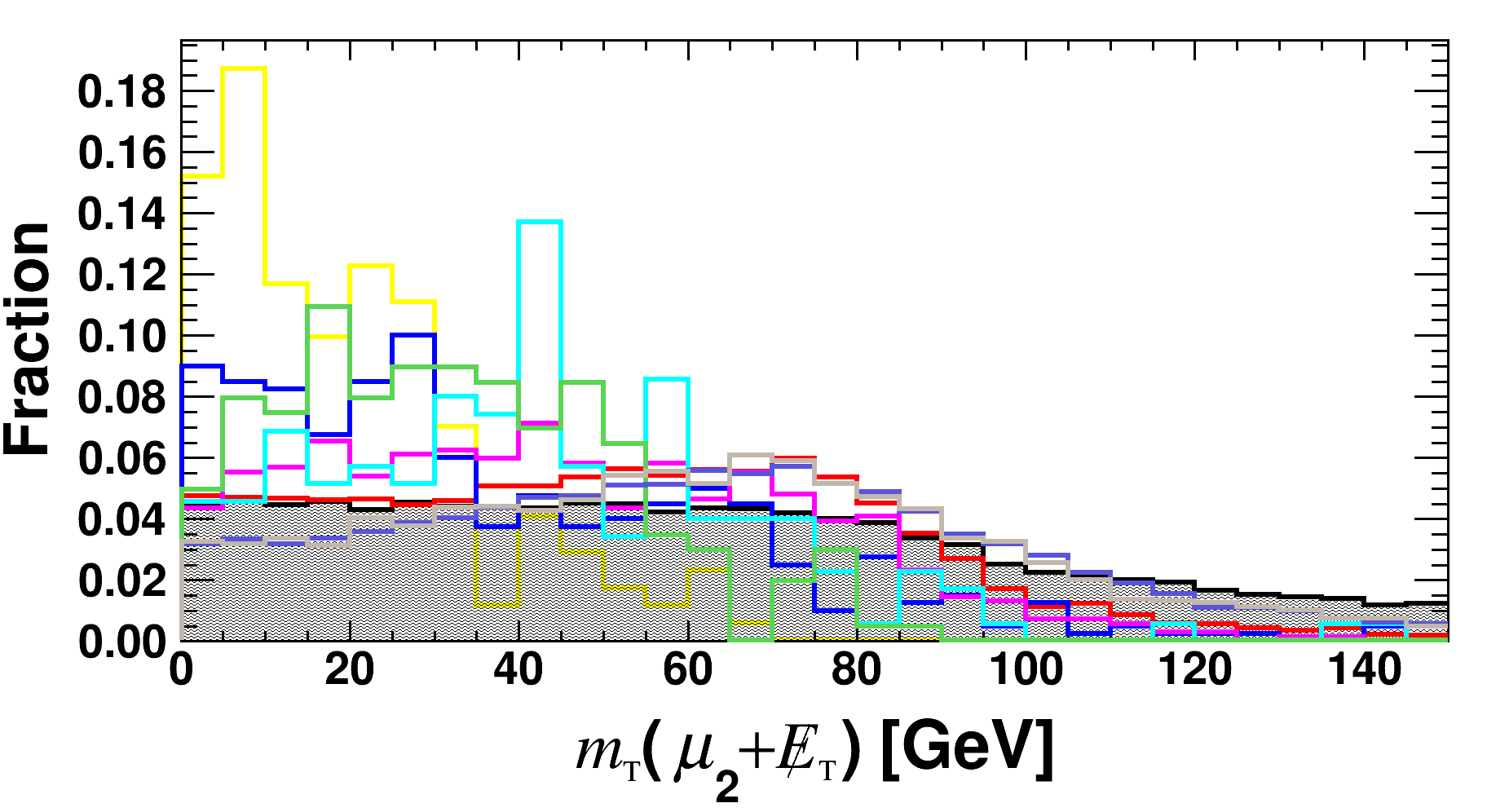}
}
\caption{
Distributions of kinematical observables for ${\mu}_{1}$ and ${\mu}_{2}$ after applying preselection criteria for the signal(black, shadow) and background processes at the HL-LHC with $\sqrt{s} =$ 14 TeV, assuming the benchmark $m_{a}$ = 400 GeV.
}
\label{Representative_Muon}
\end{figure}

%\label{Observables_Jet}
\begin{figure}[H]
\centering
\subfigure{
\includegraphics[width=7.3cm, height=4.5cm]{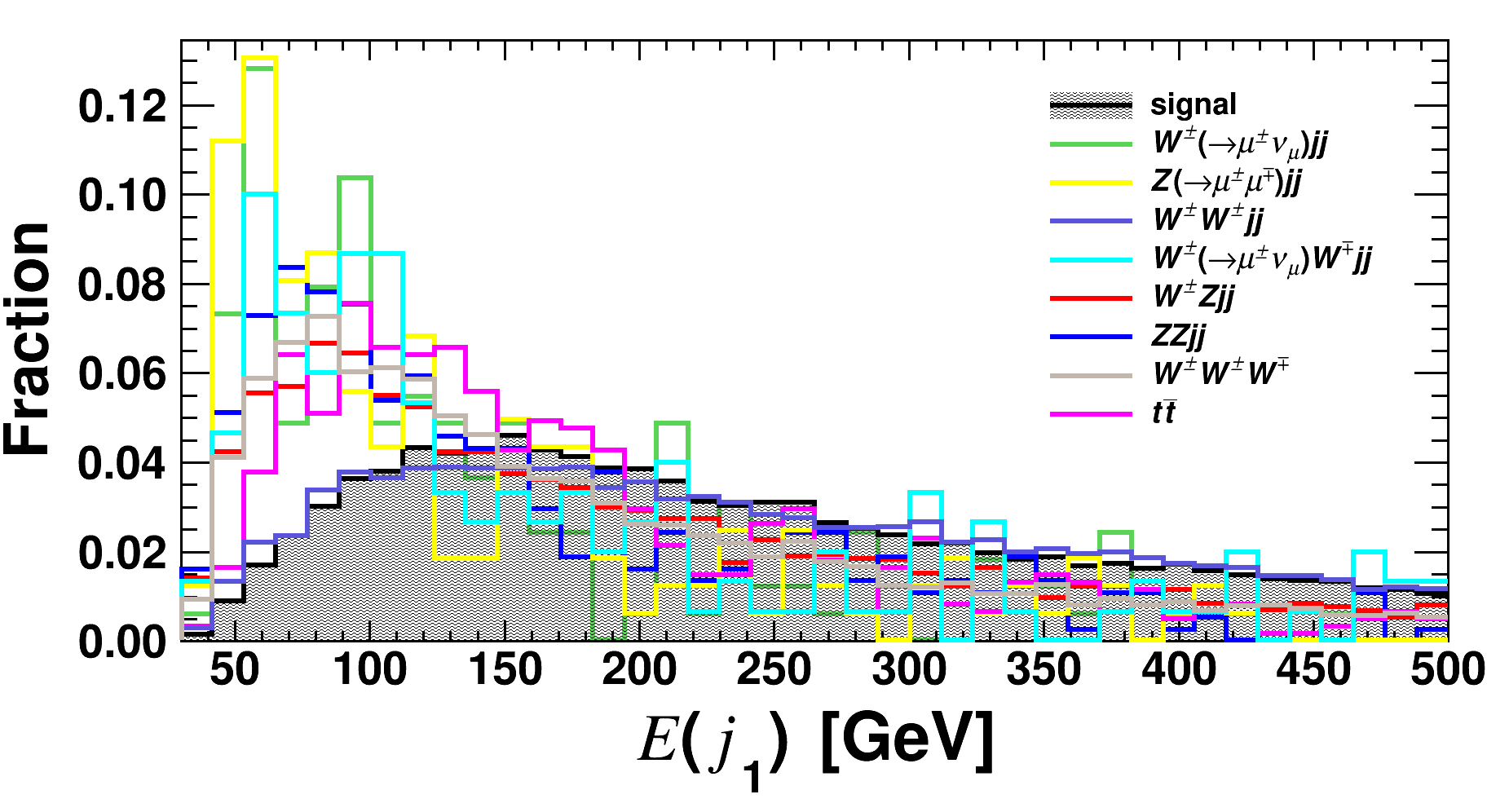}\,\,\,\,\,
\includegraphics[width=7.3cm, height=4.5cm]{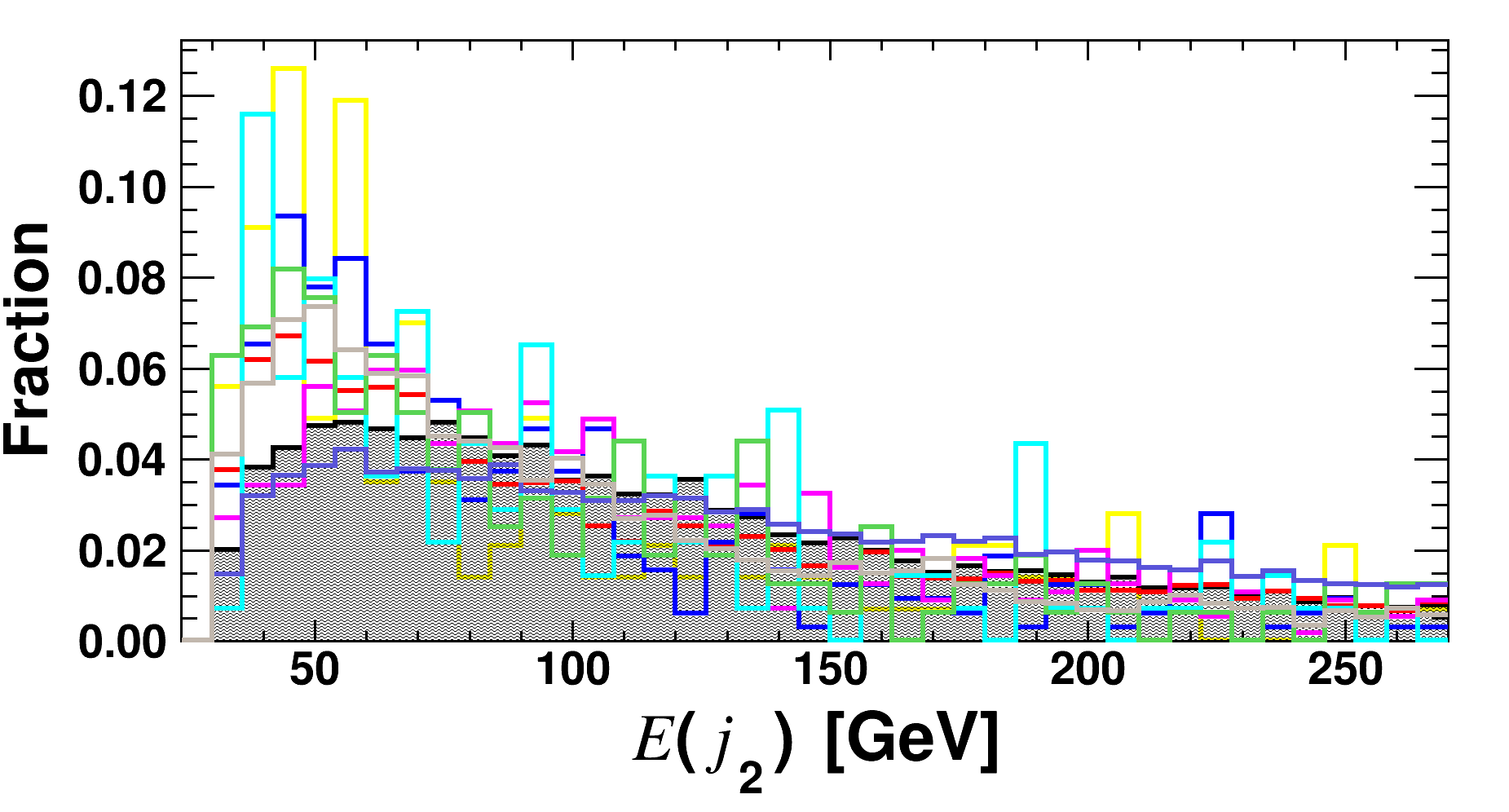}
}
\end{figure}
%\addtocounter{figure}{-1}
\vspace{-1.0cm}
\begin{figure}[H]
\centering
\subfigure{
\includegraphics[width=7.3cm, height=4.5cm]{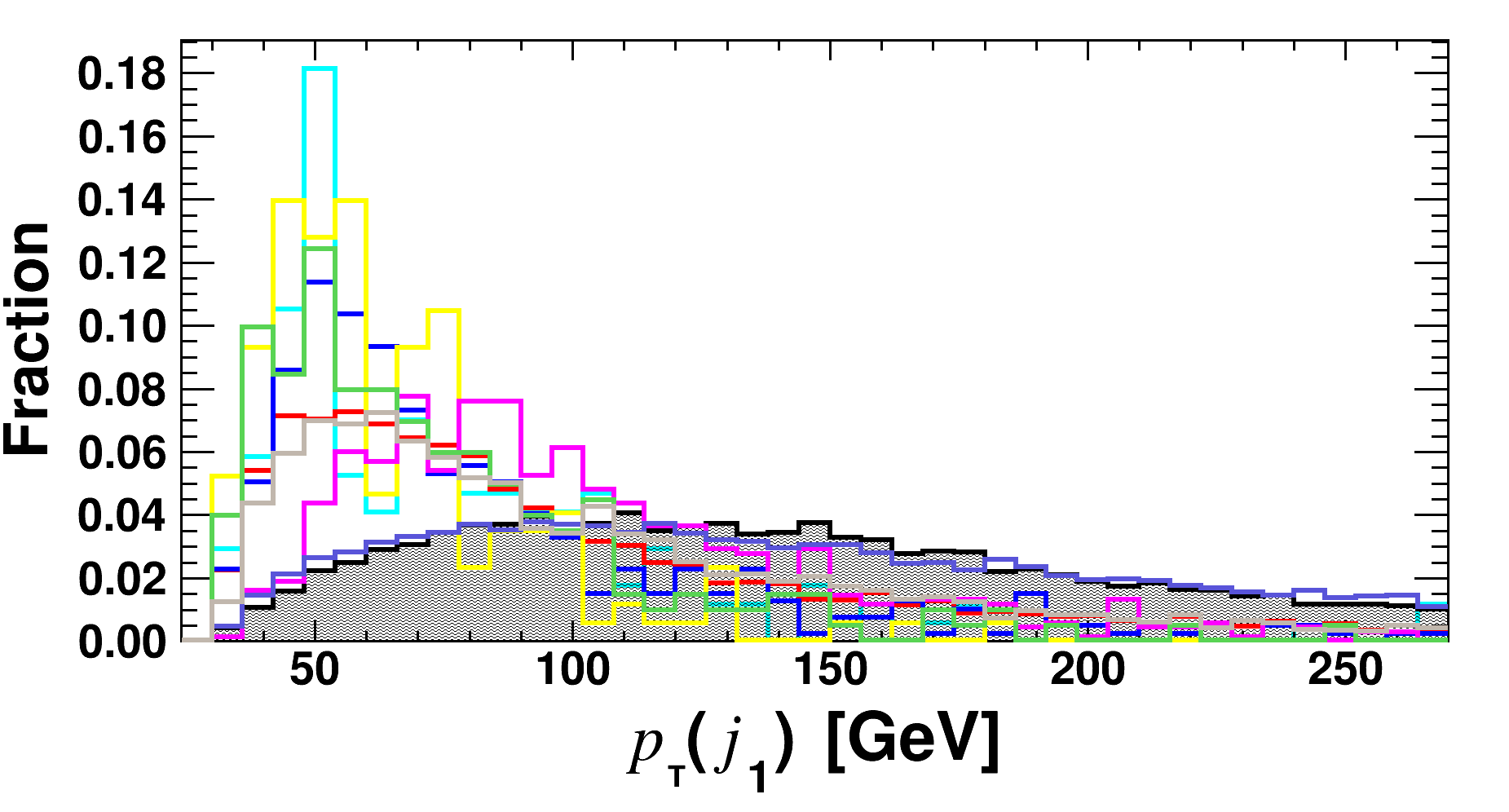}\,\,\,\,\,
\includegraphics[width=7.3cm, height=4.5cm]{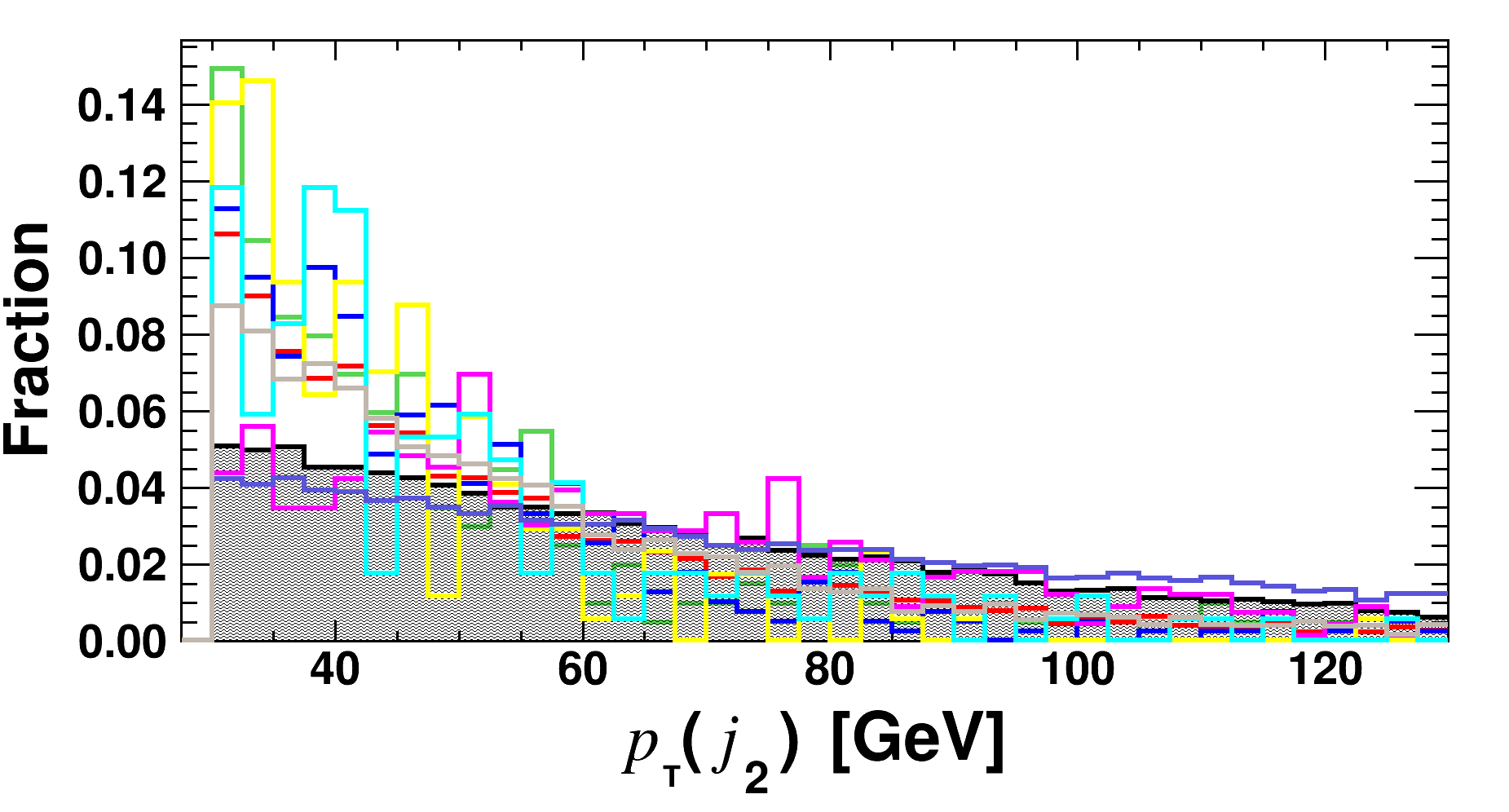}
}
\end{figure}
%\addtocounter{figure}{-1}
%
\vspace{-1.0cm}
\begin{figure}[H]
\centering
\subfigure{
\includegraphics[width=7.3cm, height=4.5cm]{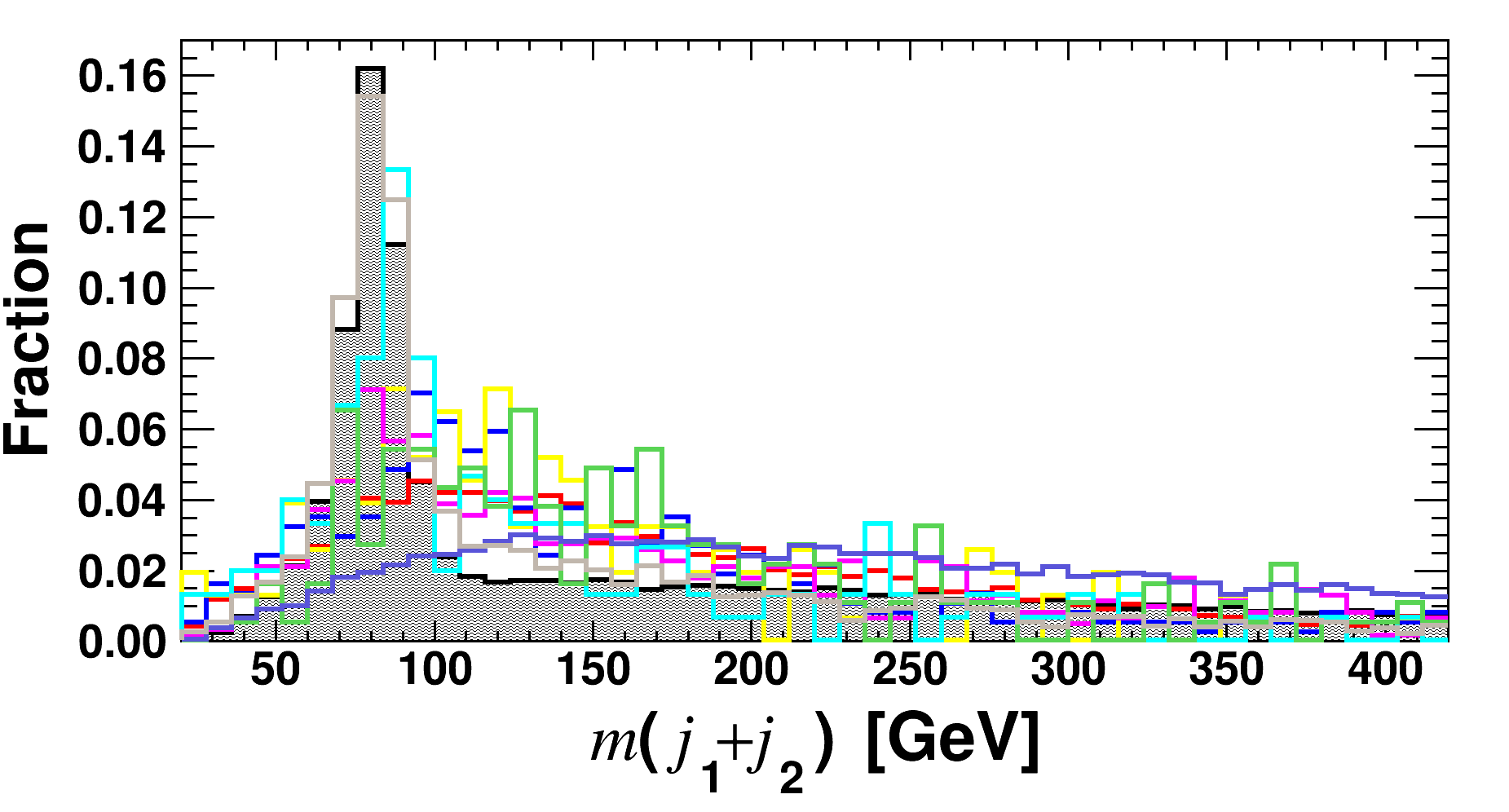}\,\,\,\,\,
\includegraphics[width=7.3cm, height=4.5cm]{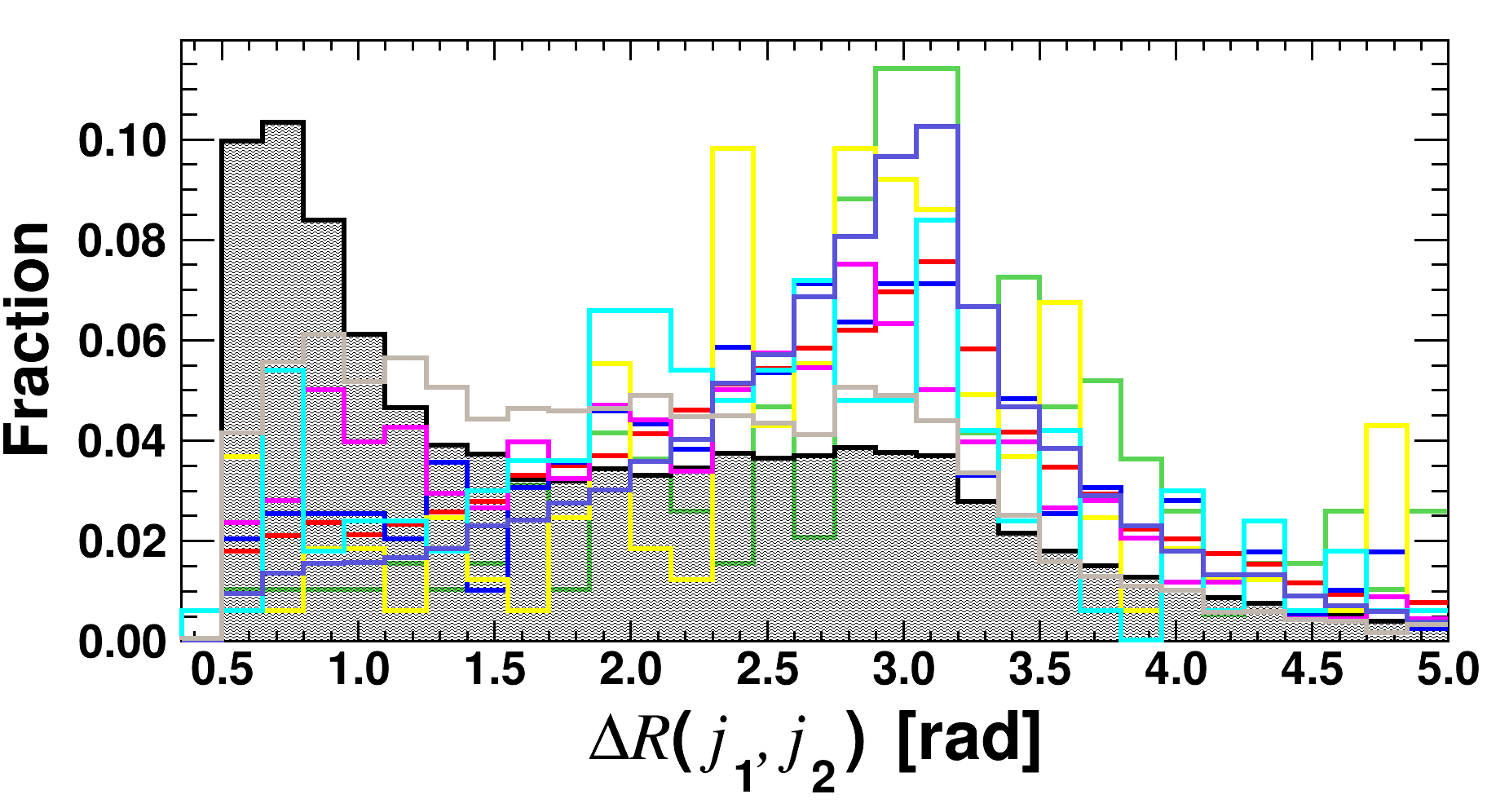}
}
\end{figure}
%\addtocounter{figure}{-1}
%
\vspace{-1.0cm}
\begin{figure}[H]
\centering
\subfigure{
\includegraphics[width=7.3cm, height=4.5cm]{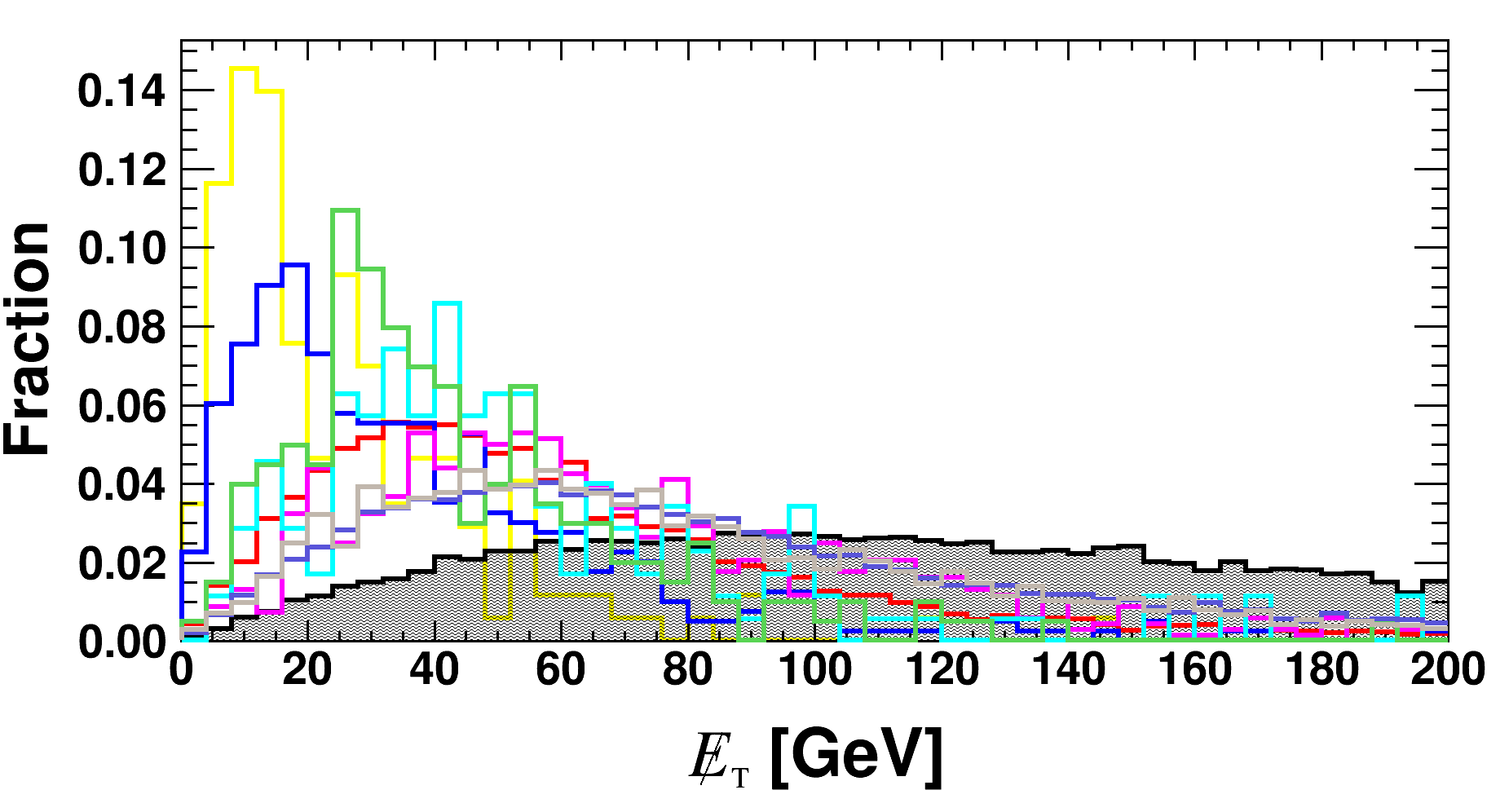}\,\,\,\,\,
\includegraphics[width=7.3cm, height=4.5cm]{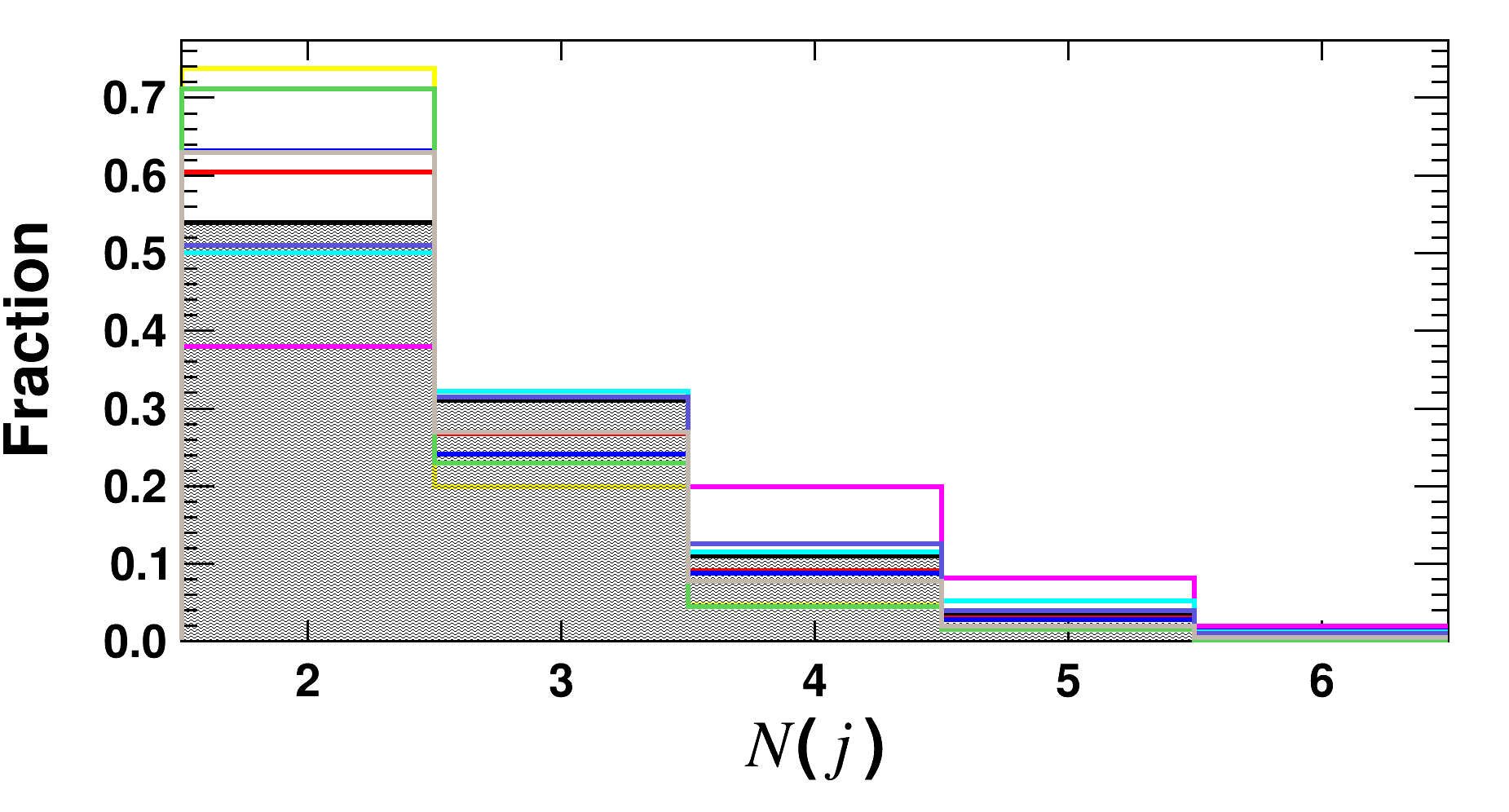}
}
\caption{
The same  as Fig.~\ref{Representative_Muon}, but for ${j}_{1}$, ${j}_{2}$, $\met$ and $N(j)$, assuming the benchmark $m_{a}$ = 400 GeV.
}
\label{Representative_Jet}
\end{figure}

%\label{Observables_JetWm}
\begin{figure}[H]
\centering
\includegraphics[width=7.3cm, height=4.5cm]{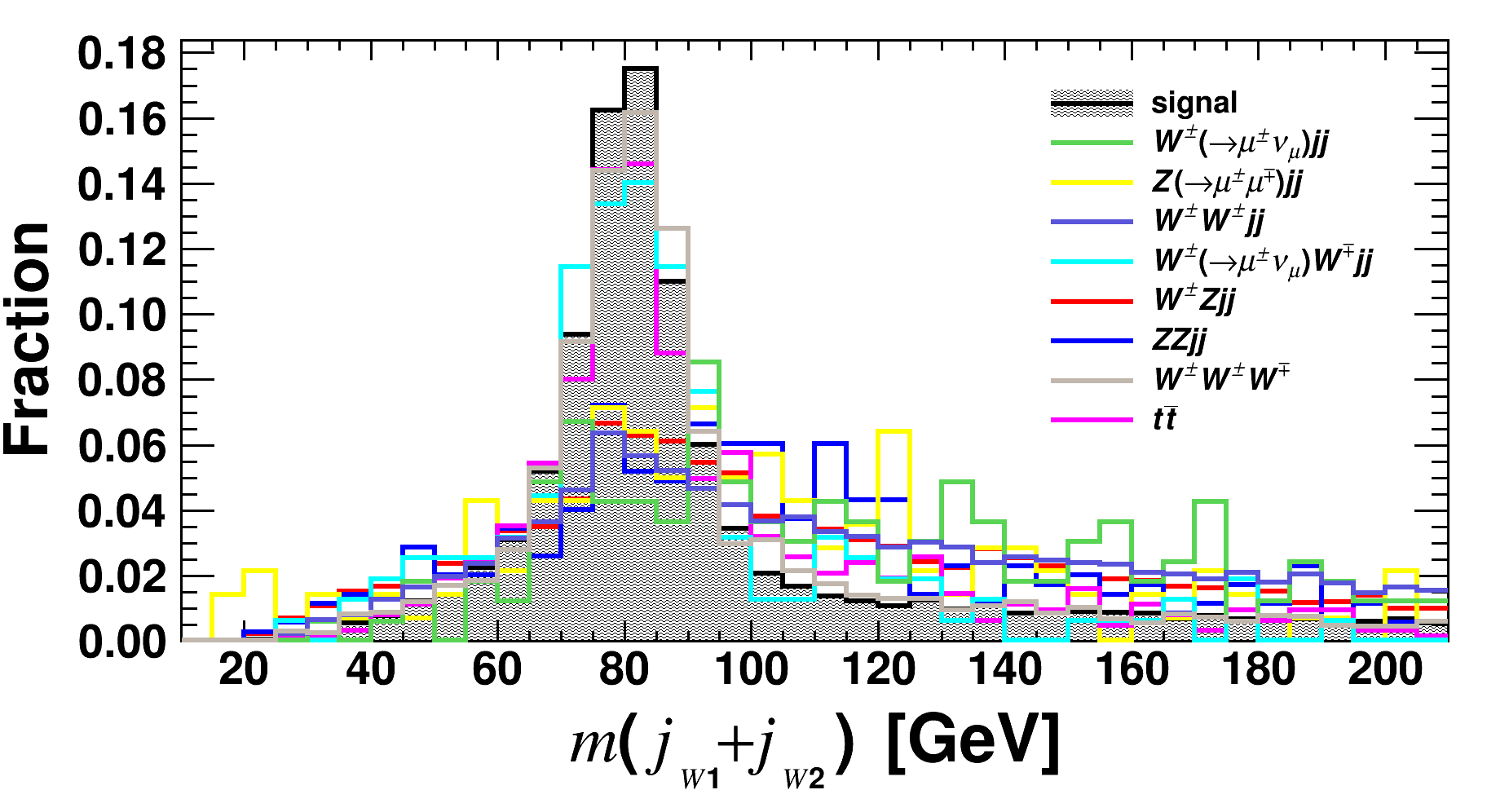}\,\,\,\,\,
\includegraphics[width=7.3cm, height=4.5cm]{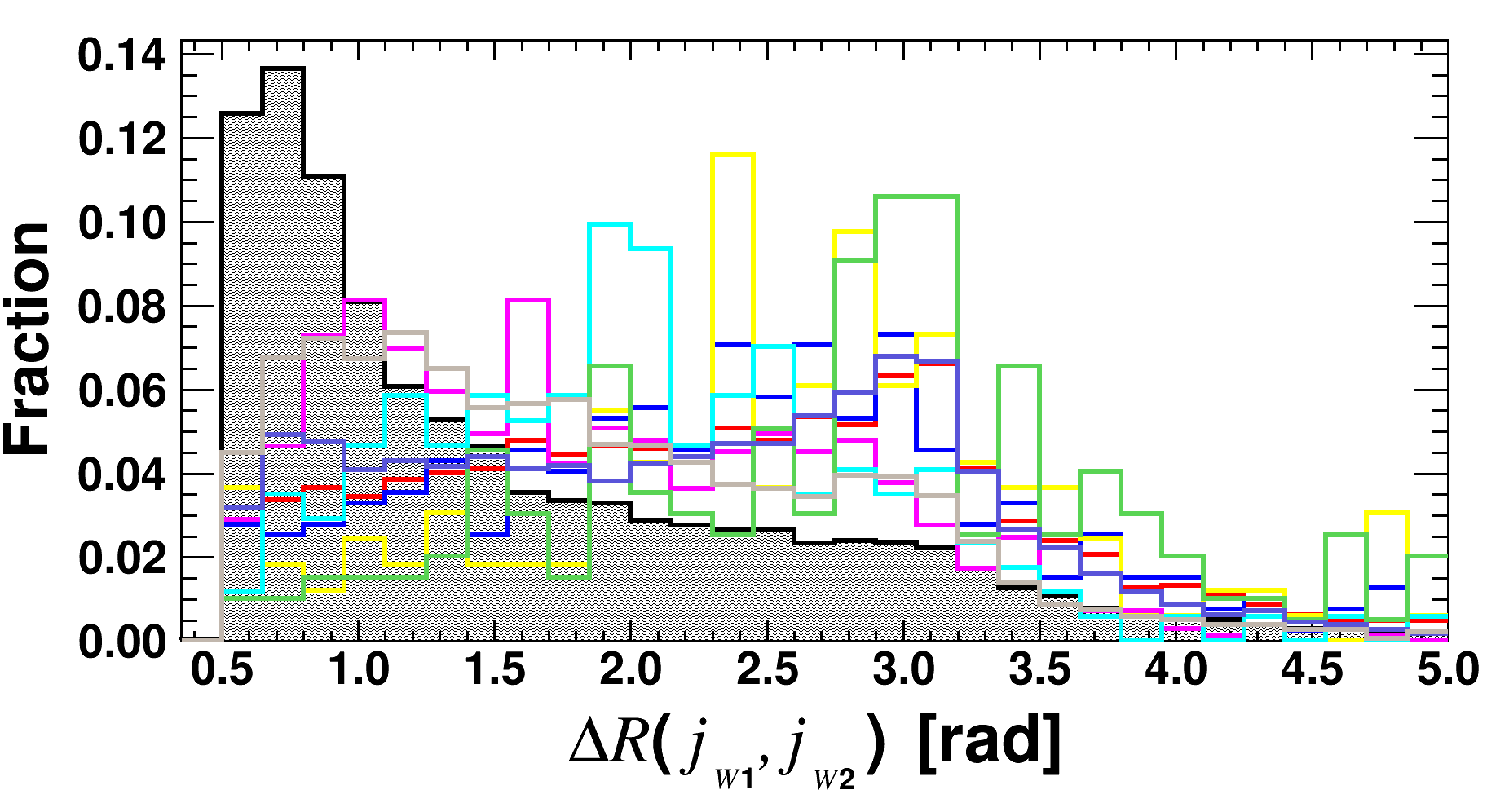}\\
\includegraphics[width=7.3cm, height=4.5cm]{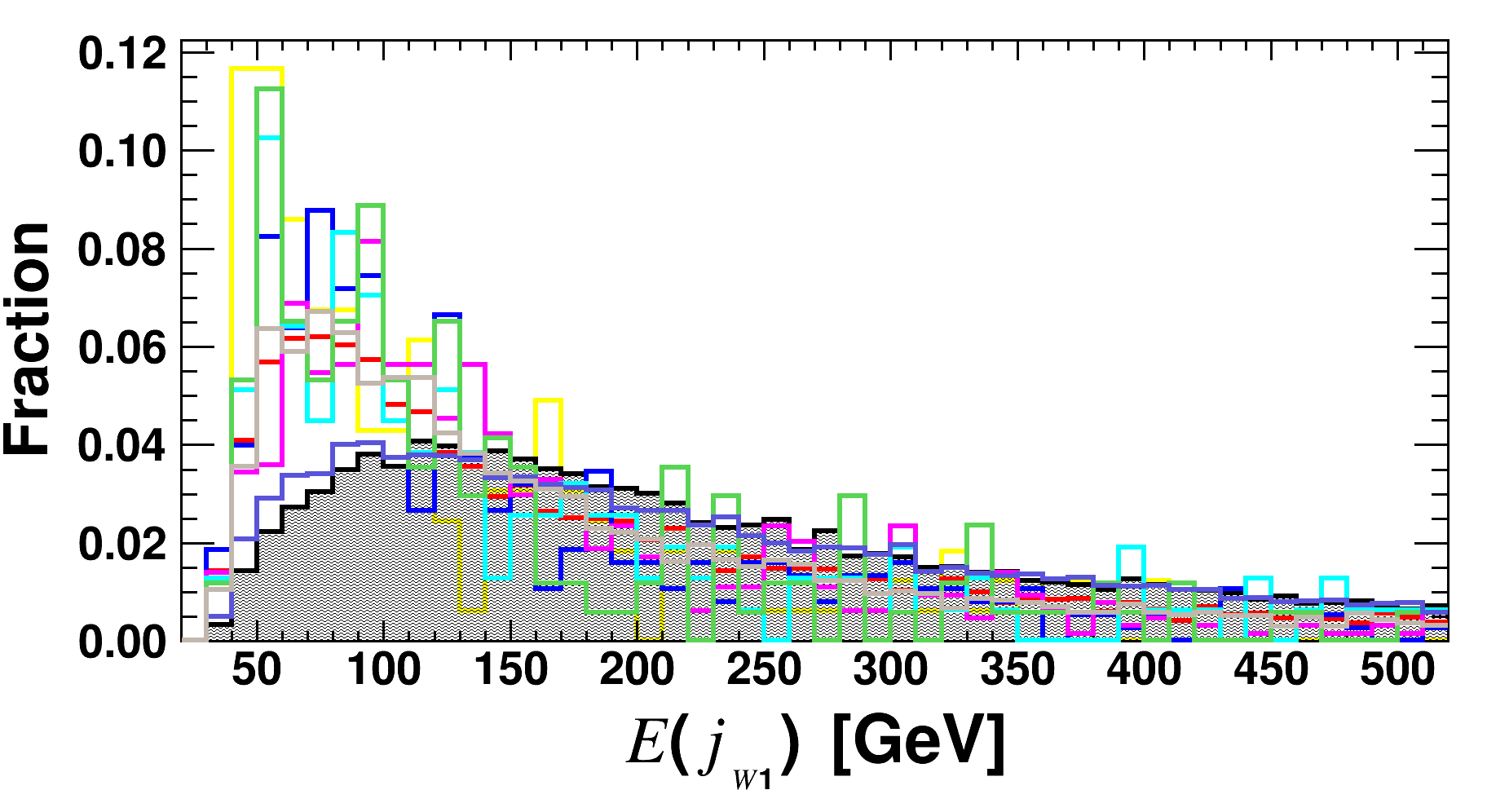}\,\,\,\,\,
\includegraphics[width=7.3cm, height=4.5cm]{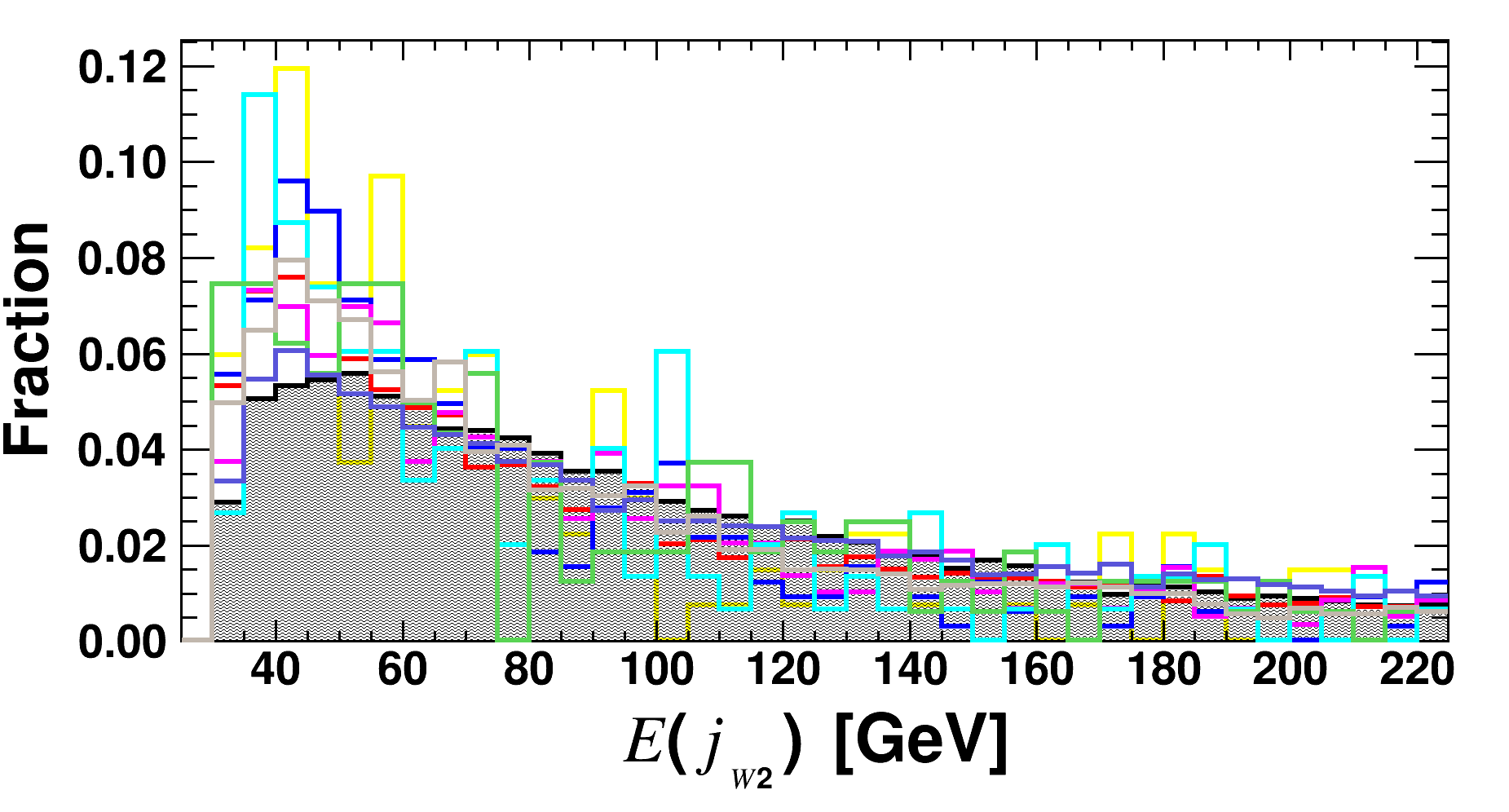}\\
\includegraphics[width=7.3cm, height=4.5cm]{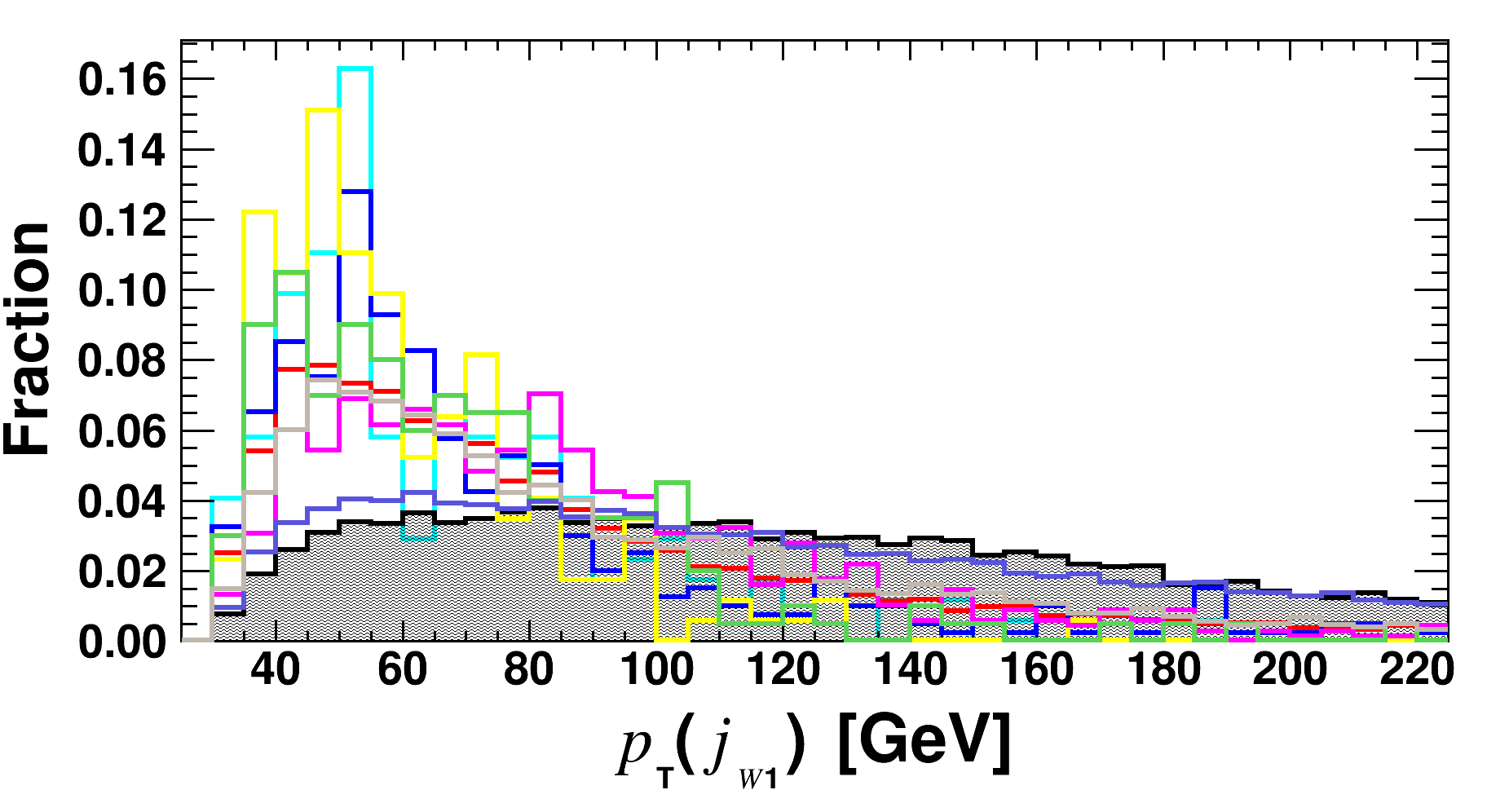}\,\,\,\,\,
\includegraphics[width=7.3cm, height=4.5cm]{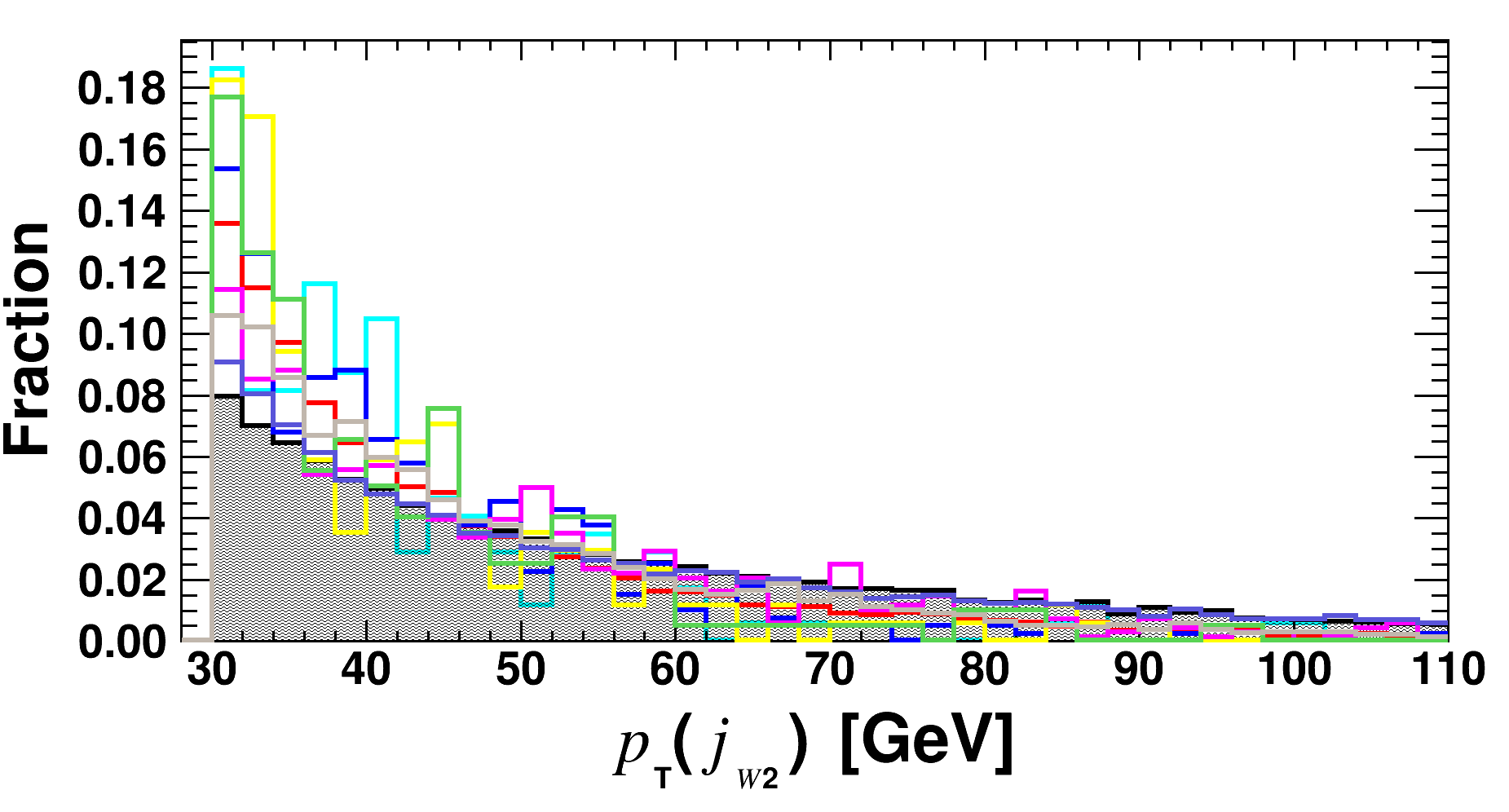}
\caption{
The same as Fig.~\ref{Representative_Muon}, but for ${j}_{_{W1}}$ and ${j}_{_{W2}}$, assuming the benchmark $m_{a}$ = 400 GeV.
}
\label{fig:jjW}
\label{Representative_JetWm}
\end{figure}

%\label{Observables_WmMuWm}
\begin{figure}[H]
\centering
\includegraphics[width=7.3cm, height=4.5cm]{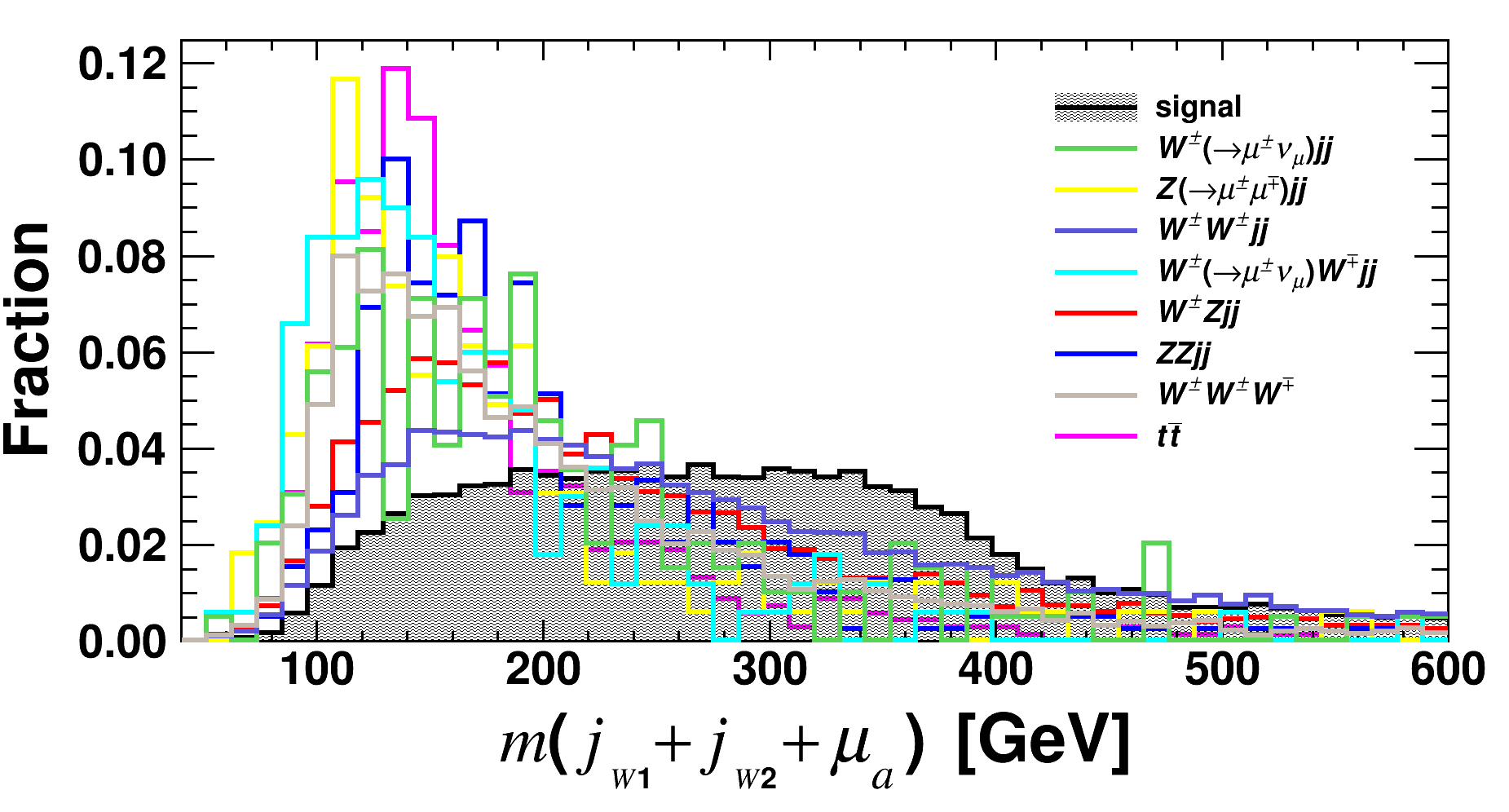}\,\,\,\,\,
\includegraphics[width=7.3cm, height=4.5cm]{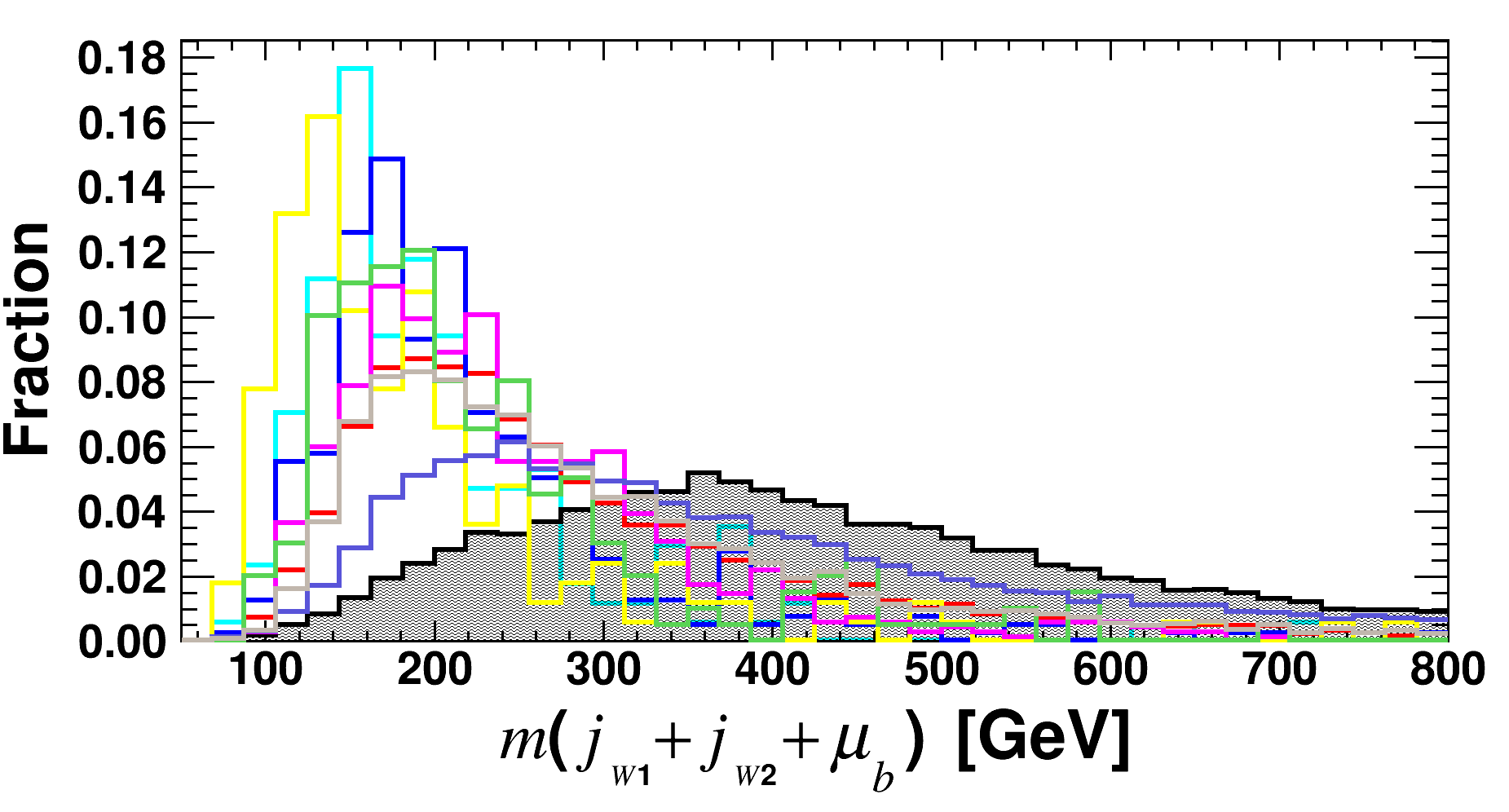}
\includegraphics[width=7.3cm, height=4.5cm]{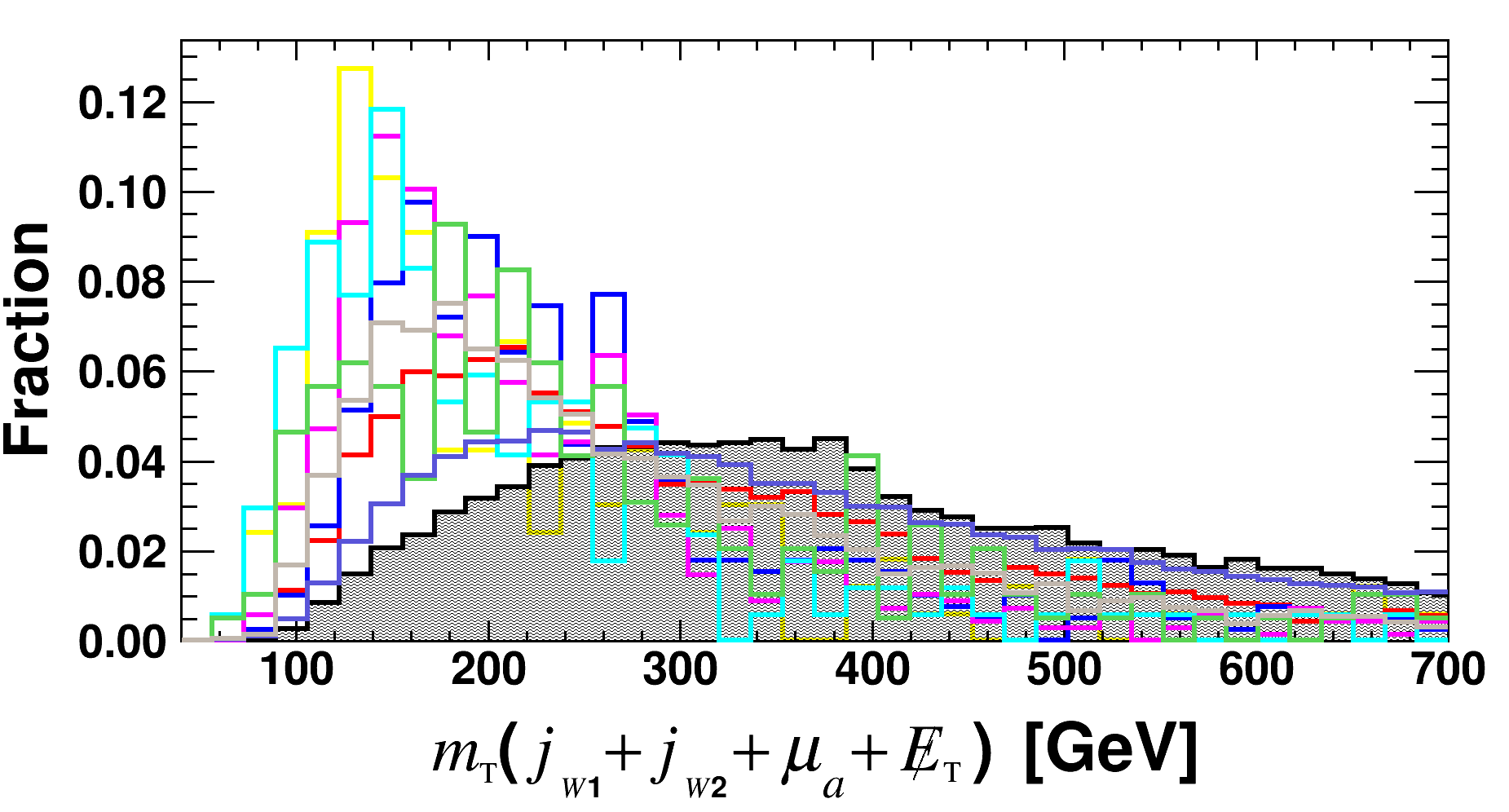}\,\,\,\,\,
\includegraphics[width=7.3cm, height=4.5cm]{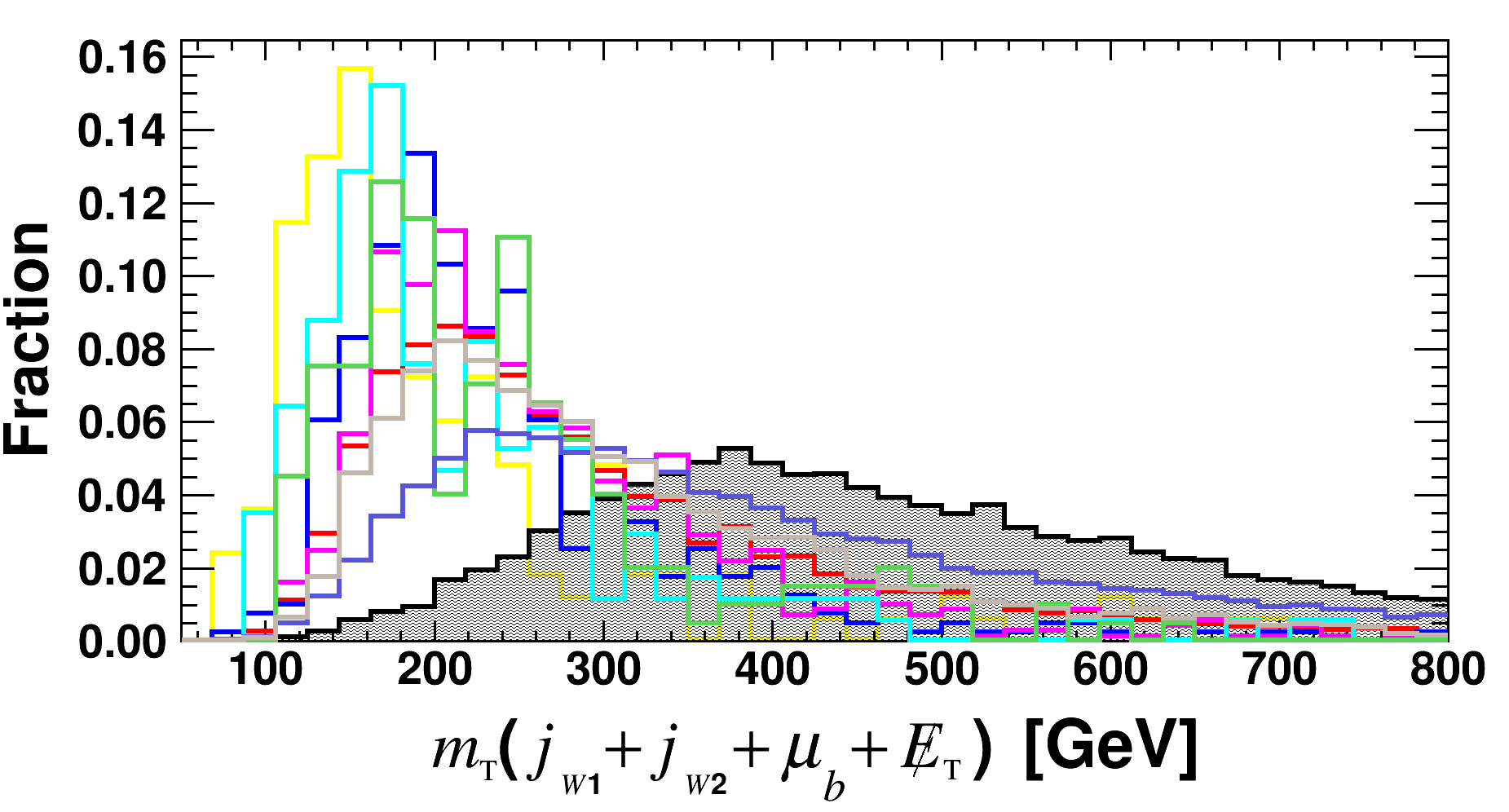}
\includegraphics[width=7.3cm, height=4.5cm]{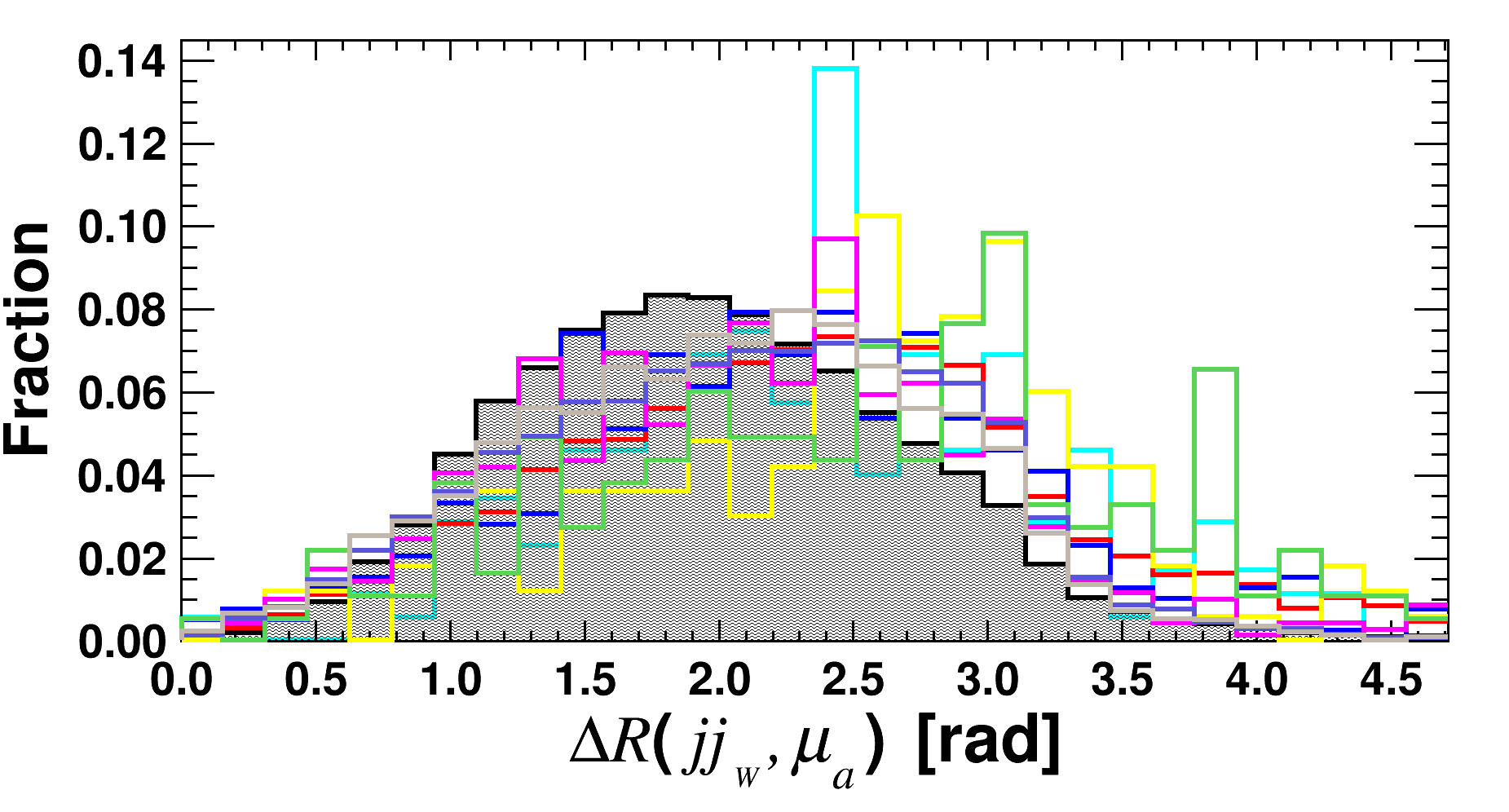}\,\,\,\,\,
\includegraphics[width=7.3cm, height=4.5cm]{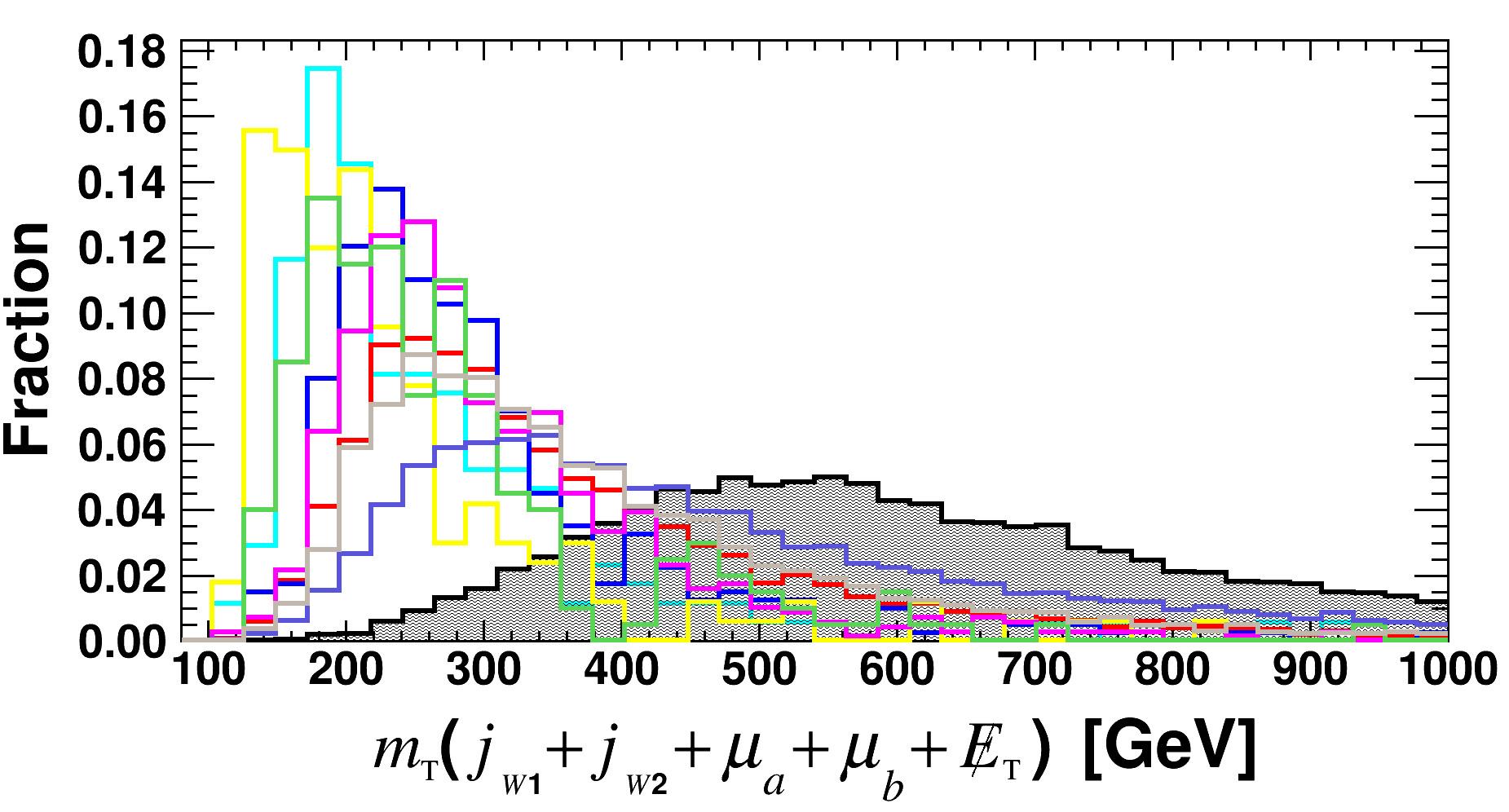}
\caption{
The same as Fig.~\ref{Representative_Muon}, but for observables related to the reconstruction of the ALP mass and the off-shell $W^{(*)\pm}$ boson, assuming the benchmark $m_{a}$ = 400 GeV.
}
\label{fig:ALPandWstar}
\end{figure}

%\label{Observables_chongjian}
\begin{figure}[h]
\centering
\subfigure{
\includegraphics[width=7.3cm, height=4.5cm]{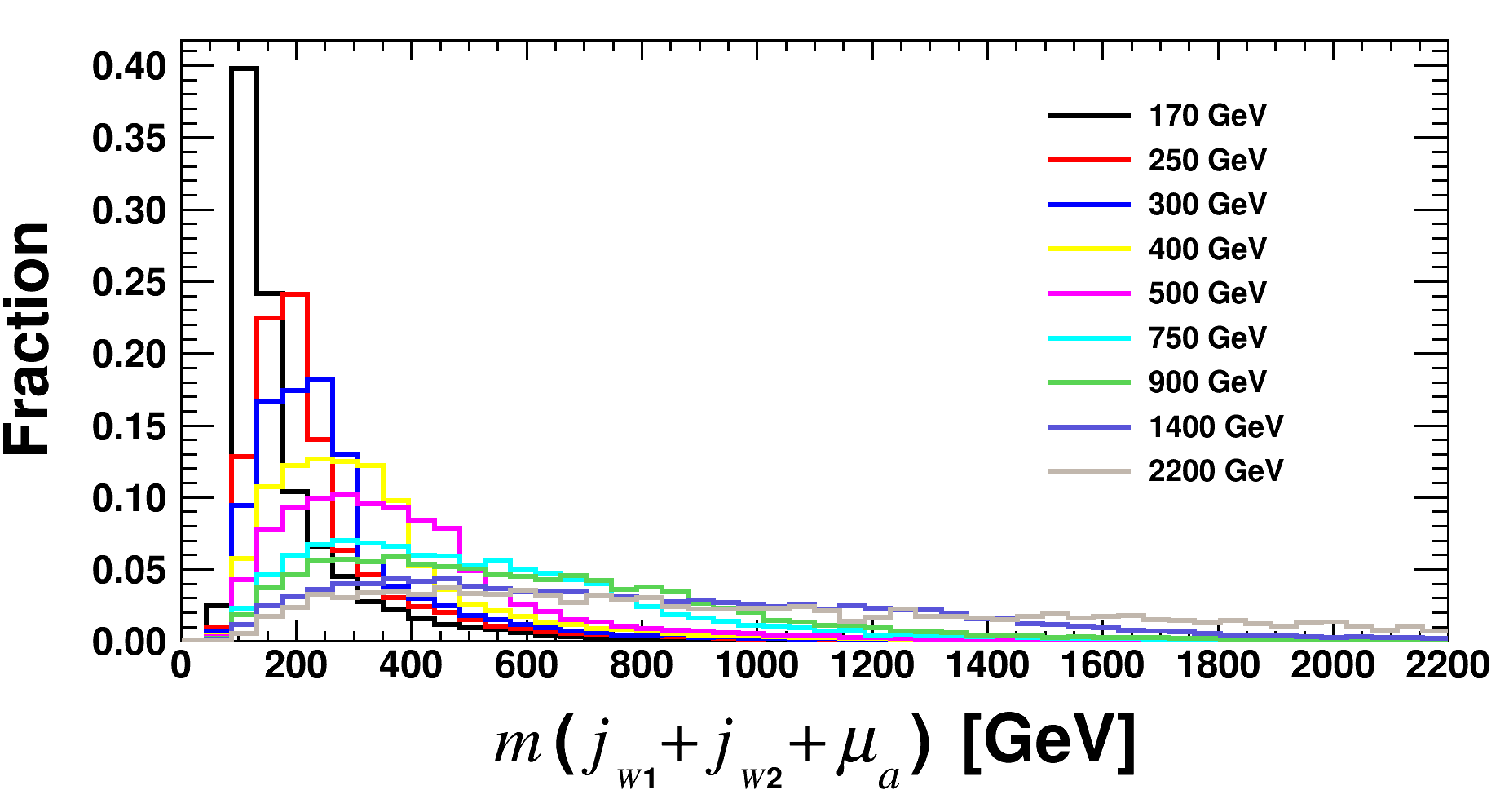}\,\,\,\,\,
\includegraphics[width=7.3cm, height=4.5cm]{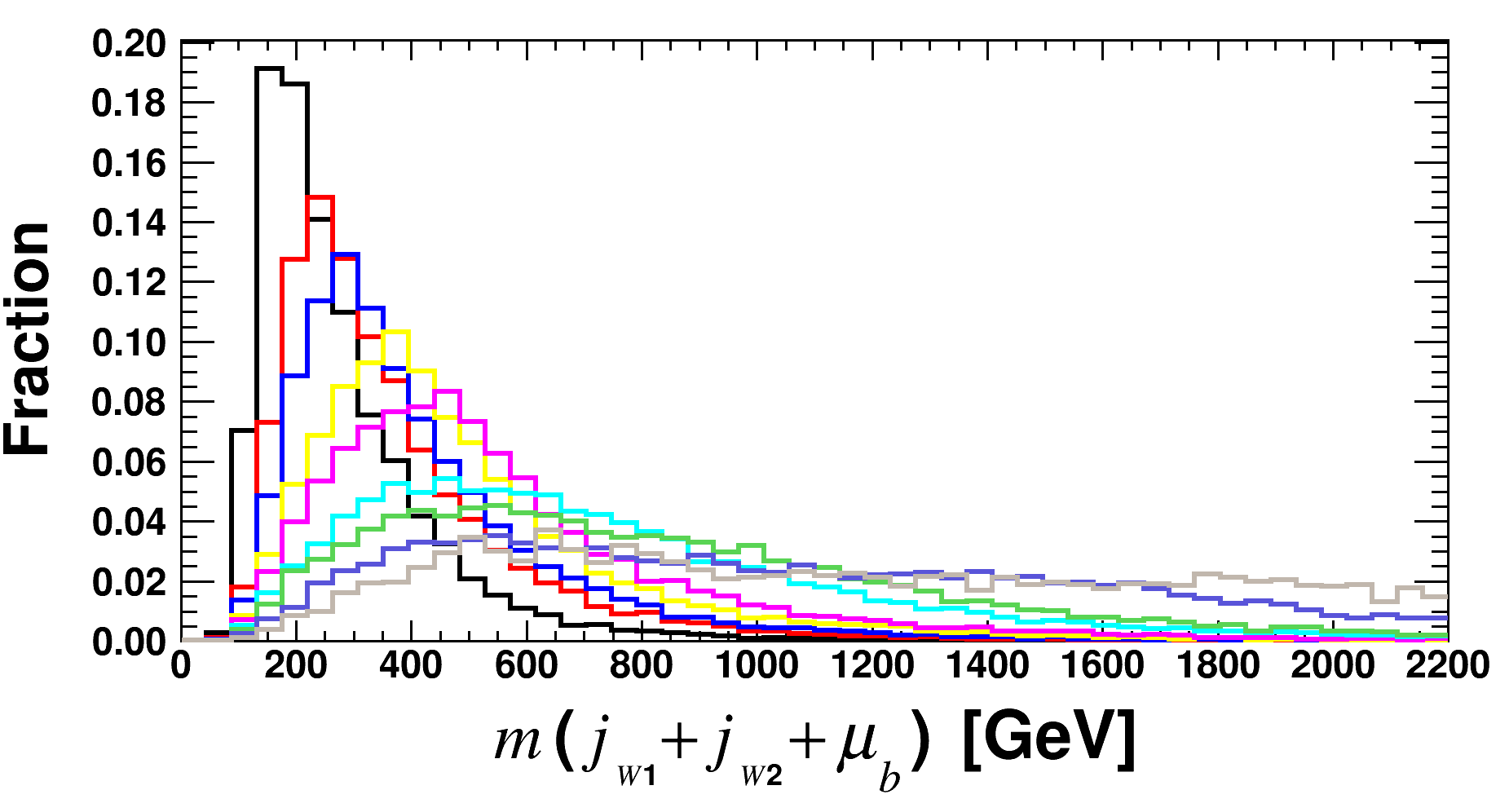}
}
\end{figure}
%\addtocounter{figure}{-1}
%
\vspace{-1.0cm}
\begin{figure}[H]
\centering
\subfigure{
\includegraphics[width=7.3cm, height=4.5cm]{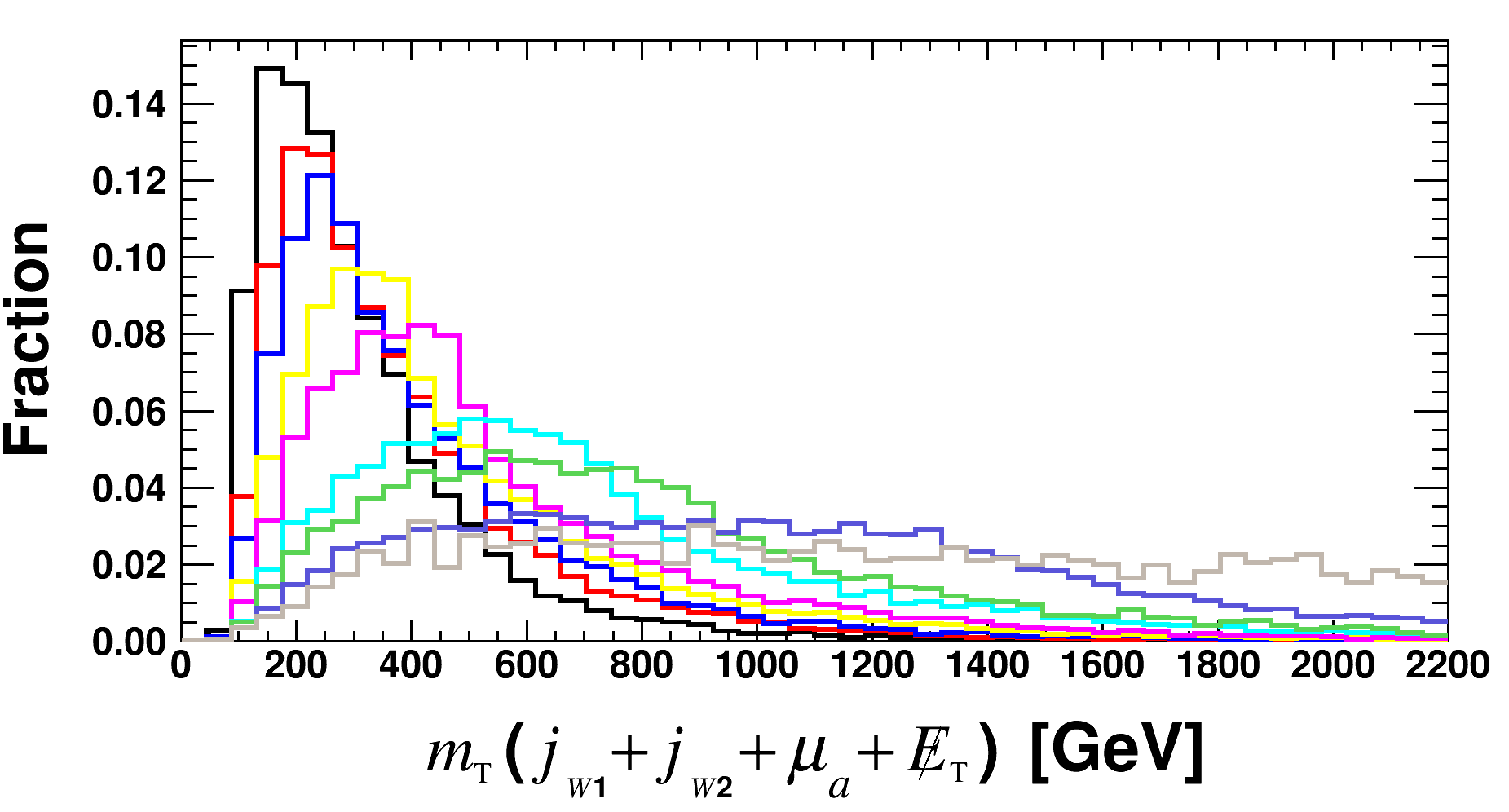}\,\,\,\,\,
\includegraphics[width=7.3cm, height=4.5cm]{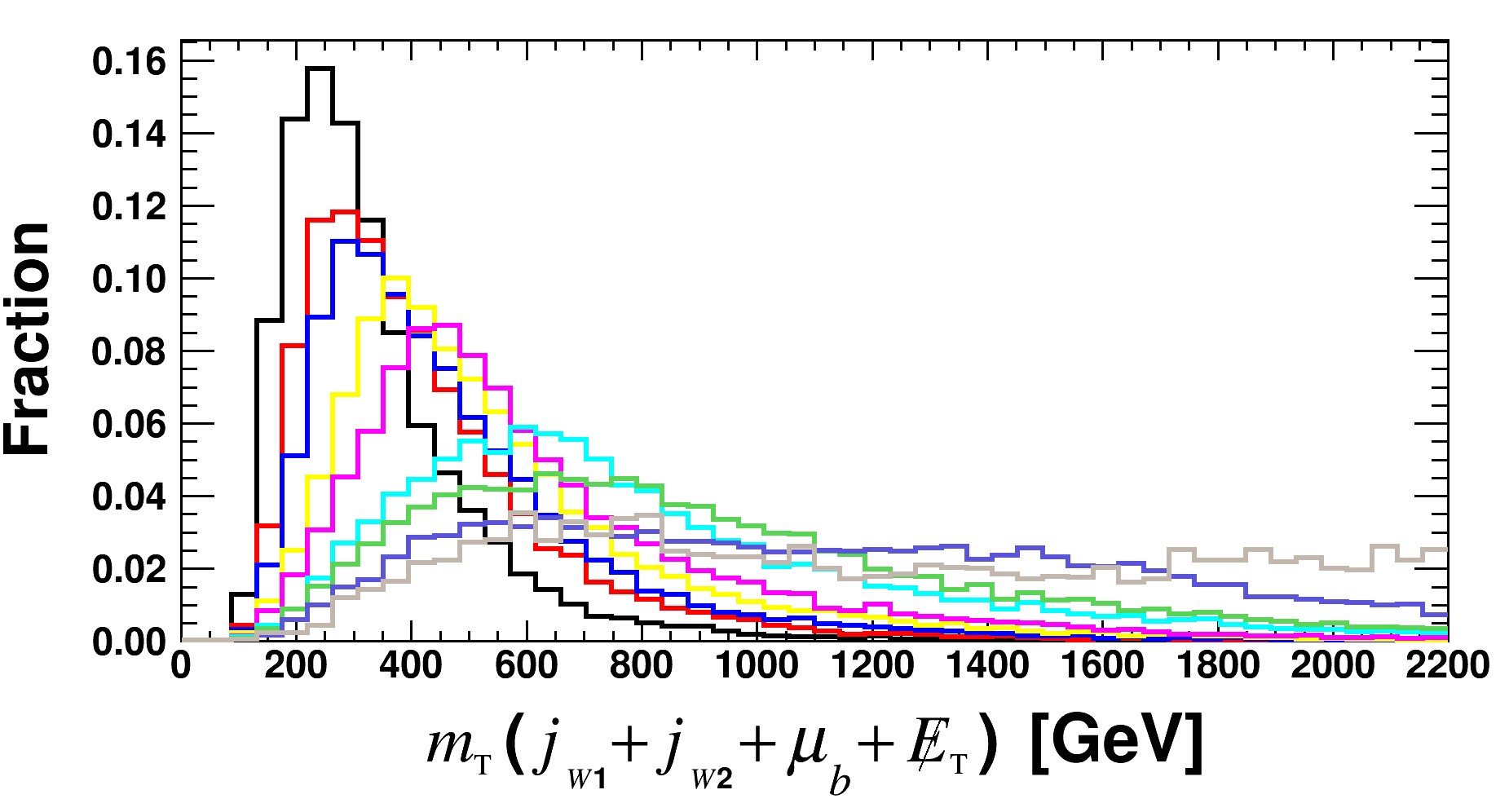}
}
\end{figure}
%\addtocounter{figure}{-1}
%
\vspace{-1.0cm}
\begin{figure}[H]
\centering
\subfigure{
\includegraphics[width=7.3cm, height=4.5cm]{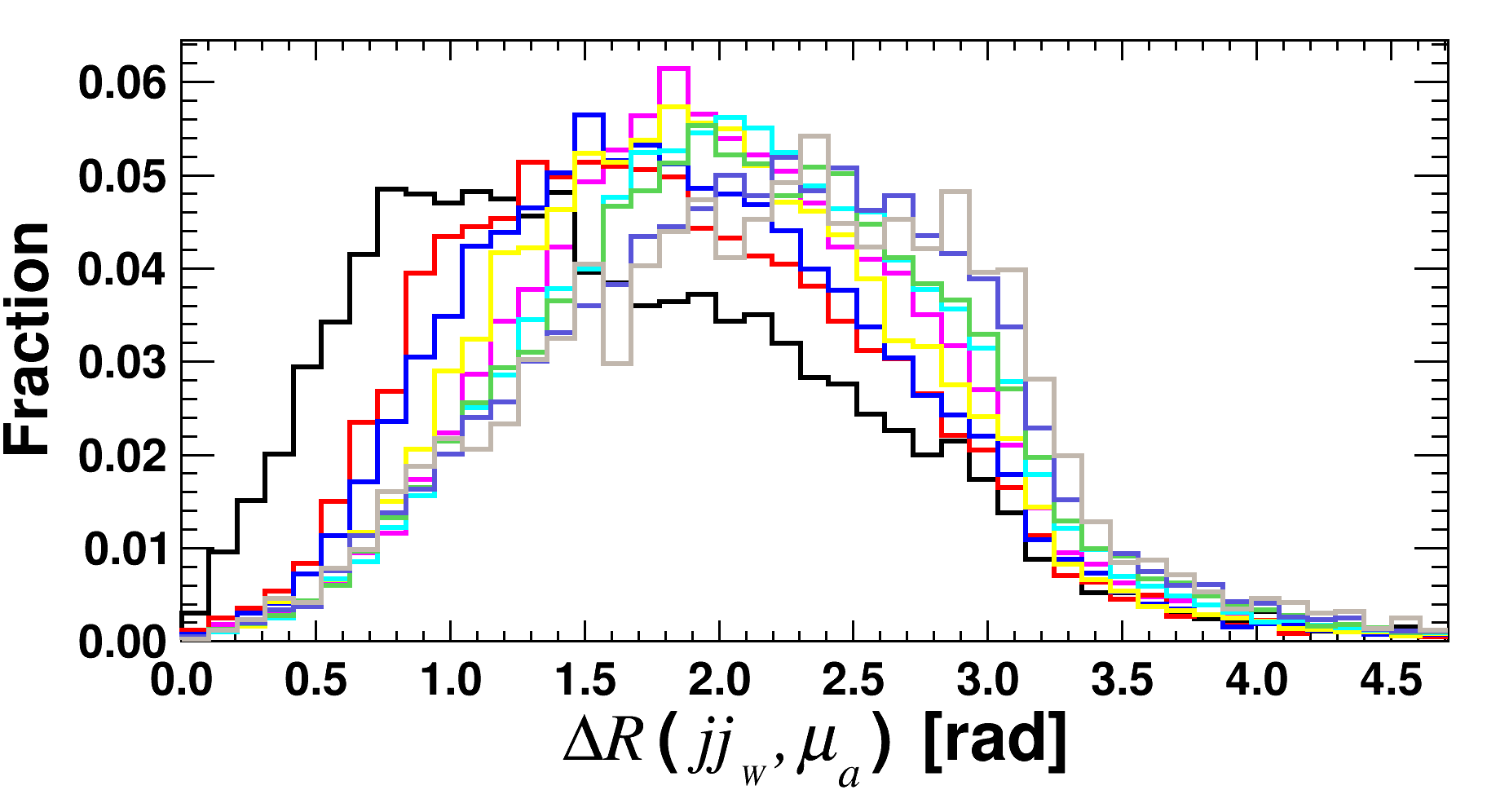}\,\,\,\,\,
\includegraphics[width=7.3cm, height=4.5cm]{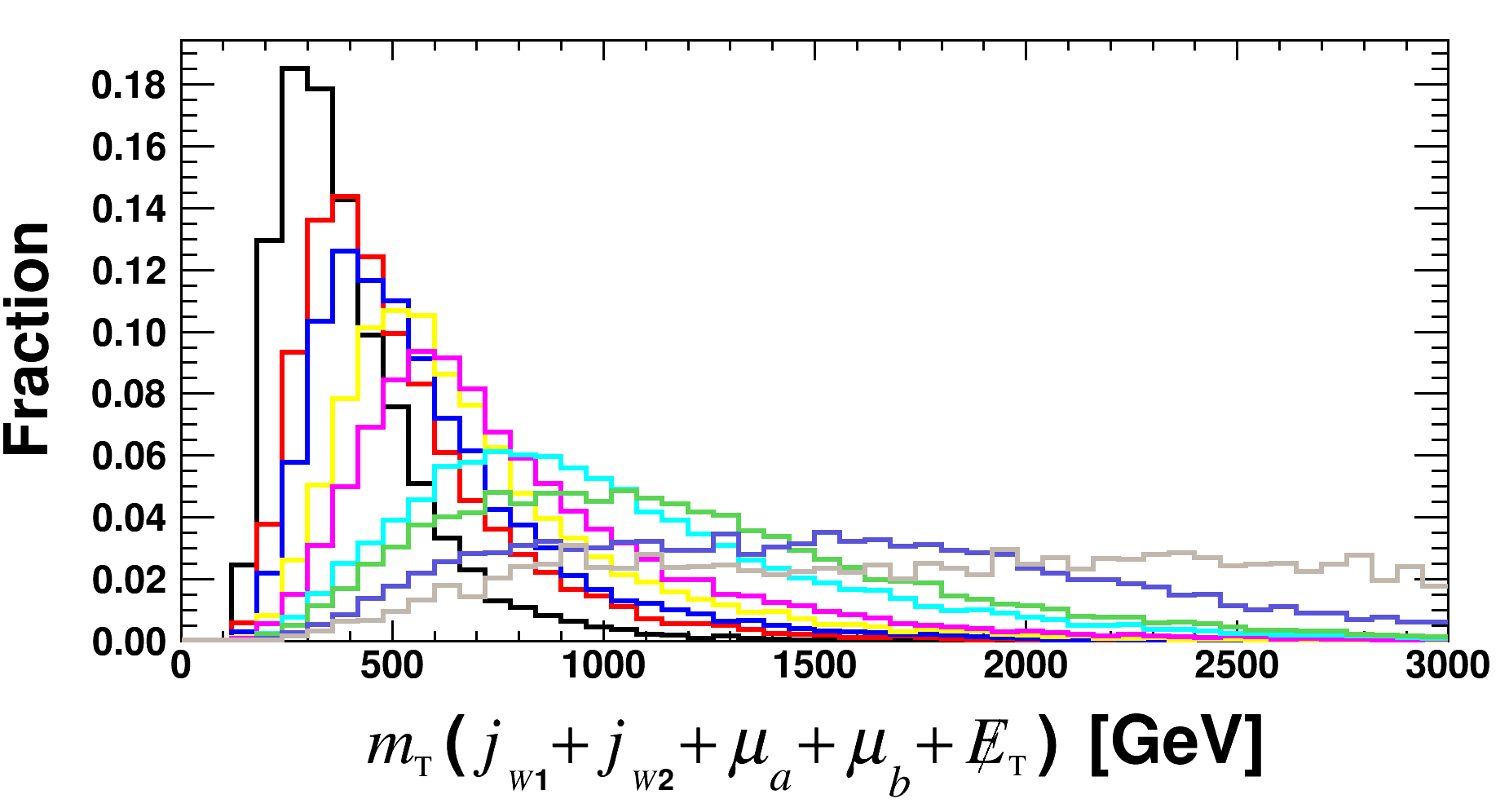}
}
\caption{
Distributions of same observables as Fig.~\ref{fig:ALPandWstar}, but for the signal only assuming different $m_a$ values.
}
\label{fig:ALPandWstarMasses}
\end{figure}

\section{Details of the MVA analyses}
\label{app:MVAdetails}

In this appendix, we provide detailed information about the MVA analyses performed using the TMVA package~\cite{TMVA:2007ngy}.
\begin{figure}[ht]
\centering
\includegraphics[width=7.3cm, height=4.5cm]{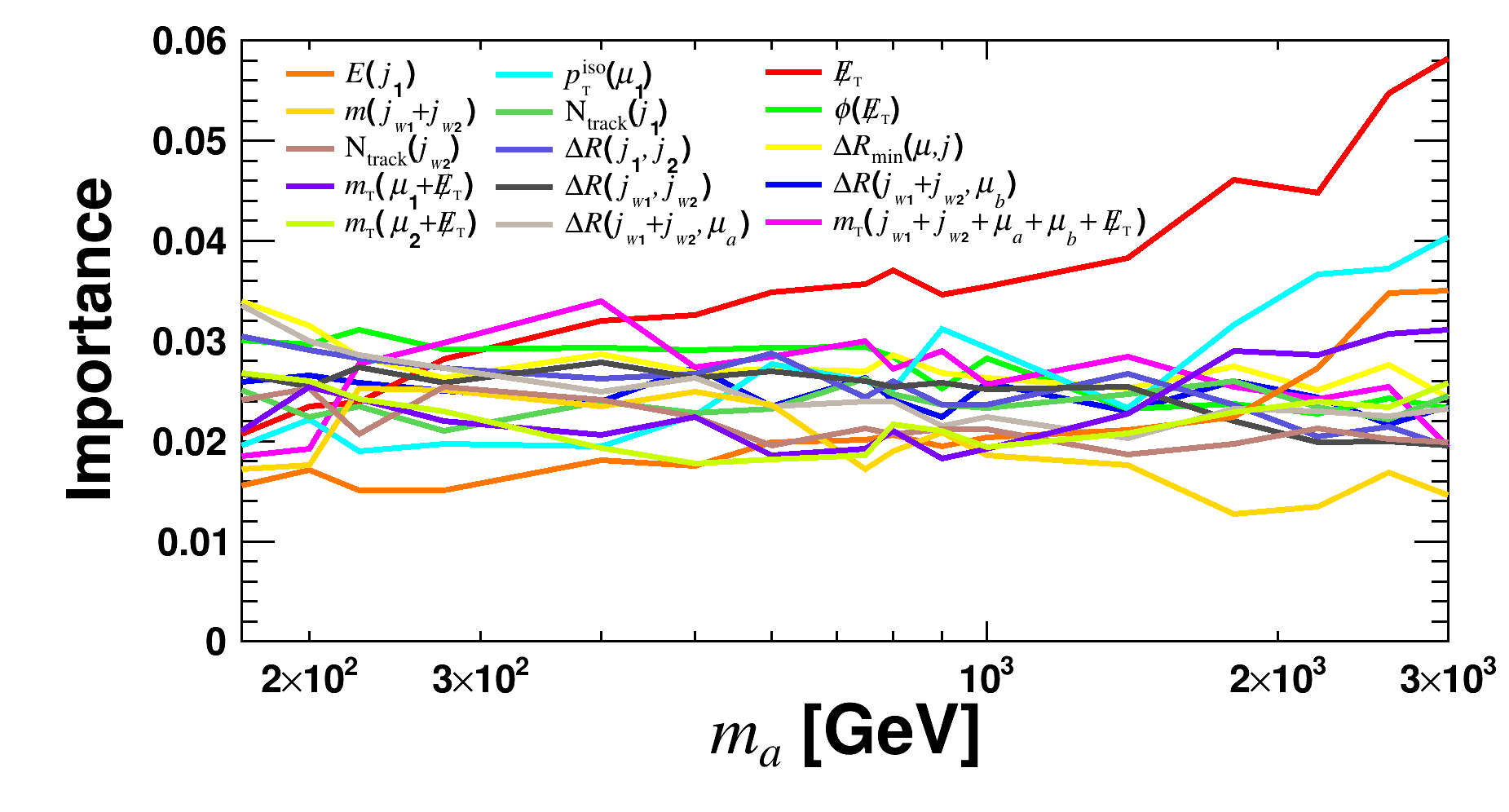}\,\,\,\,\,
\includegraphics[width=7.3cm, height=4.5cm]{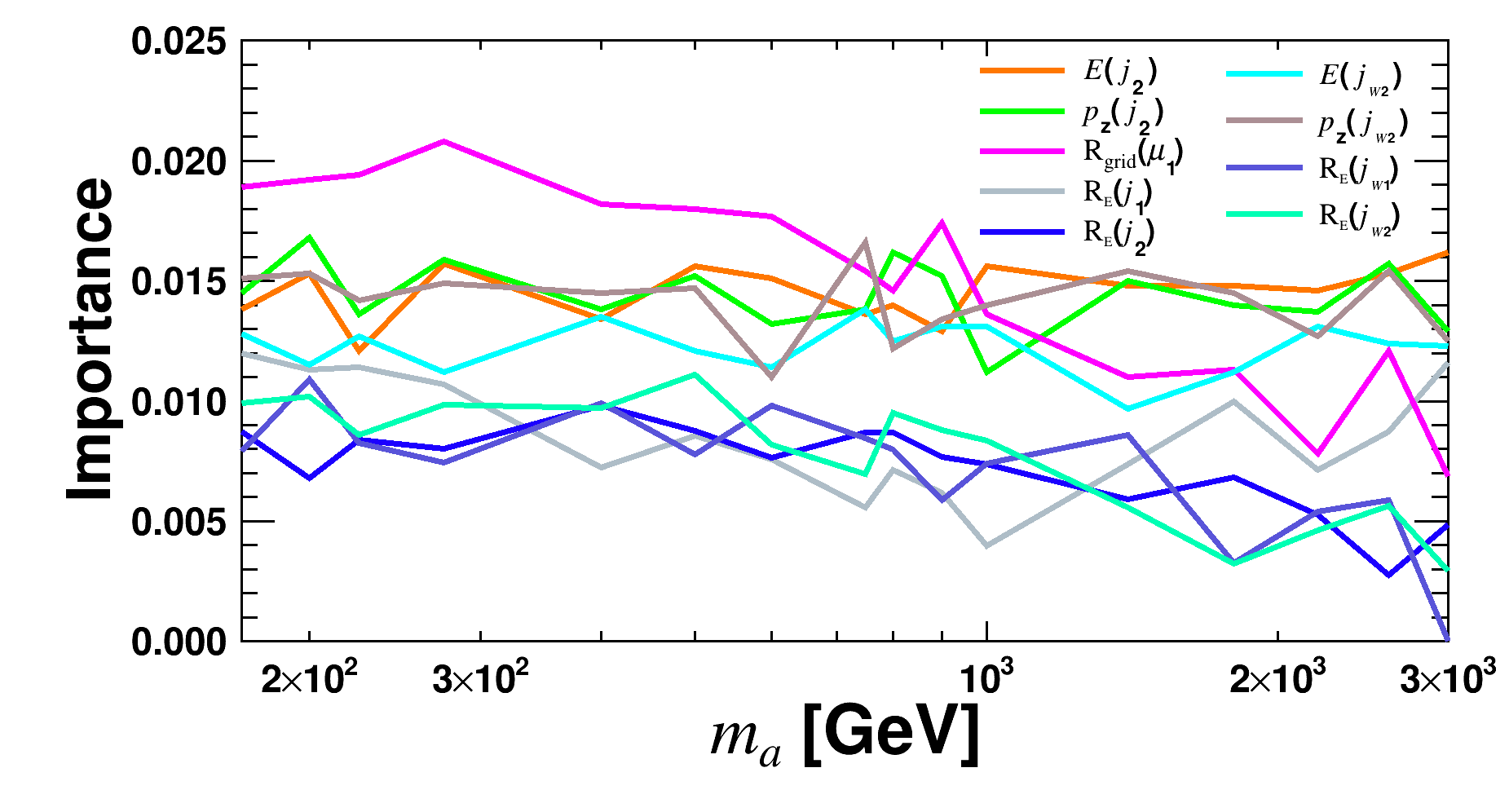}
\caption{The importance values of input observables as a function of \( m_a \), evaluated during the MVA training.
%The 15 observables in the left figure have the strongest influence on BDT at the range, and the weakest 9 observables in the left figure.
}
\label{fig:importance}
\end{figure}
Fig.~\ref{fig:importance} illustrates the importance values of input observables as a function of \( m_a \), evaluated during the MVA training.
The importance values for BDT input variables are quantified by counting how often the variables are used to split decision tree nodes, and by weighting each split occurrence by the separation gain-squared it has achieved and by the number of events in the node~\cite{TMVA:2007ngy, Breiman:2017lcz}.
The left plot highlights observables with the highest importance values, while those with the lowest importance values are displayed in the right panel.

Table~\ref{tab:correlation_400gev_bg_horizontal} presents the correlations between two observables for the total background and the signal with a benchmark mass $m_a = 400$ GeV. The listed 12 pairs have strong correlations and remain stable under variations of $m_a$.
\begin{table*}[h]
\centering
\scalebox{1}{
%\begin{ruledtabular}
\begin{tabular}{c c c c c c c c c c c}
\hline\hline
observables & signal & background \\
\hline
$p^{\rm iso}_{\rm T}(\mu_1)$ ,\,\,  $p^{\rm iso}_{\rm T,\,{\rm max}}(\mu)$&$1.000$&$0.998$\\
%\hline
$m_{\rm T}(j_{_{W1}}+j_{_{W2}}+\mu_b+\met)$,\,\,  $m_{\rm T}(j_{_{W1}}+j_{_{W2}}+\mu_a+\mu_b+\met)$&$0.919$&$0.974$\\
%\hline
$m(j_{_{W1}}+j_{_{W2}}+\mu_b)$,\,\,  $m_{\rm T}(j_{_{W1}}+j_{_{W2}}+\mu_a+\mu_b+\met)$&$0.797$&$0.955$\\
%\hline
$m(j_{_{W1}}+j_{_{W2}}+\mu_a)$,\,\,  $m(j_{_{W1}}+j_{_{W2}})$&$0.845$&$0.865$\\
%\hline
$R_{\rm E}(j_1)$,\,\,  $R_{\rm E}(j_{_{W1}})$&$0.762$&$0.896$\\
%\hline
 $E (j_1)$,\,\,  $E (j_{_{W1}})$&$0.747$&$0.850$\\
% \hline
$m(j_{_{W1}}+j_{_{W2}}+\mu_b)$,\,\,  $m_{\rm T}(j_{_{W1}}+j_{_{W2}}+\mu_b+\met)$&$0.724$&$0.926$\\
% \hline
$m(j_{_{W1}}+j_{_{W2}}+\mu_b)$,\,\,  $m(j_{_{W1}}+j_{_{W2}})$&$0.710$&$0.873$\\
% \hline
 $m_{\rm T}(j_{_{W1}}+j_{_{W2}}+\mu_b+\met)$,\,\,  $m(j_{_{W1}}+j_{_{W2}})$&$0.715$&$0.861$\\
% \hline
$p_z(j_1)$,\,\,  $p_z(j_{_{W1}})$&$0.791$&$0.838$\\
% \hline
$m(j_{_{W1}}+j_{_{W2}}+\mu_b)$,\,\,  $m(j_{_{W1}}+j_{_{W2}}+\mu_a)$&$0.740$&$0.861$\\
% \hline
$N_{\rm track}(j_1)$ ,\,\,  $N_{\rm track}(j_{_{W1}})$&$0.761$ & $0.772$ \\
\hline
\hline
\end{tabular}
%\end{ruledtabular}
}
\caption{
The correlations between two observables for the total background and the signal with a benchmark mass $m_a = 400$ GeV.
}
\label{tab:correlation_400gev_bg_horizontal}
\end{table*}
Fig.~\ref{fig:correlations} displays the correlations between pairs of observables for the signal as a function of \( m_a \). The correlations for the selected pairs exhibit significant dependence on \( m_a \).
\begin{figure}[h]
\centering
\includegraphics[width=11cm, height=6.8cm]{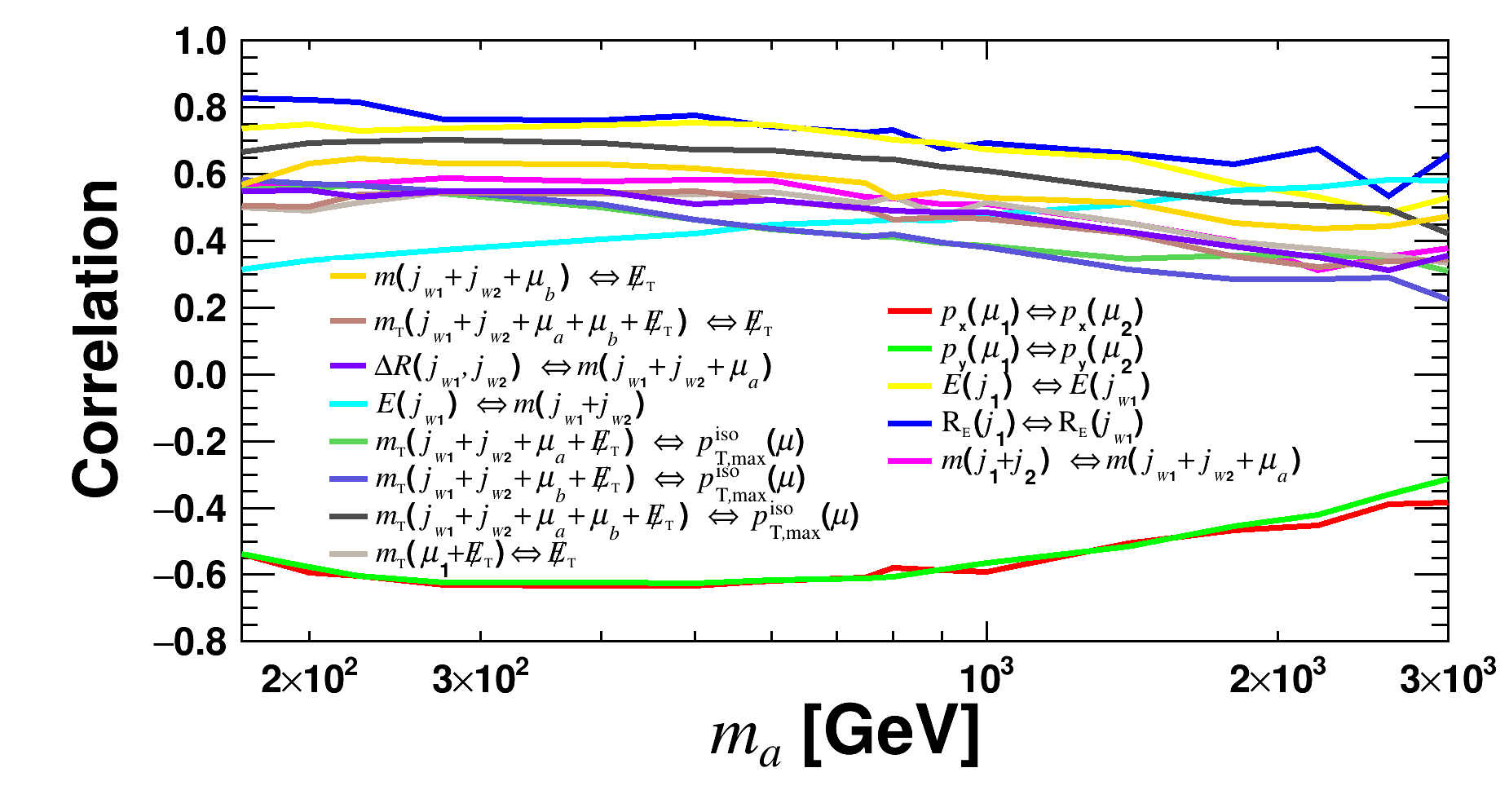}
\caption{
The correlations between pairs of observables for the signal as a function of \( m_a \).
}
\label{fig:correlations}
\end{figure}

%上面是横版，下面是竖版
%\begin{table*}[h]
%\centering
%\scalebox{0.5}{
%\begin{ruledtabular}
%\begin{tabular}{c c c c c c c c c c c}
%\hline
%\hline
%\multirow{1}{*}{ } &  & $p^{\rm iso}_{\rm T}(\mu_1)$,\,\,  $p^{\rm iso}_{\rm T,\,{\rm max}}(\mu)$& $m_{\rm T}(j_{_{W1}}+j_{_{W2}}+\mu_b+\met)$,\,\,  $m_{\rm T}(j_{_{W1}}+j_{_{W2}}+\mu_a+\mu_b+\met)$&$m(j_{_{W1}}+j_{_{W2}}+\mu_b)$,\,\,  $m_{\rm T}(j_{_{W1}}+j_{_{W2}}+\mu_a+\mu_b+\met)$&$m(j_{_{W1}}+j_{_{W2}}+\mu_a)$,\,\,  $m(j_{_{W1}}+j_{_{W2}})$\\
%& signal($m_a =$ 400 GeV) & $1.000$&$0.919$&$0.797$&$0.845$\\
%& background&$0.998$&$0.974$&$0.955$&$0.865$\\
			%\cline{2-6}
%\hline
%\multirow{3}{*}{ } &  & $R_{\rm E}(j_1)$,\,\,  $R_{\rm E}(j_{_{W1}})$& $E (j_1)$,\,\,  $E (j_{_{W1}})$&$m(j_{_{W1}}+j_{_{W2}}+\mu_b)$,\,\,  $m_{\rm T}(j_{_{W1}}+j_{_{W2}}+\mu_b+\met)$&$m(j_{_{W1}}+j_{_{W2}}+\mu_b)$,\,\,  $m(j_{_{W1}}+j_{_{W2}})$\\
%& signal($m_a =$ 400 GeV) & $0.762$&$0.747$&$0.724$&$0.710$\\
%& background&$0.896$&$0.850$&$0.926$&$0.873$\\
			%\cline{2-6}
%\hline
%\multirow{3}{*}{ } &  & $m_{\rm T}(j_{_{W1}}+j_{_{W2}}+\mu_b+\met)$,\,\,  $m(j_{_{W1}}+j_{_{W2}})$& $p_z(j_1)$,\,\,  $p_z(j_{_{W1}})$&$m(j_{_{W1}}+j_{_{W2}}+\mu_b)$,\,\,  $m(j_{_{W1}}+j_{_{W2}}+\mu_a)$&$N_{\rm track}(j_1)$,\,\,  $N_{\rm track}(j_{_{W1}})$\\
%& signal($m_a =$ 400 GeV) & $0.715$&$0.791$&$0.740$&$0.761$\\
%& background&$0.861$&$0.838$&$0.861$&$0.772$\\
			%\cline{2-6}
%\hline
%\hline
%\end{tabular}
%\end{ruledtabular}
%}
%\caption{

%}
%\label{tab:correlation_400gev&bg_Vertical}
%\end{table*}

\newpage
\section{Distributions of BDT responses}
\label{app:BDT}

\begin{figure}[h]
\centering
\includegraphics[width=7.3cm, height=4.5cm]{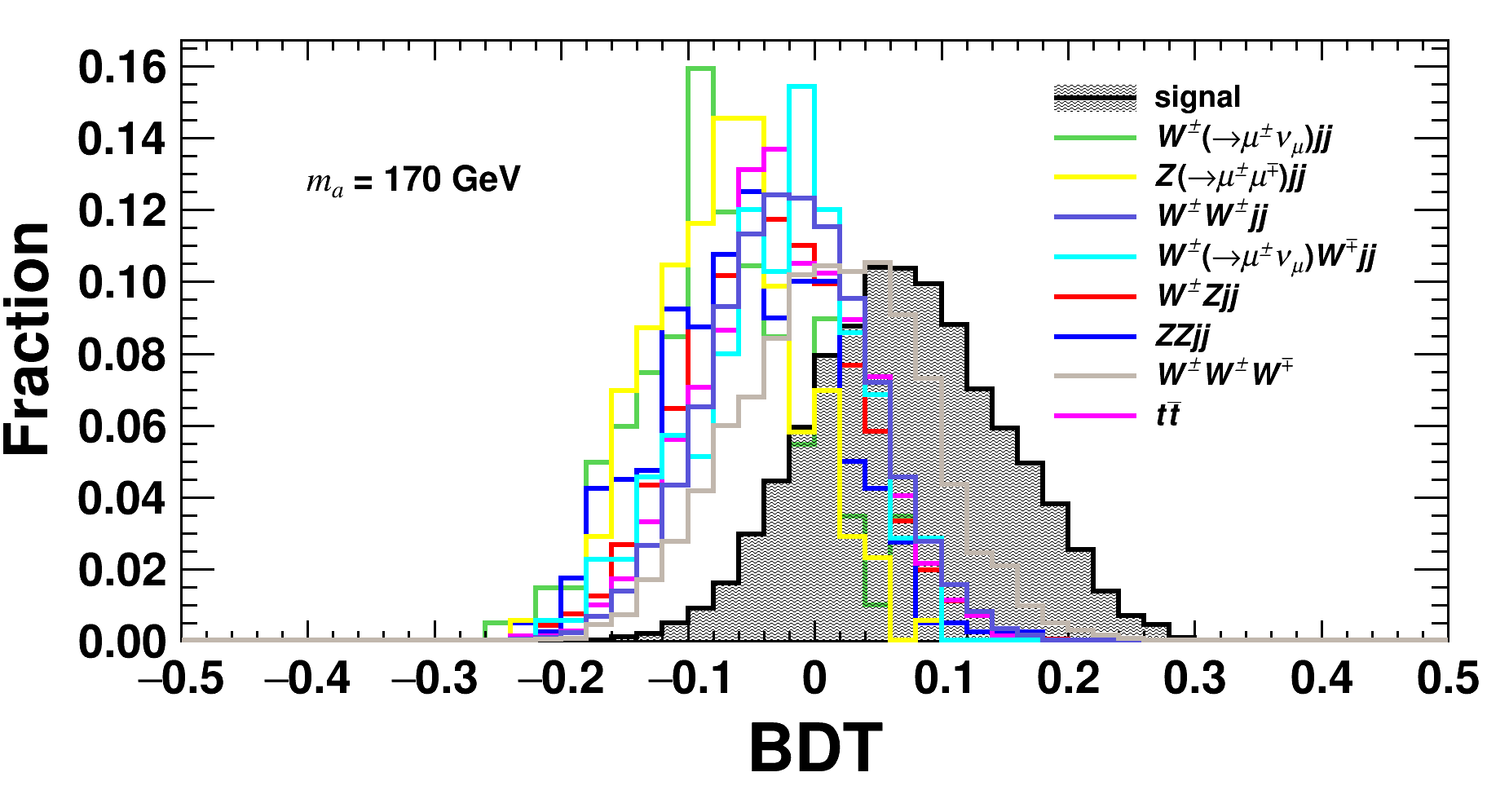}\,\,\,\,\,
\includegraphics[width=7.3cm, height=4.5cm]{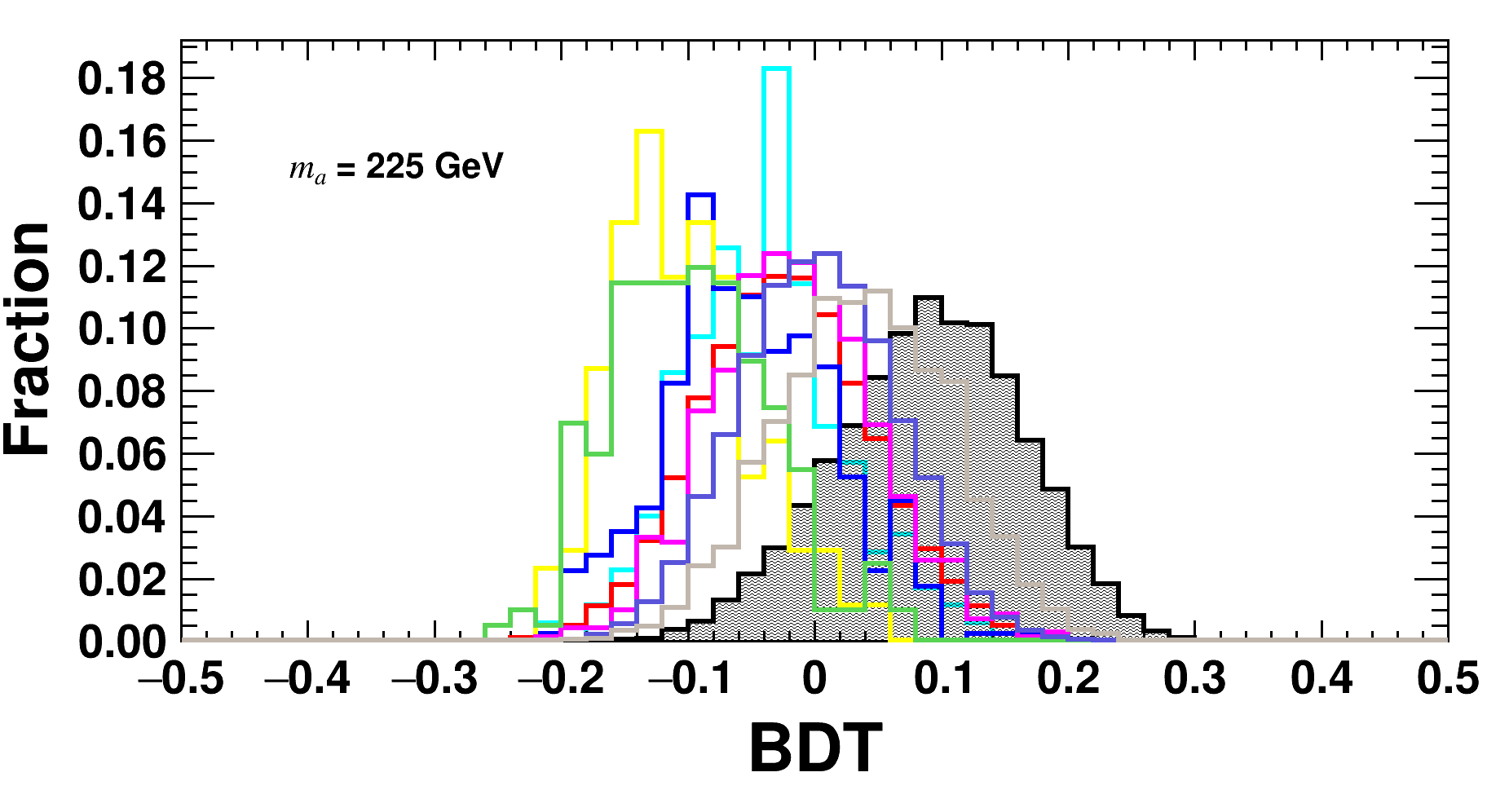}
\includegraphics[width=7.3cm, height=4.5cm]{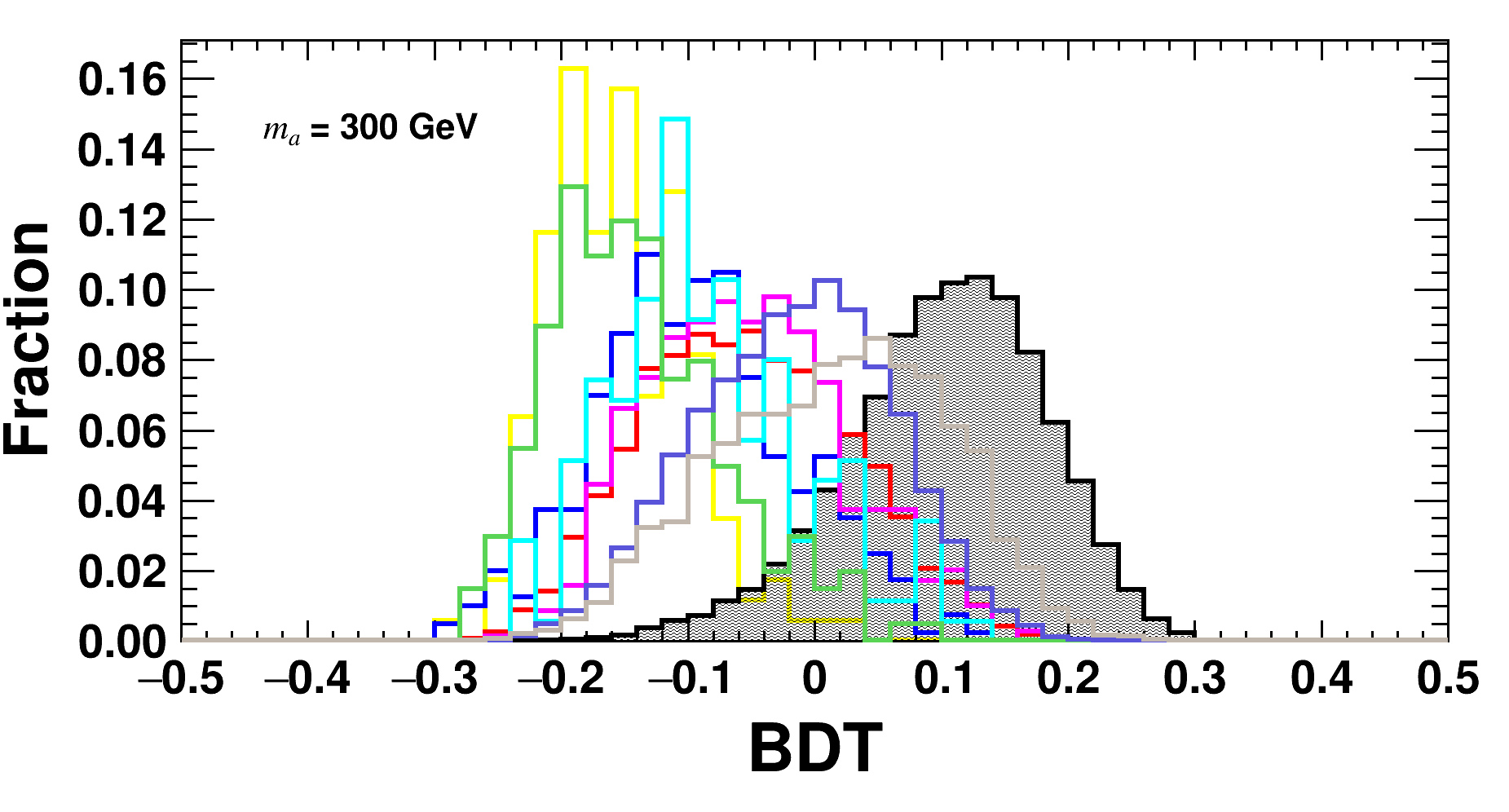}\,\,\,\,\,
\includegraphics[width=7.3cm, height=4.5cm]{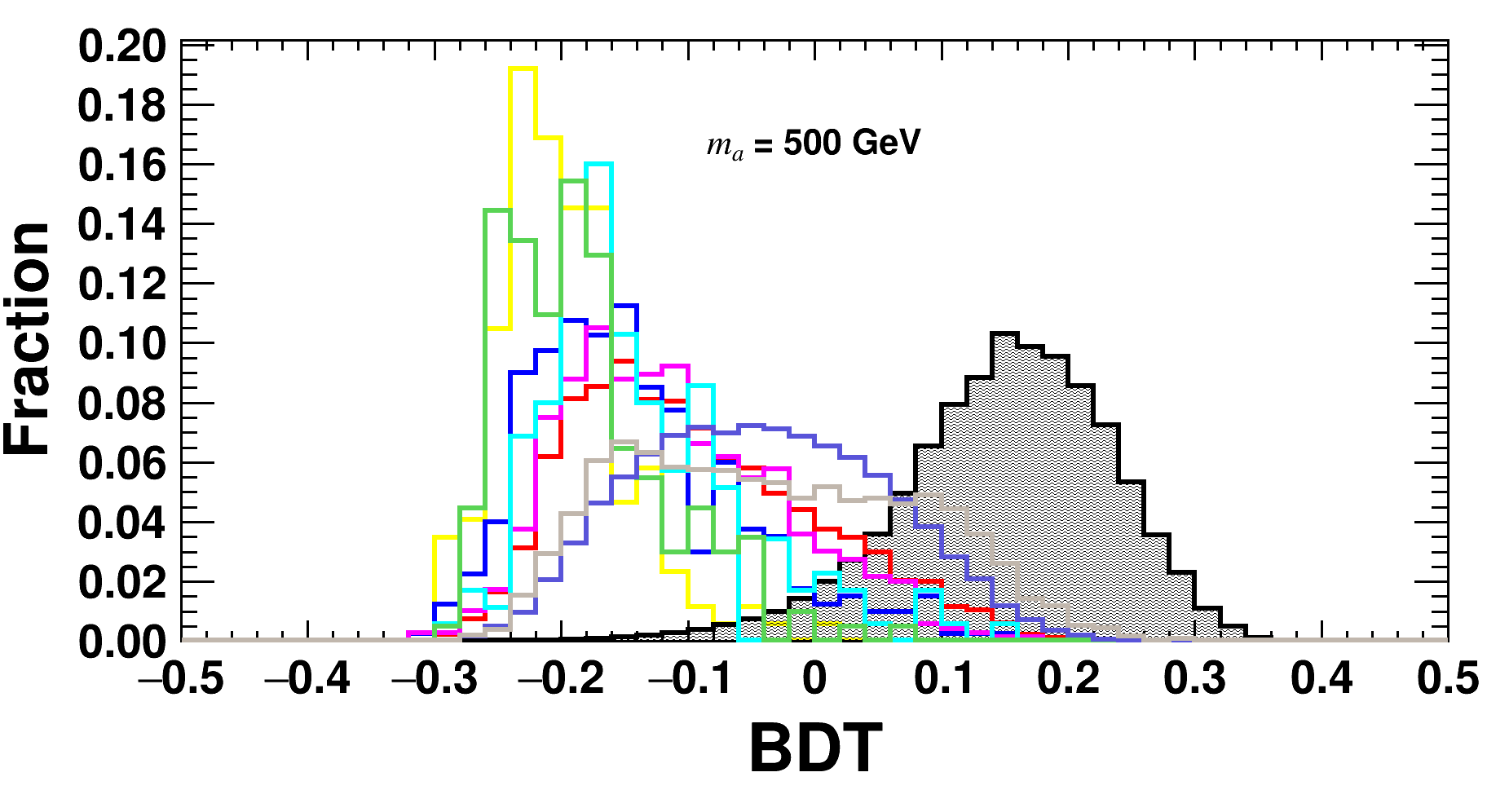}
\includegraphics[width=7.3cm, height=4.5cm]{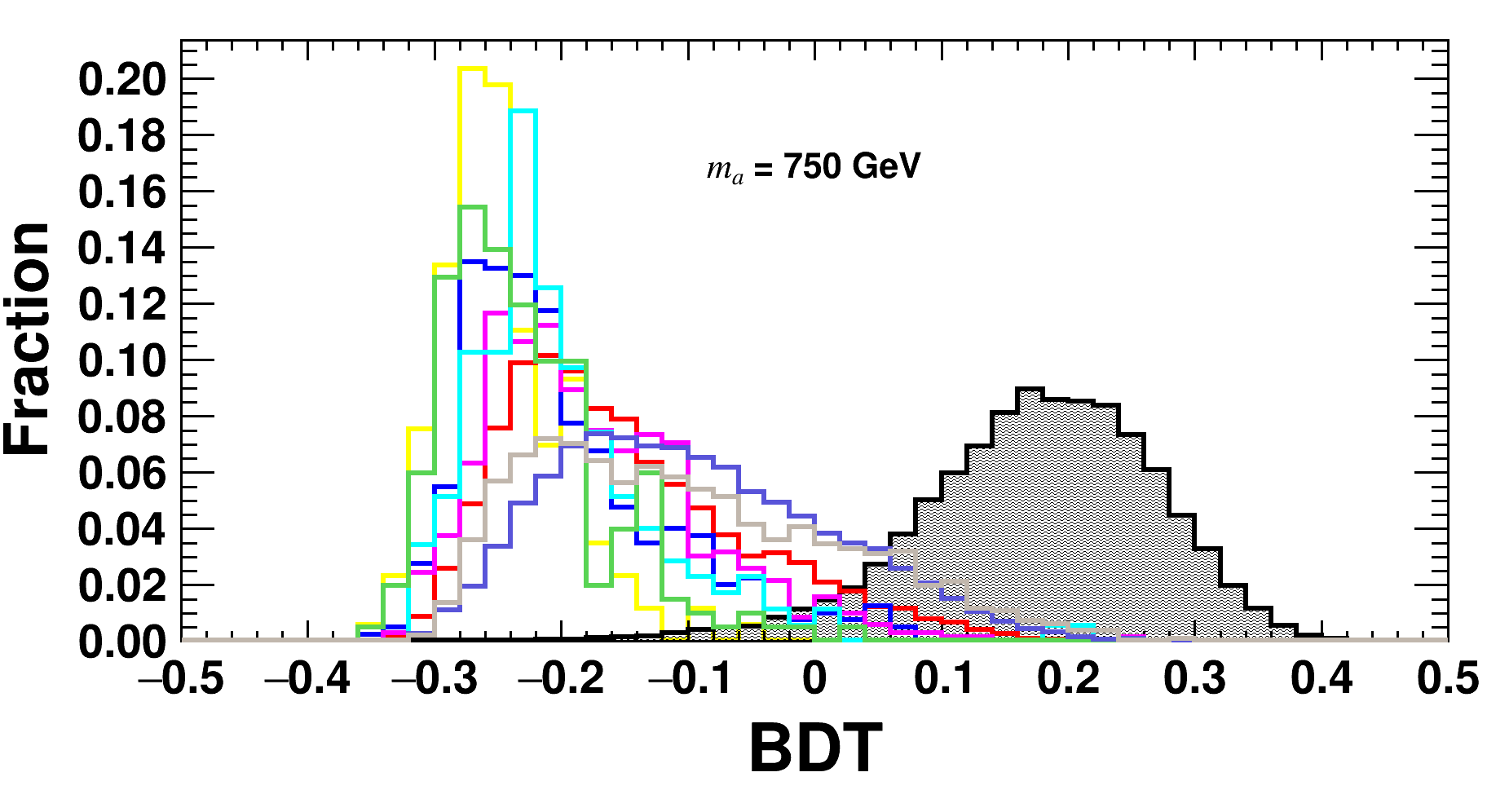}\,\,\,\,\,
\includegraphics[width=7.3cm, height=4.5cm]{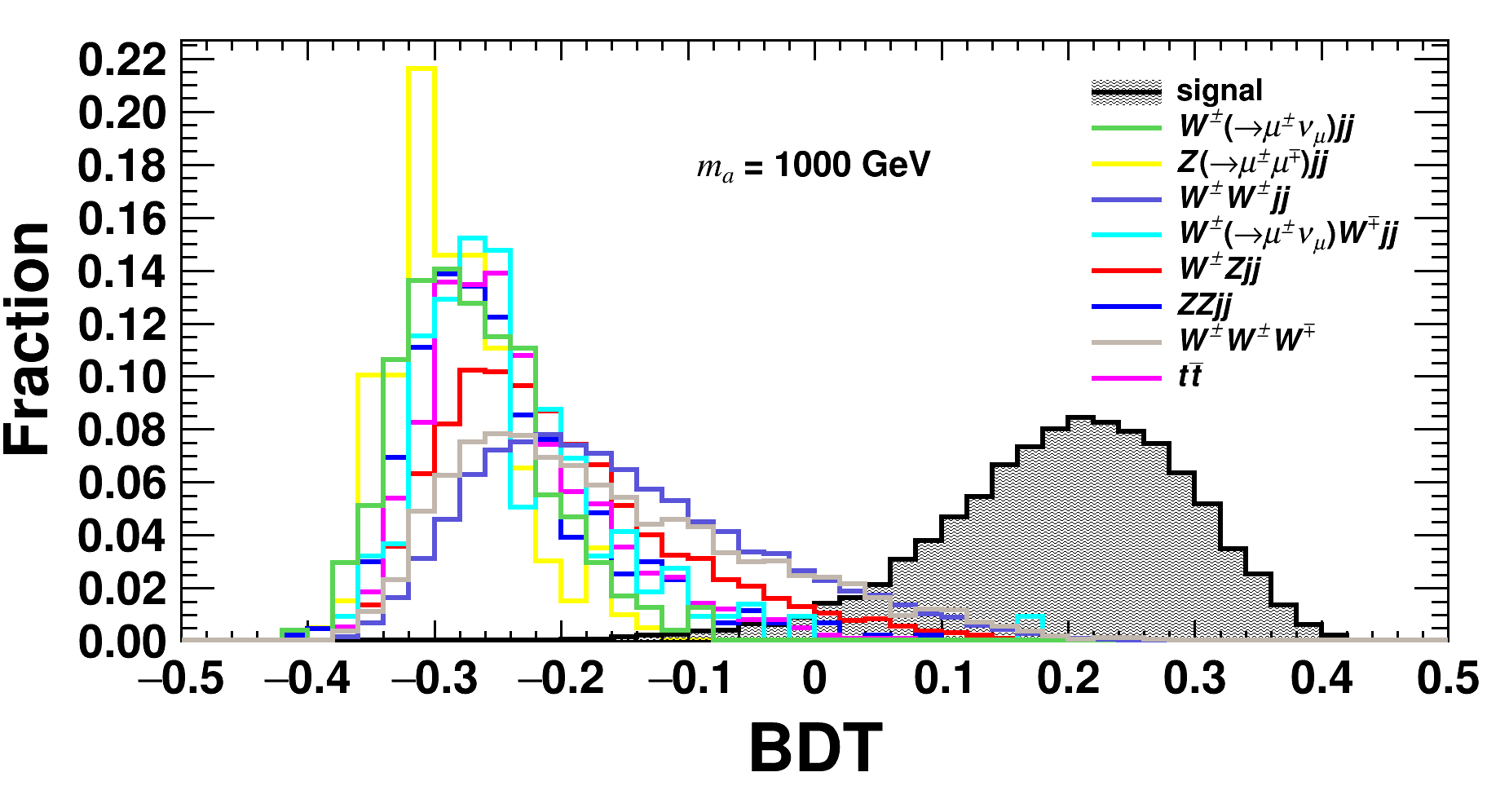}
\includegraphics[width=7.3cm, height=4.5cm]{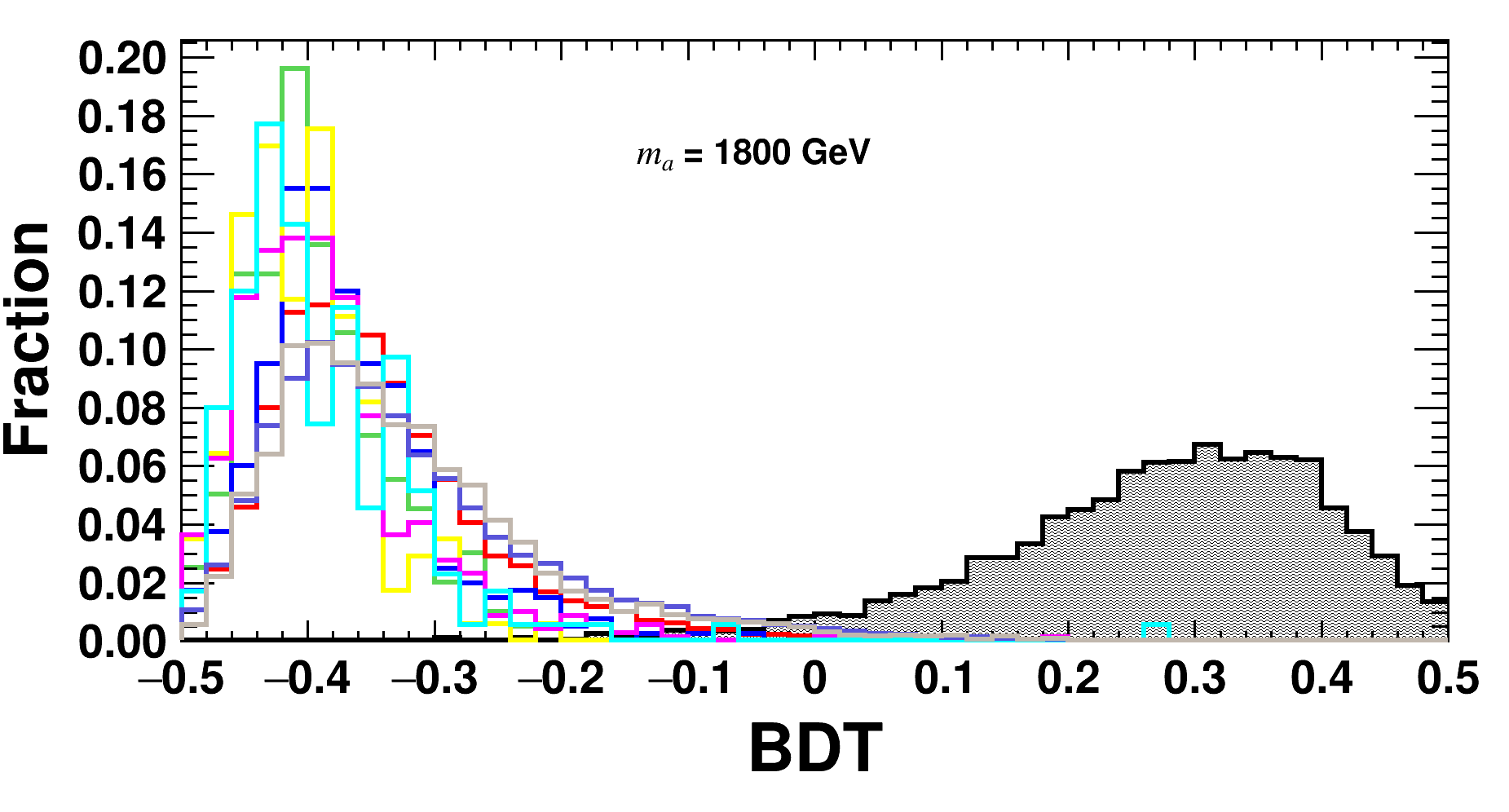}\,\,\,\,\,
\includegraphics[width=7.3cm, height=4.5cm]{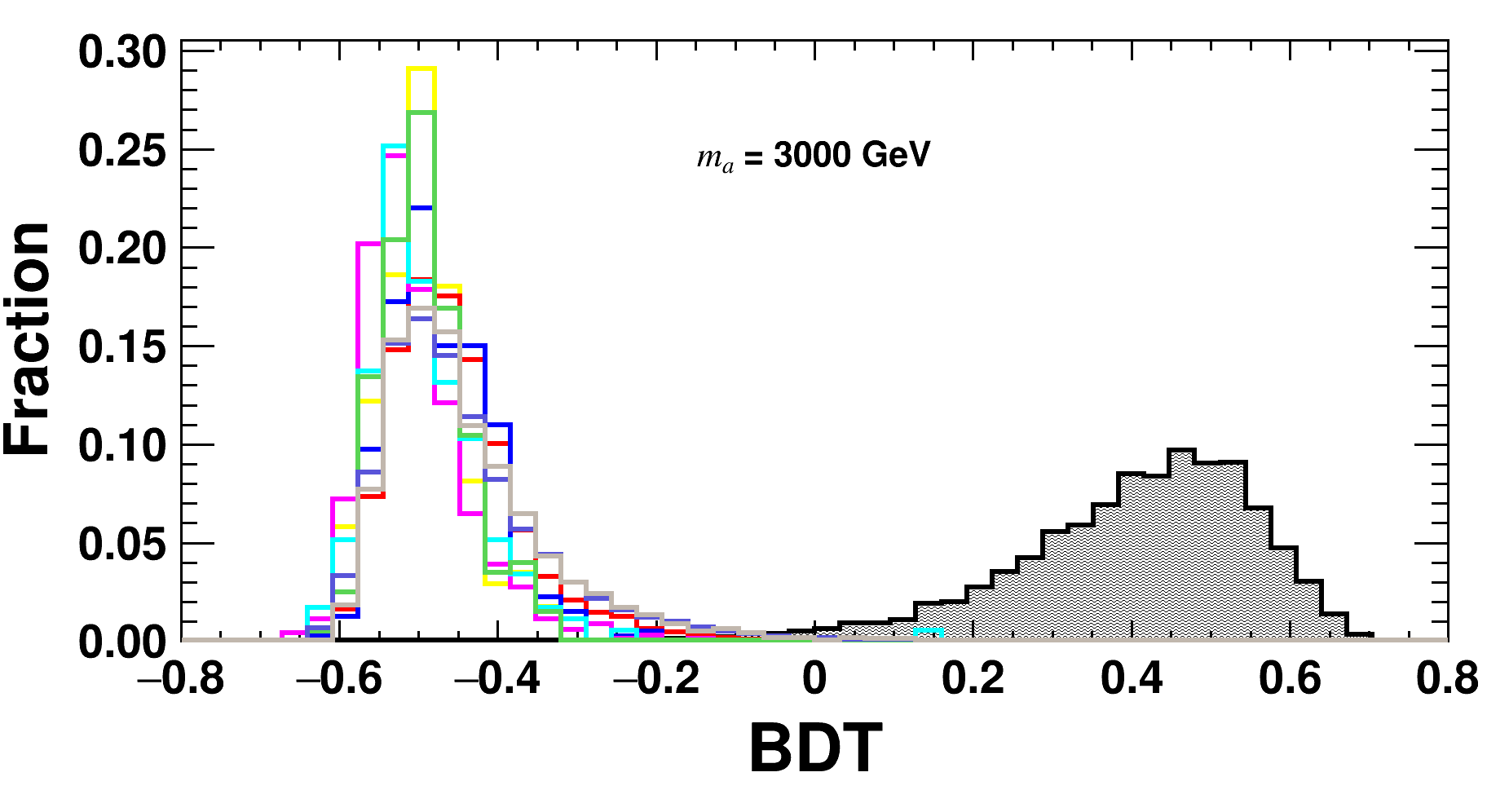}
\caption{
Distributions of BDT responses after applying preselection criteria for the signal (black, shadow) and background processes at the HL-LHC with $\sqrt{s} =$ 14 TeV in representative $m_{a}$ cases.
}
\label{fig:BDT_ALPmasses}
\end{figure}

\newpage
\section{The Selection Efficiency Table}
\label{app:Sel_Eff}

\begin{table*}[h]
\centering
\scalebox{0.55}{
%\begin{ruledtabular}
\begin{tabular}{c c c c c c c c c c c}
\hline
\hline
$m_a$ & selection & signal & $W^\pm Zjj$ & $ZZjj$ & $Z(\rightarrow\mu^{\pm}\mu^{\mp})jj$ & $t\bar{t}$ & $W^\pm(\rightarrow\mu^{\pm}\nu_{\mu}) W^\mp jj$ & $W^{\pm}(\rightarrow\mu^{\pm}\nu_{\mu}) jj$ & $W^\pm W^\pm jj$ & $W^\pm W^\pm W^\mp$\\
\hline
\multirow{2}{*}{170 GeV} & preselection & $2.18\mltp10^{-3}$&$1.93\mltp10^{-4}$&$5.00\mltp10^{-5}$&$2.01\mltp10^{-6}$&$1.63\mltp10^{-5}$&$8.97\mltp10^{-6}$ &$1.35\mltp10^{-6}$ &$3.53\mltp10^{-3}$ &$2.26\mltp10^{-3}$\\
&  BDT$>$0.127&$2.46\mltp10^{-1}$&$8.06\mltp10^{-3}$&$7.50\mltp10^{-3}$&$-$&$5.76\mltp10^{-3}$&$-$&$-$&$1.08\mltp10^{-2}$&$5.61\mltp10^{-2}$ \\
			%\cline{2-6}
\hline
\multirow{2}{*}{200 GeV}  & preselection & $2.79\mltp10^{-3}$&$1.93\mltp10^{-4}$&$5.00\mltp10^{-5}$&$2.01\mltp10^{-6}$&$1.63\mltp10^{-5}$&$8.97\mltp10^{-6}$ &$1.35\mltp10^{-6}$ &$3.53\mltp10^{-3}$ &$2.26\mltp10^{-3}$\\
& BDT$>$0.108 &$4.15\mltp10^{-1}$&$3.93\mltp10^{-2}$&$1.50\mltp10^{-2}$&$-$&$2.74\mltp10^{-2}$&$1.14\mltp10^{-2}$&$-$&$5.77\mltp10^{-2}$&$1.86\mltp10^{-1}$ \\
			%\cline{2-6}
\hline
\multirow{2}{*}{225 GeV}  & preselection & $3.24\mltp10^{-3}$&$1.93\mltp10^{-4}$&$5.00\mltp10^{-5}$&$2.01\mltp10^{-6}$&$1.63\mltp10^{-5}$&$8.97\mltp10^{-6}$ &$1.35\mltp10^{-6}$ &$3.53\mltp10^{-3}$ &$2.26\mltp10^{-3}$\\
& BDT$>$0.101  &$4.54\mltp10^{-1}$&$3.70\mltp10^{-2}$&$7.50\mltp10^{-3}$&$-$&$4.61\mltp10^{-2}$&$1.71\mltp10^{-2}$&$-$&$5.63\mltp10^{-2}$&$1.91\mltp10^{-1}$ \\
			%\cline{2-6}
\hline
\multirow{2}{*}{275 GeV} & preselection & $3.76\mltp10^{-3}$&$1.93\mltp10^{-4}$&$5.00\mltp10^{-5}$&$2.01\mltp10^{-6}$&$1.63\mltp10^{-5}$&$8.97\mltp10^{-6}$ &$1.35\mltp10^{-6}$ &$3.53\mltp10^{-3}$ &$2.26\mltp10^{-3}$\\
& BDT$>$0.115  &$4.79\mltp10^{-1}$&$3.04\mltp10^{-2}$&$1.00\mltp10^{-2}$&$-$&$1.59\mltp10^{-2}$&$5.71\mltp10^{-3}$&$-$&$4.95\mltp10^{-2}$&$1.66\mltp10^{-1}$ \\  
			%\cline{2-6}
\hline
\multirow{2}{*}{400 GeV} & preselection & $4.40\mltp10^{-3}$&$1.93\mltp10^{-4}$&$5.00\mltp10^{-5}$&$2.01\mltp10^{-6}$&$1.63\mltp10^{-5}$&$8.97\mltp10^{-6}$ &$1.35\mltp10^{-6}$ &$3.53\mltp10^{-3}$ &$2.26\mltp10^{-3}$\\
& BDT$>$0.136 &$4.99\mltp10^{-1}$&$1.12\mltp10^{-2}$&$-$&$-$&$5.76\mltp10^{-3}$&$-$&$-$&$2.49\mltp10^{-2}$&$8.76\mltp10^{-2}$ \\
\hline
\multirow{2}{*}{500 GeV}  & preselection & $4.61\mltp10^{-3}$&$1.93\mltp10^{-4}$&$5.00\mltp10^{-5}$&$2.01\mltp10^{-6}$&$1.63\mltp10^{-5}$&$8.97\mltp10^{-6}$ &$1.35\mltp10^{-6}$ &$3.53\mltp10^{-3}$ &$2.26\mltp10^{-3}$\\
& BDT$>$0.165 &$5.38\mltp10^{-1}$&$3.60\mltp10^{-3}$&$-$&$-$&$1.44\mltp10^{-3}$&$-$&$-$&$1.20\mltp10^{-2}$&$3.49\mltp10^{-2}$ \\
\hline
\multirow{2}{*}{600 GeV}  & preselection & $4.71\mltp10^{-3}$&$1.93\mltp10^{-4}$&$5.00\mltp10^{-5}$&$2.01\mltp10^{-6}$&$1.63\mltp10^{-5}$&$8.97\mltp10^{-6}$ &$1.35\mltp10^{-6}$ &$3.53\mltp10^{-3}$ &$2.26\mltp10^{-3}$\\
& BDT$>$0.173&$4.72\mltp10^{-1}$&$2.45\mltp10^{-3}$&$-$&$-$&$1.44\mltp10^{-3}$&$1.14\mltp10^{-2}$&$-$&$7.13\mltp10^{-3}$&$2.54\mltp10^{-2}$ \\                                       
\hline
\multirow{2}{*}{750 GeV} & preselection & $4.64\mltp10^{-3}$&$1.93\mltp10^{-4}$&$5.00\mltp10^{-5}$&$2.01\mltp10^{-6}$&$1.63\mltp10^{-5}$&$8.97\mltp10^{-6}$ &$1.35\mltp10^{-6}$ &$3.53\mltp10^{-3}$ &$2.26\mltp10^{-3}$\\
& BDT$>$0.167&$5.68\mltp10^{-1}$&$2.02\mltp10^{-3}$&$-$&$-$&$1.44\mltp10^{-3}$&$1.14\mltp10^{-2}$&$-$&$9.64\mltp10^{-3}$&$2.04\mltp10^{-2}$ \\                                        
\hline
\multirow{2}{*}{800 GeV} & preselection & $4.77\mltp10^{-3}$&$1.93\mltp10^{-4}$&$5.00\mltp10^{-5}$&$2.01\mltp10^{-6}$&$1.63\mltp10^{-5}$&$8.97\mltp10^{-6}$ &$1.35\mltp10^{-6}$ &$3.53\mltp10^{-3}$ &$2.26\mltp10^{-3}$\\
& BDT$>$0.165 & $7.17\mltp10^{-1}$&$3.46\mltp10^{-3}$&$-$&$-$&$1.44\mltp10^{-3}$&$1.14\mltp10^{-2}$&$-$&$1.00\mltp10^{-2}$&$2.02\mltp10^{-2}$ \\                                           
\hline
\multirow{2}{*}{900 GeV} & preselection & $4.70\mltp10^{-3}$&$1.93\mltp10^{-4}$&$5.00\mltp10^{-5}$&$2.01\mltp10^{-6}$&$1.63\mltp10^{-5}$&$8.97\mltp10^{-6}$ &$1.35\mltp10^{-6}$ &$3.53\mltp10^{-3}$ &$2.26\mltp10^{-3}$\\
&  BDT$>$0.180  	&$5.68\mltp10^{-1}$&$1.44\mltp10^{-3}$&$-$&$-$&$1.44\mltp10^{-3}$&$5.71\mltp10^{-3}$&$-$&$5.01\mltp10^{-3}$&$1.01\mltp10^{-2}$ \\                                            
\hline
\multirow{2}{*}{1000 GeV} & preselection & $4.72\mltp10^{-3}$&$1.93\mltp10^{-4}$&$5.00\mltp10^{-5}$&$2.01\mltp10^{-6}$&$1.63\mltp10^{-5}$&$8.97\mltp10^{-6}$ &$1.35\mltp10^{-6}$ &$3.53\mltp10^{-3}$ &$2.26\mltp10^{-3}$\\
&  BDT$>$0.140&$8.69\mltp10^{-1}$&$3.17\mltp10^{-3}$&$-$&$-$&$2.88\mltp10^{-3}$&$1.14\mltp10^{-2}$&$-$&$1.35\mltp10^{-2}$&$1.80\mltp10^{-2}$ \\                                            
\hline
\multirow{2}{*}{1400 GeV} & preselection & $4.77\mltp10^{-3}$&$1.93\mltp10^{-4}$&$5.00\mltp10^{-5}$&$2.01\mltp10^{-6}$&$1.63\mltp10^{-5}$&$8.97\mltp10^{-6}$ &$1.35\mltp10^{-6}$ &$3.53\mltp10^{-3}$ &$2.26\mltp10^{-3}$\\
&  BDT$>$0.150 &$8.36\mltp10^{-1}$&$2.02\mltp10^{-3}$&$-$&$-$&$1.44\mltp10^{-3}$&$5.71\mltp10^{-3}$&$-$&$8.17\mltp10^{-3}$&$9.11\mltp10^{-3}$ \\                                          \hline
\multirow{2}{*}{1800 GeV} & preselection & $4.78\mltp10^{-3}$&$1.93\mltp10^{-4}$&$5.00\mltp10^{-5}$&$2.01\mltp10^{-6}$&$1.63\mltp10^{-5}$&$8.97\mltp10^{-6}$ &$1.35\mltp10^{-6}$ &$3.53\mltp10^{-3}$ &$2.26\mltp10^{-3}$\\
&  BDT$>$0.150  &$8.31\mltp10^{-1}$&$5.76\mltp10^{-4}$&$-$&$-$&$1.44\mltp10^{-3}$&$5.71\mltp10^{-3}$&$-$&$3.50\mltp10^{-3}$&$3.20\mltp10^{-3}$ \\                                            
\hline
\multirow{2}{*}{2200 GeV} & preselection & $4.67\mltp10^{-3}$&$1.93\mltp10^{-4}$&$5.00\mltp10^{-5}$&$2.01\mltp10^{-6}$&$1.63\mltp10^{-5}$&$8.97\mltp10^{-6}$ &$1.35\mltp10^{-6}$ &$3.53\mltp10^{-3}$ &$2.26\mltp10^{-3}$\\
&  BDT$>$0.080&$9.27\mltp10^{-1}$&$7.20\mltp10^{-4}$&$-$&$-$&$-$&$5.71\mltp10^{-3}$&$-$&$5.01\mltp10^{-3}$&$3.20\mltp10^{-3}$ \\                                            
\hline
\multirow{2}{*}{2600 GeV} & preselection & $4.68\mltp10^{-3}$&$1.93\mltp10^{-4}$&$5.00\mltp10^{-5}$&$2.01\mltp10^{-6}$&$1.63\mltp10^{-5}$&$8.97\mltp10^{-6}$ &$1.35\mltp10^{-6}$ &$3.53\mltp10^{-3}$ &$2.26\mltp10^{-3}$\\
&  BDT$>$0.004  &$9.34\mltp10^{-1}$&$4.32\mltp10^{-4}$&$-$&$-$&$-$&$5.71\mltp10^{-3}$&$-$&$3.07\mltp10^{-3}$&$2.22\mltp10^{-3}$ \\                                            
\hline
\multirow{2}{*}{3000 GeV} & preselection & $4.61\mltp10^{-3}$&$1.93\mltp10^{-4}$&$5.00\mltp10^{-5}$&$2.01\mltp10^{-6}$&$1.63\mltp10^{-5}$&$8.97\mltp10^{-6}$ &$1.35\mltp10^{-6}$ &$3.53\mltp10^{-3}$ &$2.26\mltp10^{-3}$\\
&  BDT$>$0.003  &$9.69\mltp10^{-1}$&$8.64\mltp10^{-4}$&$-$&$-$&$-$&$5.71\mltp10^{-3}$&$-$&$4.82\mltp10^{-3}$&$5.17\mltp10^{-3}$ \\               \hline
\hline
\end{tabular}
%\end{ruledtabular}
}
\caption{
Selection efficiencies of preselection and BDT criteria for signal and background processes at the HL-LHC with $\sqrt{s} =$ 14 TeV assuming different ALP masses,
where ``$-$" means the number of events can be reduced to be negligible with $\mathcal{L} =$ 3 ab$^{-1}$.
}
\label{tab:allEfficiencies}
\end{table*}

%\newpage
\acknowledgments
We thank Bin Diao and Zilong Ding for helpful discussions.
Y.N.M. and Y.X. are supported by the National Natural Science Foundation of China under grant no.~12205227.
K.W. is supported by the National Natural Science Foundation of China under grant no.~11905162,
the Excellent Young Talents Program of the Wuhan University of Technology under grant no.~40122102, and the research program of the Wuhan University of Technology under grant no.~2020IB024.
The simulation and analysis work of this article was completed with the computational cluster provided by the Theoretical Physics Group at the Department of Physics, School of Physics and Mechanics, Wuhan University of Technology.

%%%%%%%%%%%%%%%%%%%%%%%%%%%%%%%%%%%%%%%%%

\bibliographystyle{JHEP}
\bibliography{Refs}

\end{document}